\documentclass[%
reprint,
superscriptaddress,
 amsmath,amssymb,
 aps,
]{revtex4-1}

\usepackage{graphicx}
\usepackage{float}
\usepackage{braket}
\usepackage{subcaption}
\usepackage{color, ulem}
\usepackage{gensymb}

\begin{document}

\title{Effect of bilayer stacking on the atomic and electronic structure of twisted double bilayer graphene}

\author{Xia Liang}
\affiliation{Departments of Materials and Physics and the Thomas Young Centre for Theory and Simulation of Materials, Imperial College London, South Kensington Campus, London SW7 2AZ, UK\\}
\author{Zachary A. H. Goodwin}
\affiliation{Departments of Materials and Physics and the Thomas Young Centre for Theory and Simulation of Materials, Imperial College London, South Kensington Campus, London SW7 2AZ, UK\\}
\author{Valerio Vitale}
\affiliation{Departments of Materials and Physics and the Thomas Young Centre for Theory and Simulation of Materials, Imperial College London, South Kensington Campus, London SW7 2AZ, UK\\}
\author{Fabiano Corsetti}
\affiliation{Departments of Materials and Physics and the Thomas Young Centre for Theory and Simulation of Materials, Imperial College London, South Kensington Campus, London SW7 2AZ, UK\\}
\author{Arash A. Mostofi}
\affiliation{Departments of Materials and Physics and the Thomas Young Centre for Theory and Simulation of Materials, Imperial College London, South Kensington Campus, London SW7 2AZ, UK\\}
\author{Johannes Lischner}
\affiliation{Departments of Materials and Physics and the Thomas Young Centre for Theory and Simulation of Materials, Imperial College London, South Kensington Campus, London SW7 2AZ, UK\\}

\date{\today}

\begin{abstract}
Twisted double bilayer graphene has recently emerged as an interesting moir\'e material that exhibits strong correlation phenomena that are tunable by an applied electric field. Here we study the atomic and electronic properties of three different graphene double bilayers: double bilayers composed of two AB stacked bilayers (AB/AB), double bilayers composed of two AA stacked bilayers (AA/AA) as well as heterosystems composed of one AB and one AA bilayer (AB/AA). The atomic structure is determined using classical force fields. We find that the inner layers of the double bilayer exhibit significant in-plane and out-of-plane relaxations, similar to twisted bilayer graphene. The relaxations of the outer layers depend on the stacking: atoms in AB bilayers follow the relaxations of the inner layers, while atoms in AA bilayers attempt to avoid higher-energy AA stacking. For the relaxed structures, we calculate the electronic band structures using the tight-binding method. All double bilayers exhibit flat bands at small twist angles, but the shape of the bands depends sensitively on the stacking of the outer layers. To gain further insight, we study the evolution of the band structure as the outer layers are rigidly moved away from the inner layers, while preserving their atomic relaxations. This reveals that the hybridization with the outer layers results in an additional flattening of the inner-layer flat band manifold. Our results establish AA/AA and AB/AA twisted double bilayers as interesting moir\'e materials with different flat band physics compared to the widely studied AB/AB system.
\end{abstract}

\maketitle

\section{Introduction}

Introducing a twist between two stacked graphene sheets creates a moir\'e pattern which is characterized by a spatially varying stacking configuration between the two layers~\cite{GBWT,Bistritzer12233,LDE,NSCS,PhysRevB.82.121407,Carr_2017,Tritsaris_2020}. Specifically, AA stacked regions are surrounded by AB or BA stacked regions~\cite{AC,JainSandeepK2017Sota,SETLA}. Importantly, the properties of twisted bilayer graphene (tBLG) can be controlled via the twist angle between the two layers~\cite{PDTBLG,KDP,KL,PHD_3,PTBLG,CCRPA,PHD_2,WC,ECM,EE,PHD_4,OMACM,PHD_1}. Near the first magic angle (approximately 1.1$\degree$), tBLG exhibits flat electronic bands~\cite{Bistritzer12233}, correlated insulator states and superconductivity at low temperatures~\cite{YuanCao2018Usim,YuanCao2018Ciba,TSTBLG,SOM,NAT_SS,NAT_MEI,NAT_CO,SCHFC,PCIS,SCDID}. 

Besides tBLG, other moir\'e materials have been explored, including bilayers of transition metal dichalcogenides~\cite{UFSS,TITMD,WangQing2012Eaoo,JariwalaDeep2014Edaf,KinFaiMak2016Paoo} and graphene systems consisting of more than two sheets. A prominent member of the latter class is twisted double bilayer graphene (tDBLG), which is obtained by introducing a twist between two \textit{bilayers} of graphene~\cite{KoshinoMikito2019Bsat,BIBI,haddadi2019moir,PhysRevLett.123.197702,STSCTDB,Samajdar2020}. Several groups~\cite{BIBI,cao2019electric,TCT,PhysRevLett.123.197702,RickhausPeter2019GOiT} used the tear and stack method to fabricate tDBLG consisting of two AB stacked bilayers (which we denote as AB/AB tDBLG). They observed that this system also exhibits flat bands~\cite{haddadi2019moir,ChoiY.W.2019Ibga,ChebroluNarasimha2019Fbit,CulchacF.J.2020Fbag} which can be tuned by applying an electric field. In addition, correlated insulator states in applied electric fields have been experimentally observed~\cite{BIBI,cao2019electric,TCT,AdakPratap2020Tbag,PhysRevLett.123.197702}.


To understand the electronic structure of AB/AB tDBLG, several groups calculated the band structure using the tight-binding approach~\cite{haddadi2019moir,ChoiY.W.2019Ibga,CulchacF.J.2020Fbag,RickhausPeter2019GOiT} or continuum models~\cite{haddadi2019moir,STSCTDB,ChebroluNarasimha2019Fbit,KoshinoMikito2019Bsat,WuFengcheng2020Fasi} and found flat bands near a magic angle of 1.3$\degree$, which is somewhat larger than the magic angle in tBLG~\cite{Bistritzer12233}. The flat bands exhibit nonvanishing Chern numbers suggesting that tDBLG might serve as a platform for studying the interplay of topological properties and strong electron correlations~\cite{STSCTDB}. Haddadi~\textit{et al.}~\cite{haddadi2019moir} carried out density-functional theory (DFT) calculations for AB/AB tDBLG and observed a gap opening at large twist angles, which was interpreted as the consequence of an intrinsic symmetric polarization in the system. Rickhaus~\textit{et al.}~\cite{RickhausPeter2019GOiT} verified this prediction experimentally and demonstrated that the band gap can be closed by the application of an external electric field. 

To date, all studies of tDBLG have focused on systems formed from bilayers with AB stacking, which is the lowest-energy stacking configuration~\cite{META_AAp}. However, it is also possible to fabricate bilayers with AA stacking~\cite{RakhmanovAL2012IotA, RozhkovA.V2016Epog,WangDali2012Tetc,META_AAp} which corresponds to a metastable configuration. Interestingly, AA stacked bilayer graphene exhibits a fundamentally different electronic structure to AB stacked bilayer graphene~\cite{RozhkovA.V2016Epog,TsaiSing-Jyun2012GLli,AbdullahHasanM.2018EtaK}. While the bands in the undoped AB bilayer are parabolic and touch at the Fermi level~\cite{RozhkovA.V2016Epog,EPG}, the band structure of the AA bilayer approximately consists of two copies of the monolayer band structure that are shifted in energy relative to each other~\cite{RozhkovA.V2016Epog}. In the undoped AA bilayer, the electron and hole Fermi surfaces are nested making this system unstable to phase transitions induced by electron interactions. Several studies~\cite{RakhmanovAL2012IotA,RozhkovA.V2016Epog} have analyzed the interplay of the various competing phases and suggest that undoped AA bilayer graphene should exhibit an antiferromagnetic ground state. This suggests that AA stacked bilayer graphene is an interesting building block for novel moir\'e materials.

In this paper we study the atomic and electronic structure of various tDBLG systems consisting of AA and AB bilayers. In particular, we investigate both AA/AA and AB/AB ``homodouble bilayers" as well as the AB/AA ``heterodouble bilayer". First, the atomic structure of these systems is obtained from relaxations using classical force fields. We find that in all tDLBG systems, the atomic structure of the inner two layers is similar to that of tBLG with significant in-plane and out-of-plane displacements. The atomic structure of the outer layers is influenced by that of the inner layers, but also depends on the bilayer stacking: Outer layers in AB systems display similar relaxations as the inner layers, while there are significant differences for the outer layers in AA systems as the atoms attempt to avoid the higher-energy AA stacking. Next, the electronic structure is determined using a tight-binding approach that includes the intrinsic symmetric polarization. We find that all systems exhibit flat bands, whose widths depend sensitively on the twist angle. While the flat bands in AB/AB tDBLG are separated from higher energy bands by finite-energy gaps, this is not the case in AA/AA and AA/AB tDBLG. Interestingly, the band structures of the various tDBLG systems exhibit features that are reminiscent of both the constituent bilayers and also of tBLG. To understand the flat band formation in more detail, we study the evolution of the band structure as the distance of the outer layers from the inner tBLG unit is increased rigidly while maintaining the atomic relaxation of each layer. This demonstrates that tDBLG inherits the flat bands of the inner tBLG unit and suggests that it is useful to think of tDBLG as tBLG that has been ``functionalized" by adding the outer layers. 

\section{Methods} 

We study commensurate moir\'e units of tDBLG. Similarly to tBLG~\cite{SAVINI201162,PhysRevB.81.165105}, these can be described by two integers $n$ and $m$ and the corresponding primitive moir\'e lattice vectors can be expressed in terms of $n$ and $m$ according to
\begin{equation}
\textbf{t}_1 = n\textbf{a}_1 + m\textbf{a}_2 \: ; \; \textbf{t}_2 = -m \textbf{a}_{1}+(n+m) \textbf{a}_{2},
\end{equation}
\noindent where $\textbf{a}_1 = a/2(\sqrt{3}, - 1)$ and $\textbf{a}_2 = a/2(\sqrt{3}, 1)$ denote the graphene primitive lattice vectors with $a = 2.42~\text{\AA}$ being the lattice constant of graphene~\cite{EPG}. 

Starting from tDBLG composed of flat graphene sheets, we determine the relaxed atomic structure using a classical force field model as implemented in the LAMMPS software package~\cite{LAMMPS}. To describe interactions between carbon atoms belonging to the same graphene layer, the AIREBO-morse potential~\cite{AIREBO} was used with a cutoff distance of 2.5~$\text{\AA}$~\cite{GuineaF.2019Cmft,Wijk_2015}. For interactions between carbon atoms in different graphene sheets, we used the Kolmogorov-Crespi potential~\cite{KC} with a cutoff distance of 20~$\text{\AA}$~\cite{AngeliM2018ED6s,JainSandeepK2017Sota,Wijk_2015}. The fast inertial relaxation engine (FIRE) is used for the relaxations with an energy tolerance of $10^{-10}$~eV per atom and a displacement tolerance of $10^{-7}~\text{\AA}$~\cite{PhysRevLett.97.170201}. 

To calculate the electronic band structure of tDBLG, we employ an atomistic tight-binding approach. The Hamiltonian is given by
\begin{equation}
\mathcal{\hat{H}} = \sum_{i}\epsilon_{i}\hat{c}^{\dagger}_{i}\hat{c}_{i} + \sum_{ij}[ t(\textbf{r}_{i} - \textbf{r}_{j})\hat{c}^{\dagger}_{j}\hat{c}_{i} + \text{h.c.}],
\end{equation}
\noindent where $\epsilon_{i}$ denotes the on-site energy of the $p_{z}$-orbital on carbon atom $i$ at position $\textbf{r}_{i}$ and $\hat{c}^{\dagger}_{i}$ ($\hat{c}_{i}$) creates (annihilates) an electron in this orbital. Spin indices have been suppressed for clarity of notation. The hopping parameter $t(\textbf{r}_{i} - \textbf{r}_{j})$ between atoms $i$ and $j$ is calculated using the Slater-Koster rules~\cite{SK}
\begin{equation}
t(\textbf{r}) = V_{pp\sigma}(\textbf{r})\bigg(\dfrac{\textbf{r}\cdot\hat{\textbf{e}}_{z}}{|\textbf{r}|}\bigg)^{2} + V_{pp\pi}(\textbf{r})\bigg(1 - \dfrac{\textbf{r}\cdot\hat{\textbf{e}}_{z}}{|\textbf{r}|}\bigg)^{2},
\end{equation}
with $V_{pp\sigma}(\textbf{r}) = V_{pp\sigma}^{0}\exp\{q_{\sigma}(1 - |\textbf{r}|/d_{\textrm{AB}})\}\Theta(R_\textrm{c}-|\mathbf{r}|)$ and $V_{pp\pi}(\textbf{r}) = V_{pp\pi}^{0}\exp\{q_{\pi}(1 - |\textbf{r}|/a^{\prime})\}\Theta(R_\textrm{c}-|\mathbf{r}|)$. We use the following parameters: $V_{pp\sigma}^{0} = 0.48$ eV, which is the $\sigma$ hopping between $p_{z}$-orbitals, and $V_{pp\pi}^{0} = -2.7$ eV for the $\pi$ hopping ~\cite{FC,SK,EPG}; $a^{\prime} = 1.397~\text{\AA}$ is the carbon-carbon bond length and $d_{\textrm{AB}} = 3.35~\text{\AA}$ is the interlayer spacing of an AB stacked bilayer graphene. Finally, $q_{\sigma} = 7.43$ and $q_{\pi} = 3.14$ are dimensionless the decay parameters~\cite{LDE,NSCS}. Hoppings between carbon atoms whose distance is larger than the cutoff $R_\textrm{c}=10~\text{\AA}$ are neglected~\cite{AngeliM2018ED6s}.

To obtain tight-binding band structures that agree with DFT calculations of AB/AB stacked tDBLG~\cite{RickhausPeter2019GOiT,haddadi2019moir} an on-site energy of -30~meV must be included for all atoms in the inner layers. We use the same on-site energy for the calculations of AA/AA and AB/AA tDBLG. This is reasonable as charge transfer, implied by the presence of an intrinsic symmetric polarization~\cite{haddadi2019moir}, is driven by the work function difference between the central tBLG unit and the outer graphene monolayers and should only weakly depend on the outer layer stacking. To assess the accuracy of this assumption, we have compared the resulting tight-binding band structures with \textit{ab initio} DFT results (see Appendix A) and found good agreement.

We also present results for the density of states (DOS) of tDBLG. For all systems, the DOS was calculated using a 31$\times$31 Monkhorst-Pack kpoint grid centered at the $\Gamma$ point of the first Brillouin zone. To obtain smooth curves, a Gaussian broadening was employed that was optimized for each twist angle. The values of the Gaussian broadening parameters are as follows: $\sigma_g$ = 4~meV for $\theta$ = 2.45$\degree$, $\sigma_g$ = 3~meV for $\theta$ = 2.13$\degree$, $\sigma_g$ = 2.5~meV for $\theta$ = 1.89$\degree$, and $\sigma_g$ = 1.6~meV for $\theta$ = 1.70$\degree$.

\begin{figure*}[ht]
    \centering
    \begin{subfigure}[b]{0.425\textwidth}
        \centering
        \includegraphics[width=\textwidth]{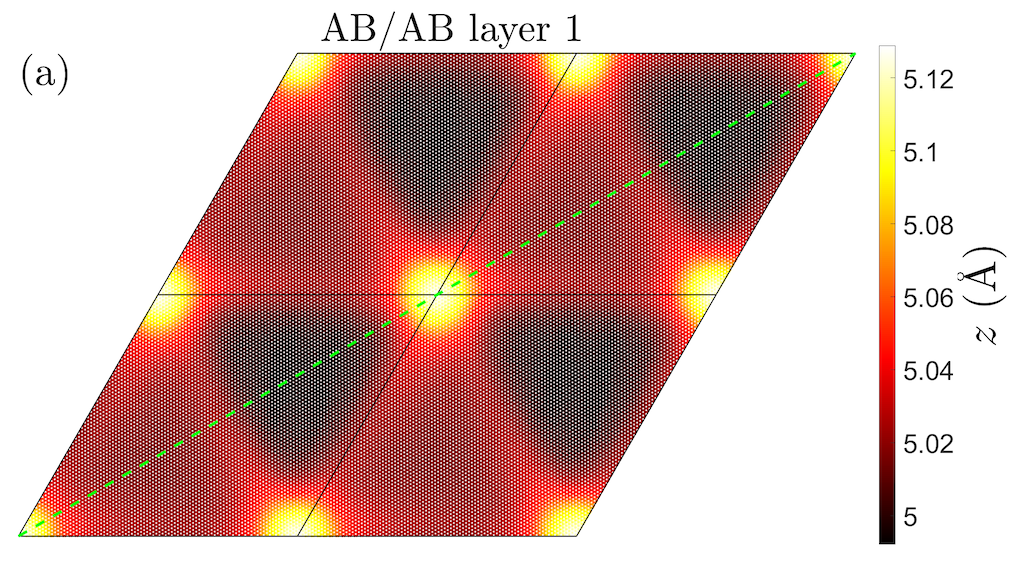}
    \end{subfigure}
    \begin{subfigure}[b]{0.425\textwidth}  
        \centering 
        \includegraphics[width=\textwidth]{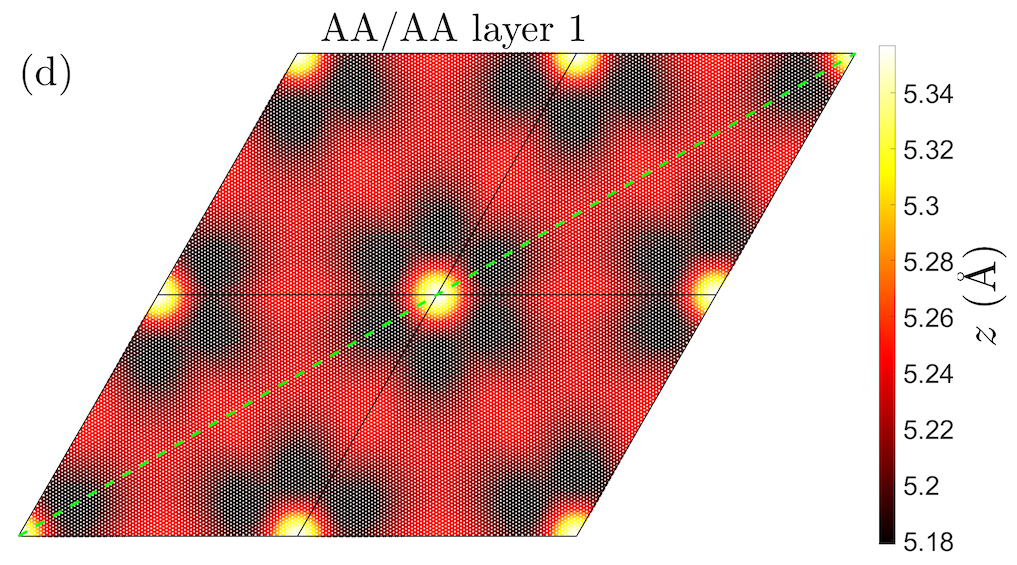}
    \end{subfigure}
    \hfill
    \vskip\baselineskip
    \begin{subfigure}[b]{0.425\textwidth}  
        \centering 
        \includegraphics[width=\textwidth]{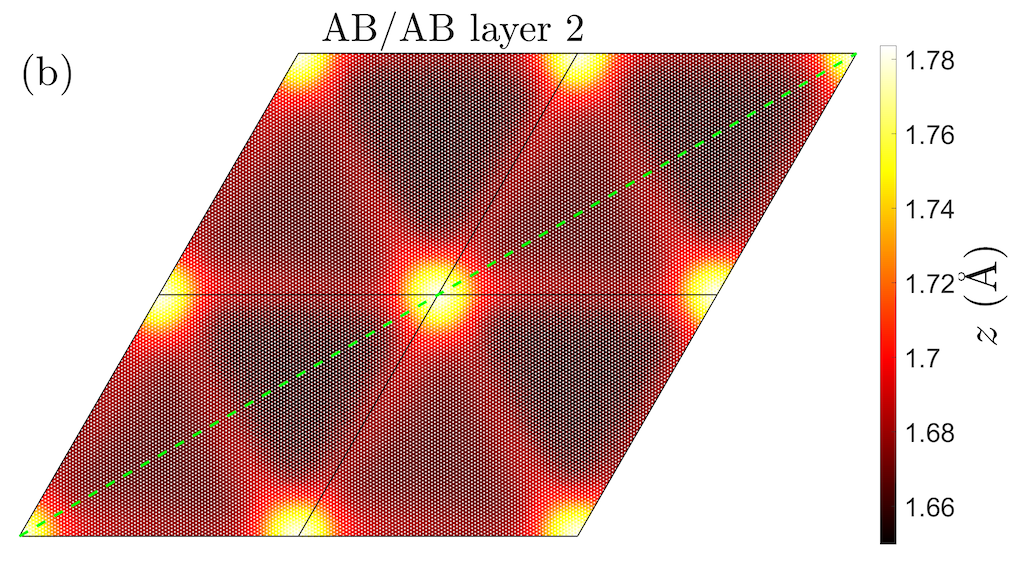}
    \end{subfigure}
    \begin{subfigure}[b]{0.425\textwidth}   
        \centering 
        \includegraphics[width=\textwidth]{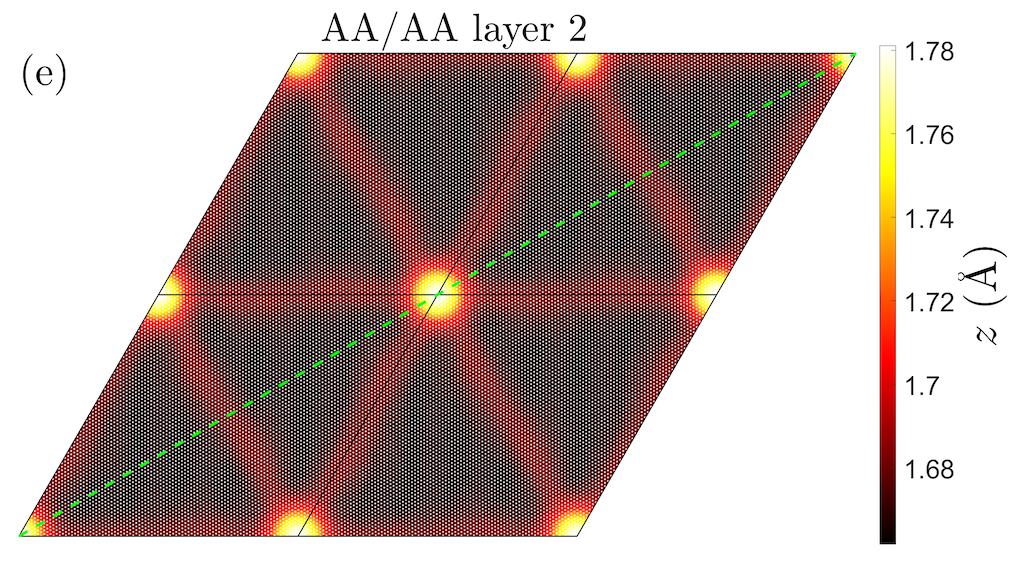}
    \end{subfigure}
    \vskip\baselineskip
    \begin{subfigure}[b]{0.425\textwidth}   
        \centering 
        \includegraphics[width=\textwidth]{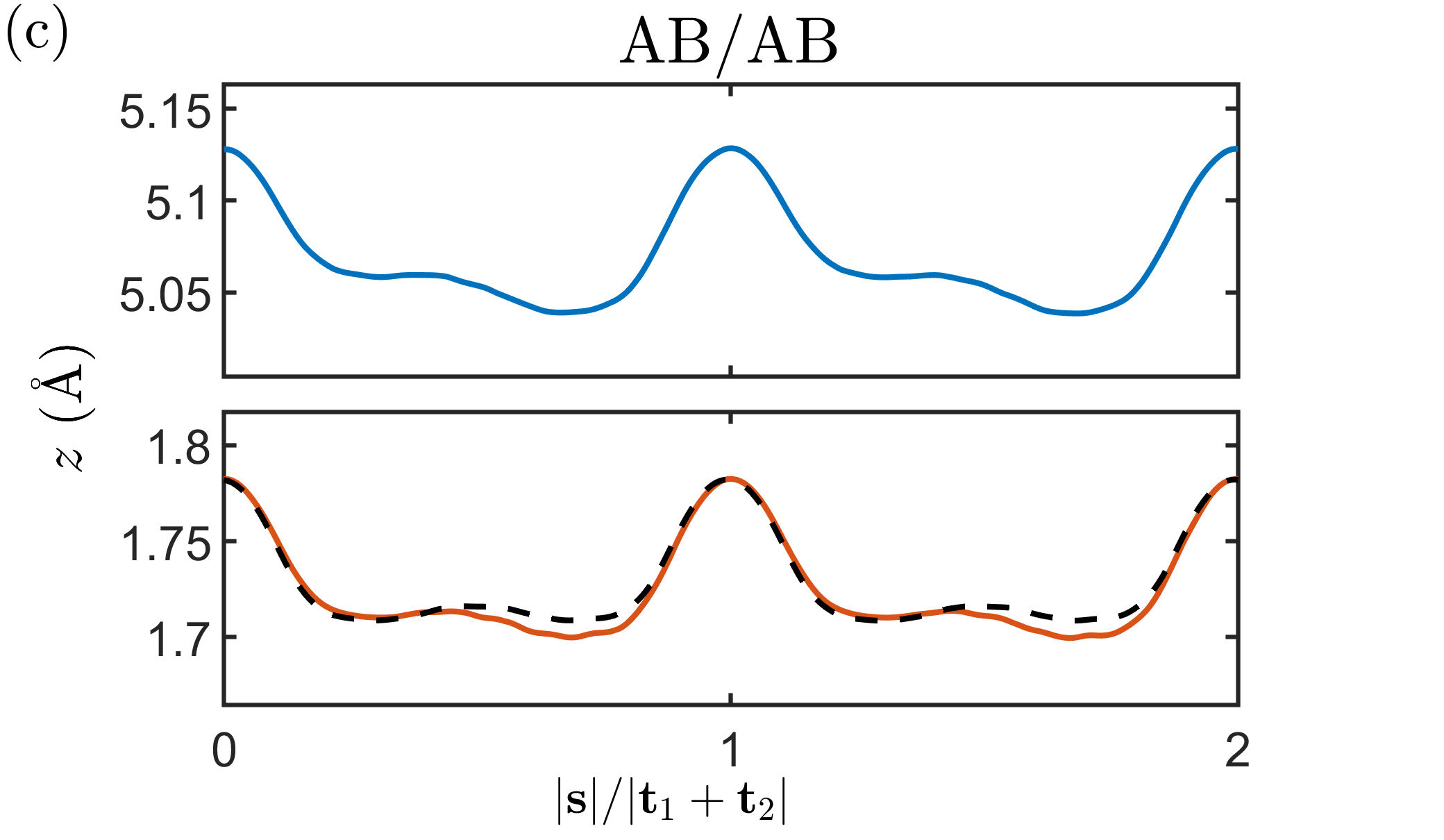}
    \end{subfigure}
    \begin{subfigure}[b]{0.425\textwidth}  
        \centering 
        \includegraphics[width=\textwidth]{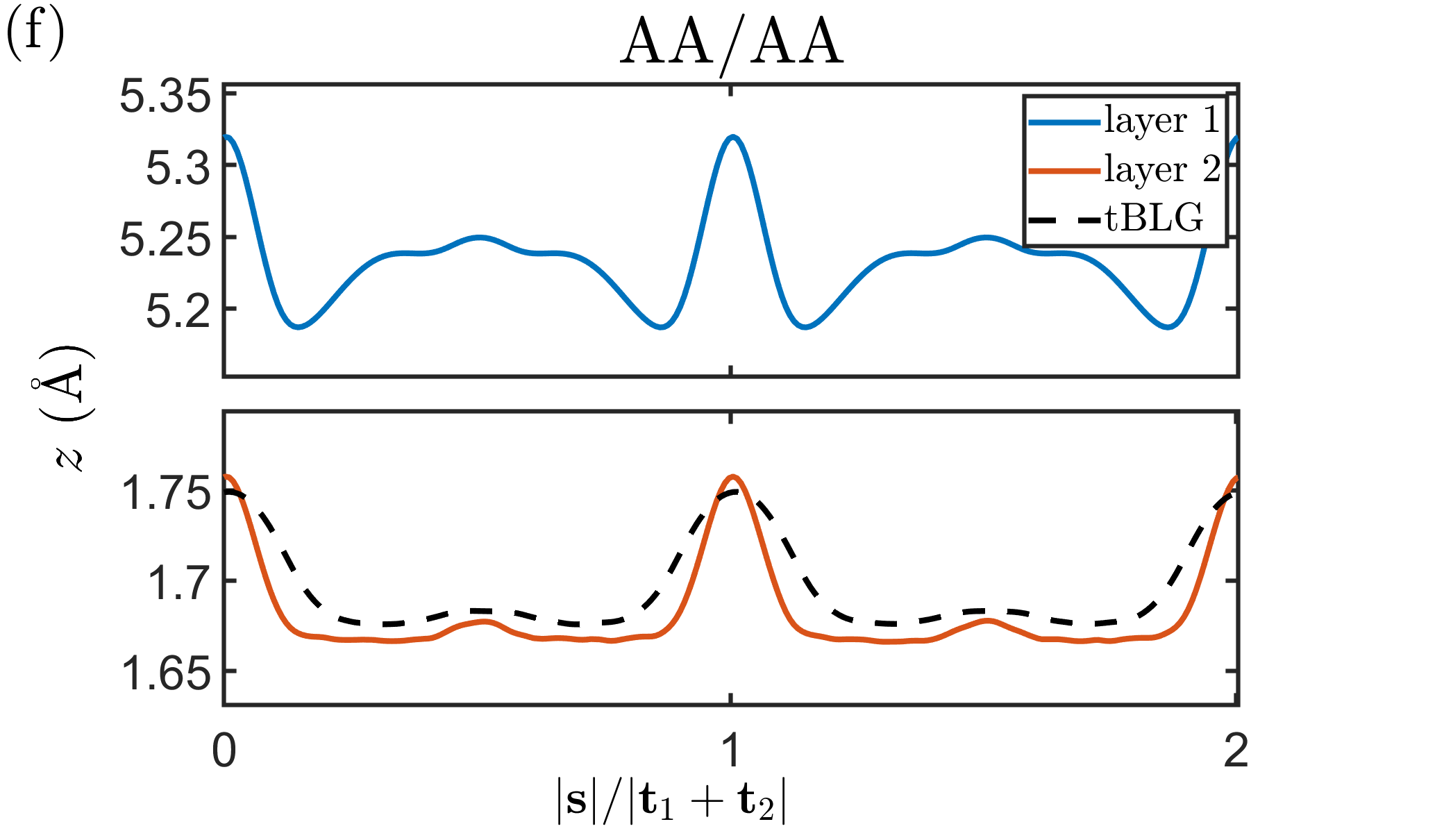}
    \end{subfigure}
    \caption{Out-of-plane atomic displacements of tDBLG at $\theta=0.73\degree$ measured as the distance from the central plane (whose $z$ coordinate is equal to the average $z$ coordinate of all atoms in the unit cell). (a) Atoms in layer 1 of AB/AB tDBLG; (b) atoms in layer 2 of AB/AB tDBLG; (c) out-of-plane displacements along the moir\'e cell diagonal [indicated by the dashed green line in (a) and (b) corresponding to $\mathbf{s}=\alpha(\mathbf{t}_1+\mathbf{t}_2)$ with $\alpha$ ranging from 0 to 2] for atoms in layer 1 (upper subplot) and layer 2 (lower subplot); the black dashed line shows the out-of-plane displacement of the top layer in tBLG. [(d)-(f)] Analogous plots for AA/AA tDBLG.} 
    \label{Z-surface}
\end{figure*}

\begin{figure*}[ht]
    \centering
    \begin{subfigure}[b]{0.425\textwidth}
        \centering
        \includegraphics[width=\textwidth]{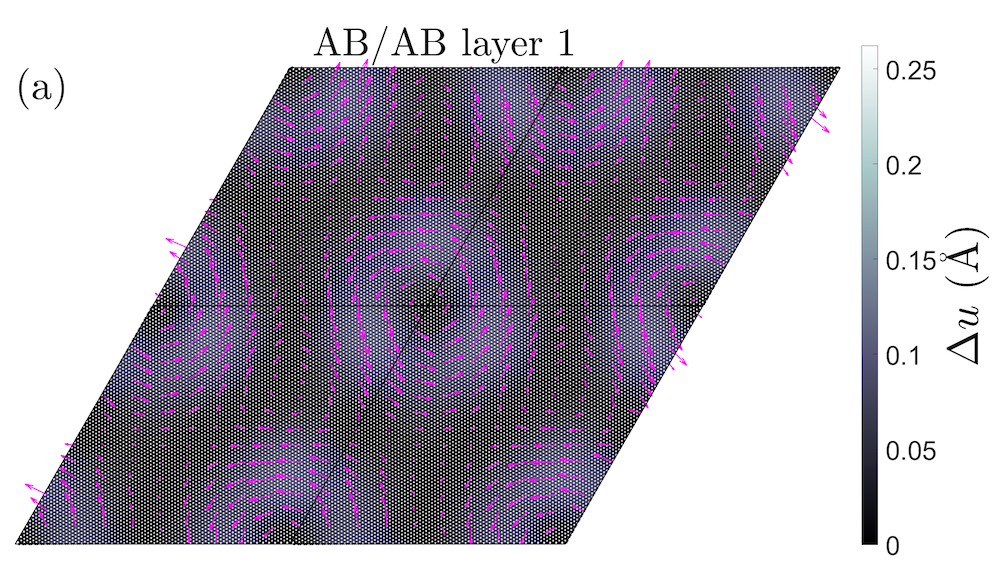}
    \end{subfigure}
    \begin{subfigure}[b]{0.425\textwidth}  
        \centering 
        \includegraphics[width=\textwidth]{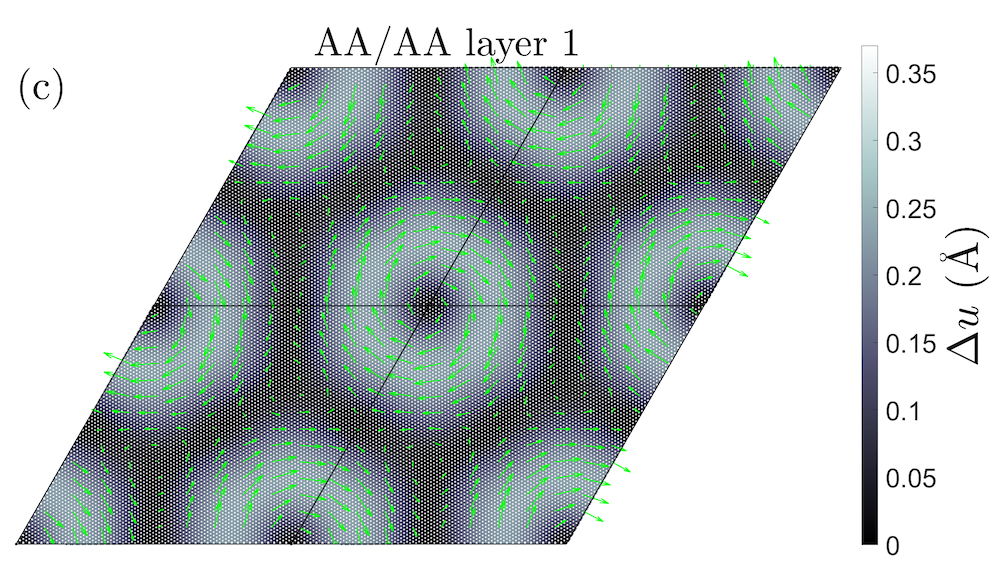}
    \end{subfigure}
    \vskip\baselineskip
    \begin{subfigure}[b]{0.425\textwidth}   
        \centering 
        \includegraphics[width=\textwidth]{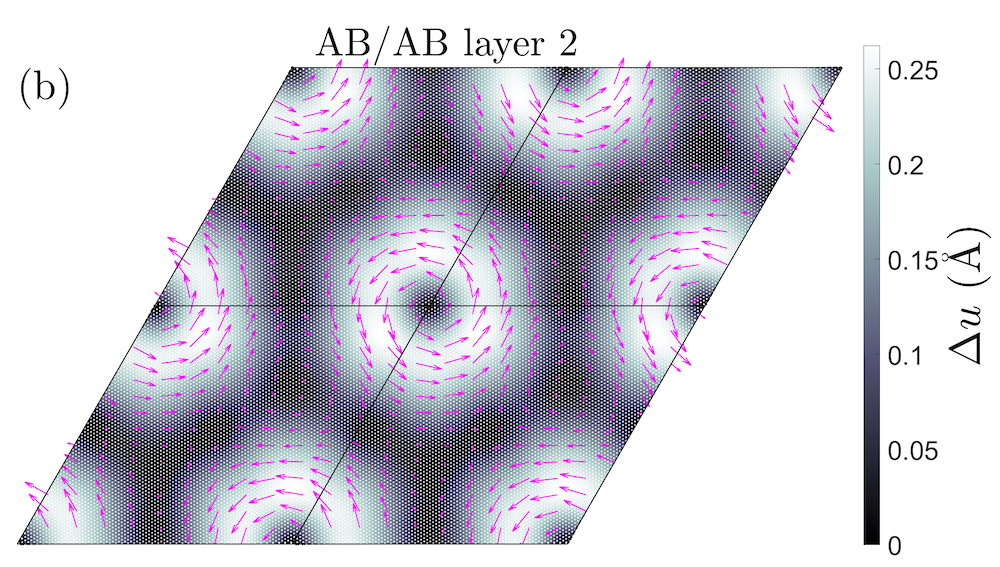}
    \end{subfigure}
    \begin{subfigure}[b]{0.425\textwidth}   
        \centering 
        \includegraphics[width=\textwidth]{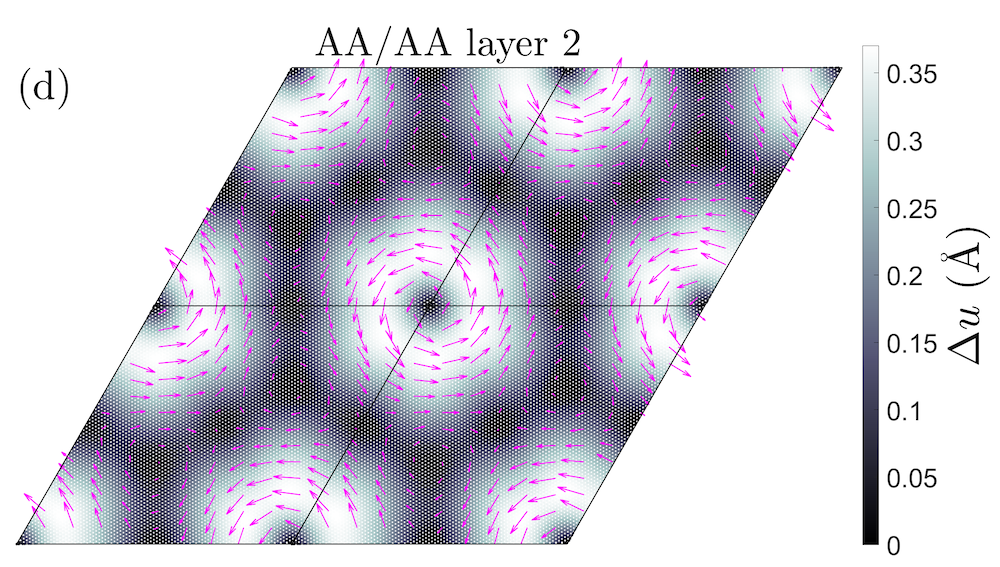}
    \end{subfigure}
    \caption{Atomic in-plane displacements of tDBLG with a twist angle of $\theta=0.73\degree$. The arrows connect the initial unrelaxed atomic positions to the final relaxed ones. Purple arrows indicate an anticlockwise rotation around the AA regions, while green arrows indicate a clockwise rotation. (a) Layer 1 of AB/AB tDBLG, (b) layer 2 of AB/AB tDBLG. [(c) and (d)] Analogous plots for AA/AA tDBLG.} 
    \label{inplane-disp}
\end{figure*}

\begin{figure*}[ht]
    \centering
    \begin{subfigure}[b]{0.425\textwidth}
        \centering
        \includegraphics[width=\textwidth]{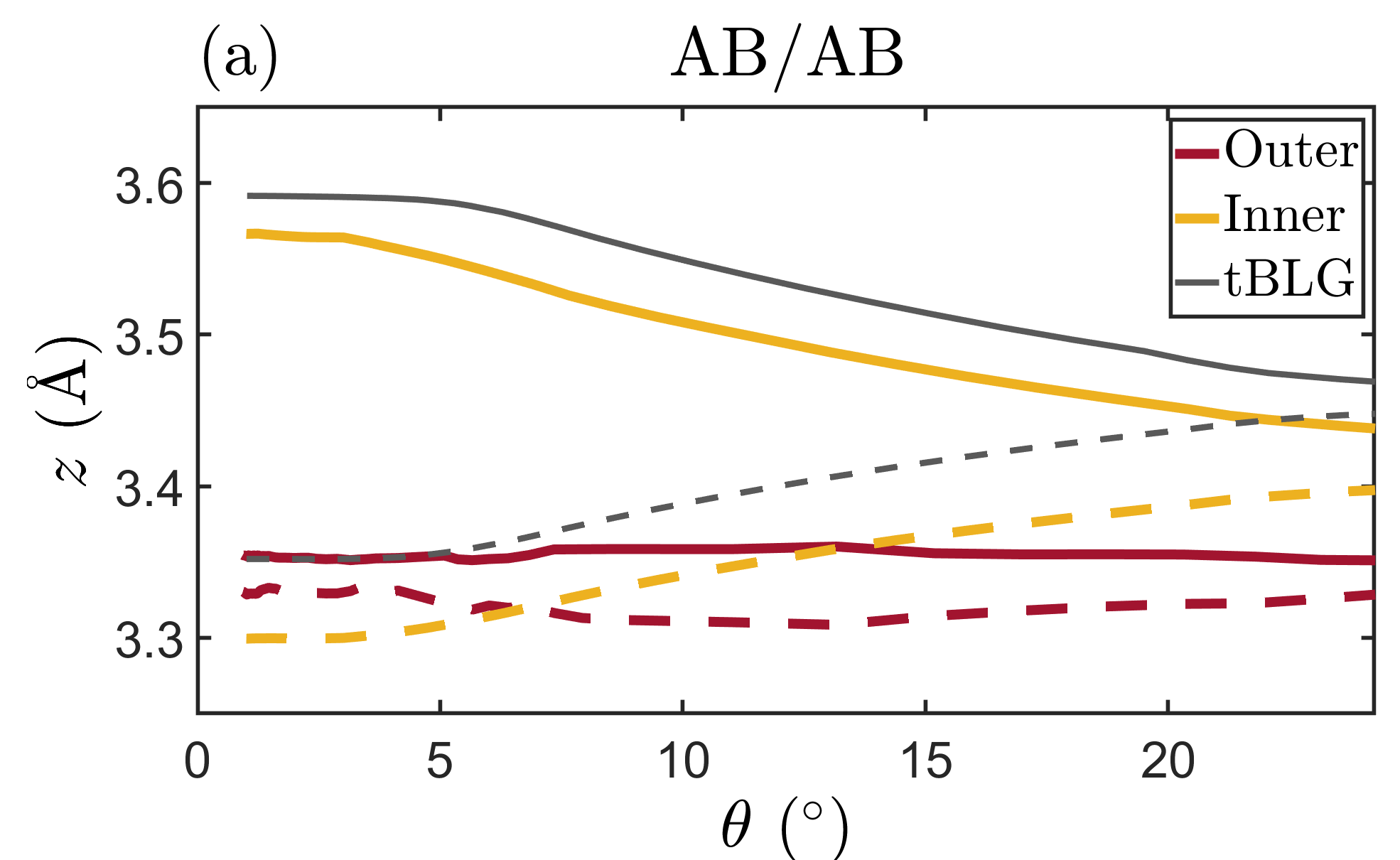}
    \end{subfigure}
    \begin{subfigure}[b]{0.425\textwidth}  
        \centering 
        \includegraphics[width=\textwidth]{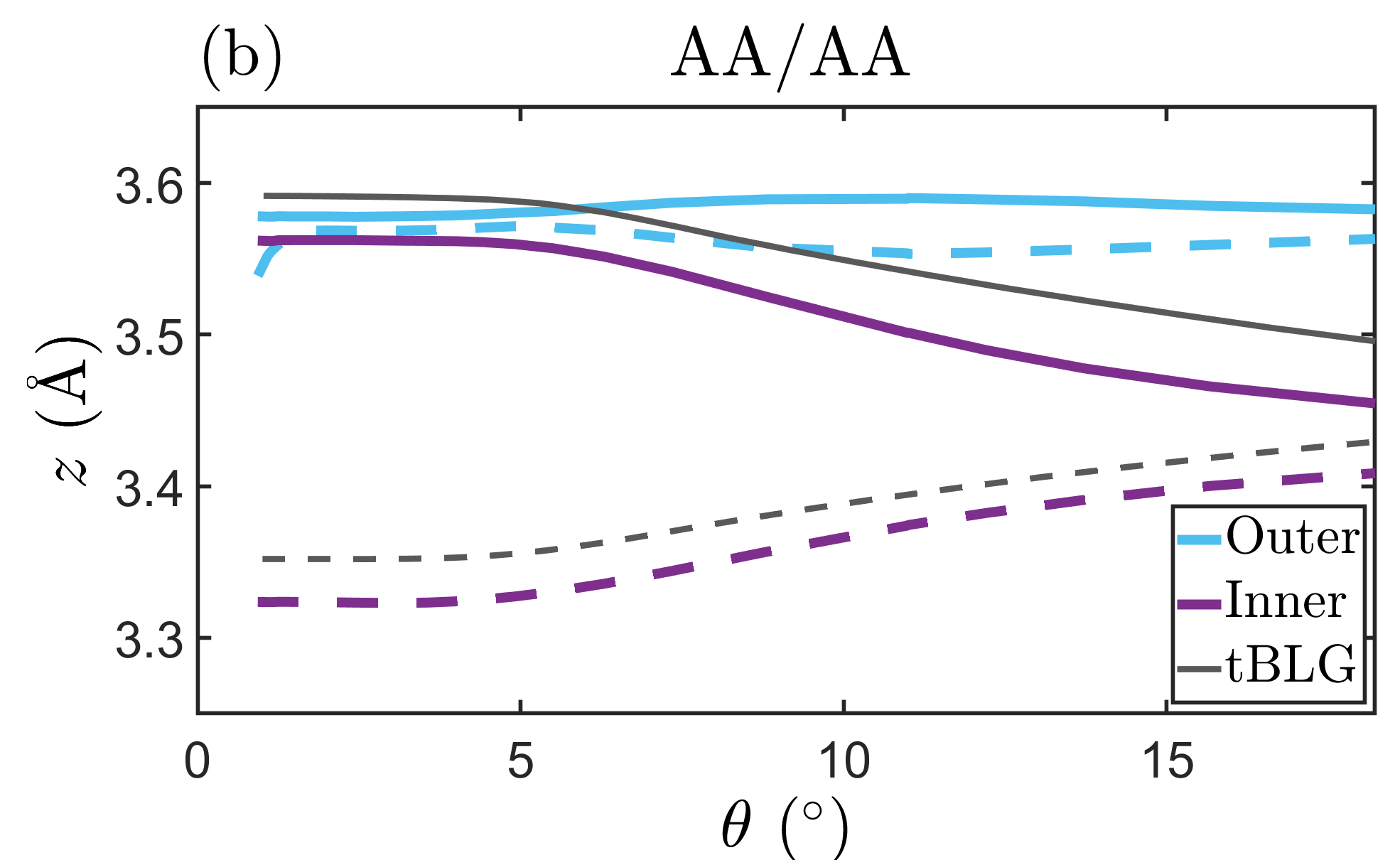}
    \end{subfigure}
    \vskip\baselineskip
    \begin{subfigure}[b]{0.425\textwidth}   
        \centering 
        \includegraphics[width=\textwidth]{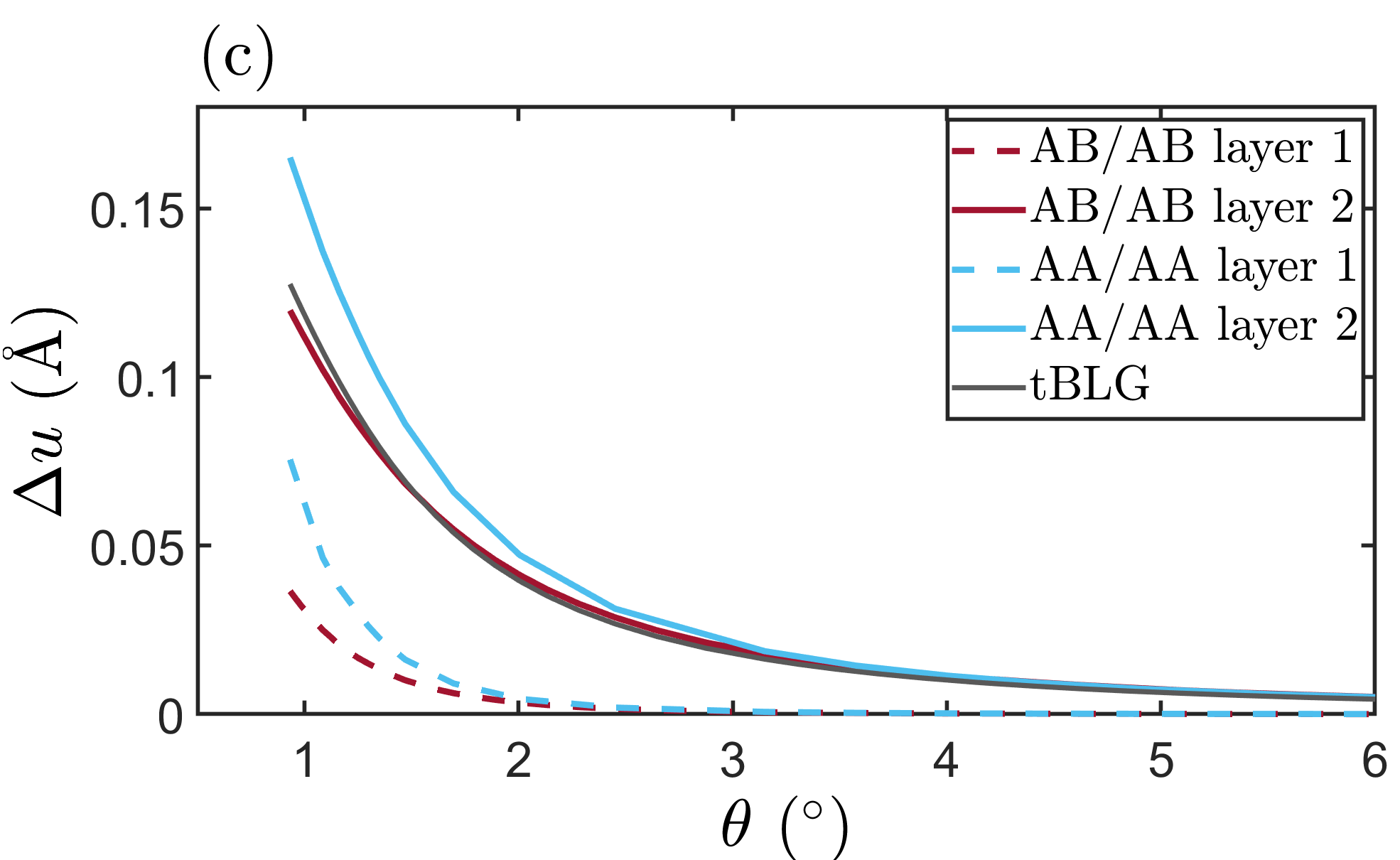}
    \end{subfigure}
    \begin{subfigure}[b]{0.425\textwidth}  
        \centering 
        \includegraphics[width=\textwidth]{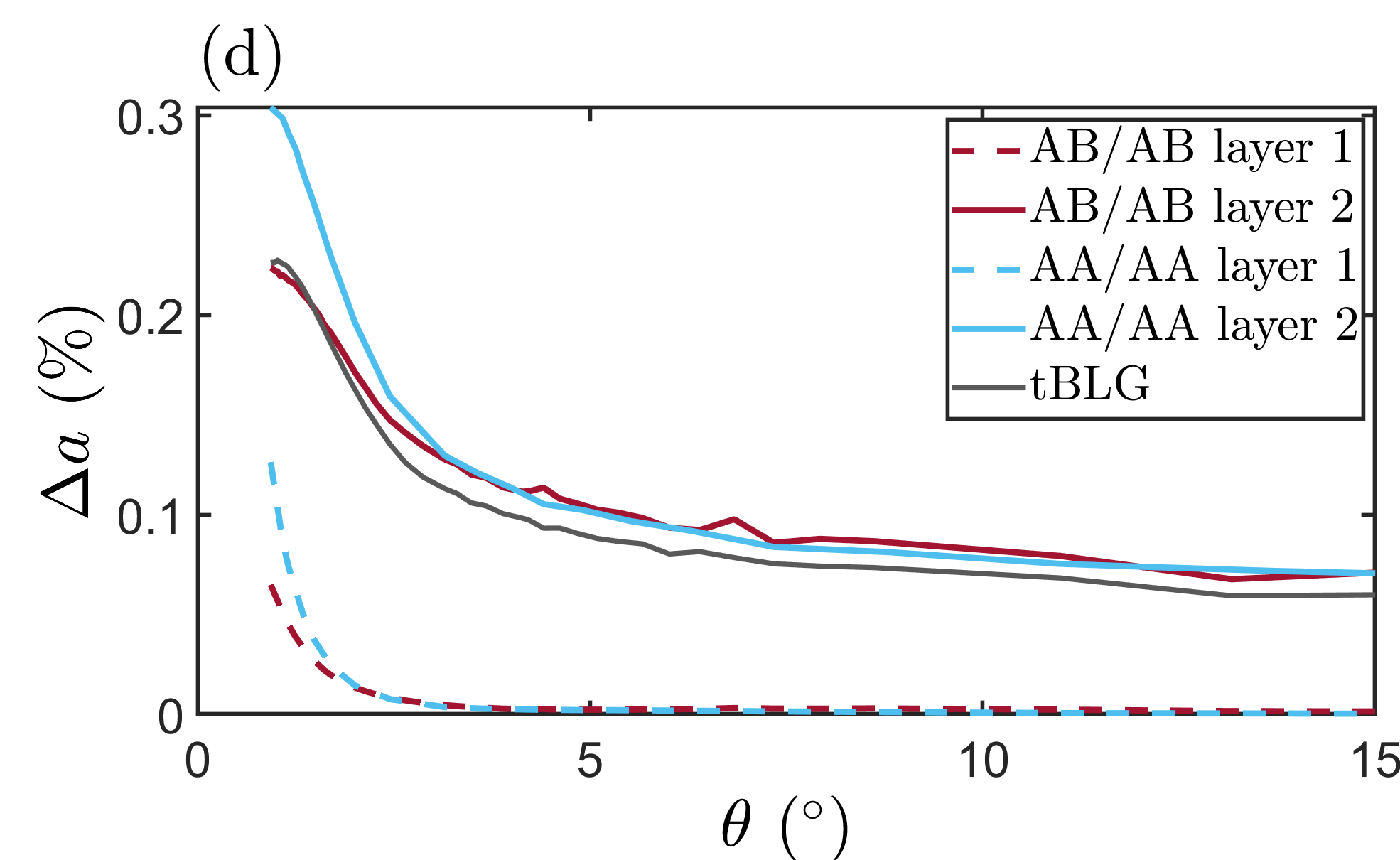}
    \end{subfigure}
    \caption{Twist angle dependence of atomic relaxations in tDBLG. (a) Maximum (solid lines) and minimum (dashed lines) interlayer separations as a function of twist angle for AB/AB tDBLG (between outer layers shown in red and between inner layers in yellow) and tBLG (black). (b) Analogous plot for AA/AA stacked tDBLG with interlayer separation between outer layers shown in blue and between inner layers in purple. (c) Average in-plane displacement, $\Delta u$, as a function of twist angle for AB/AB tDBLG, AA/AA tDBLG, and tBLG. (d) Difference between the maximum and minimum carbon-carbon bond lengths, $\Delta a$, in percentage of the equilibrium bond length as a function of twist angle.}
    \label{Theta}
\end{figure*}

\begin{figure*}[ht]
\begin{subfigure}{0.36780856\textwidth}
  \centering
  \includegraphics[width=1\linewidth]{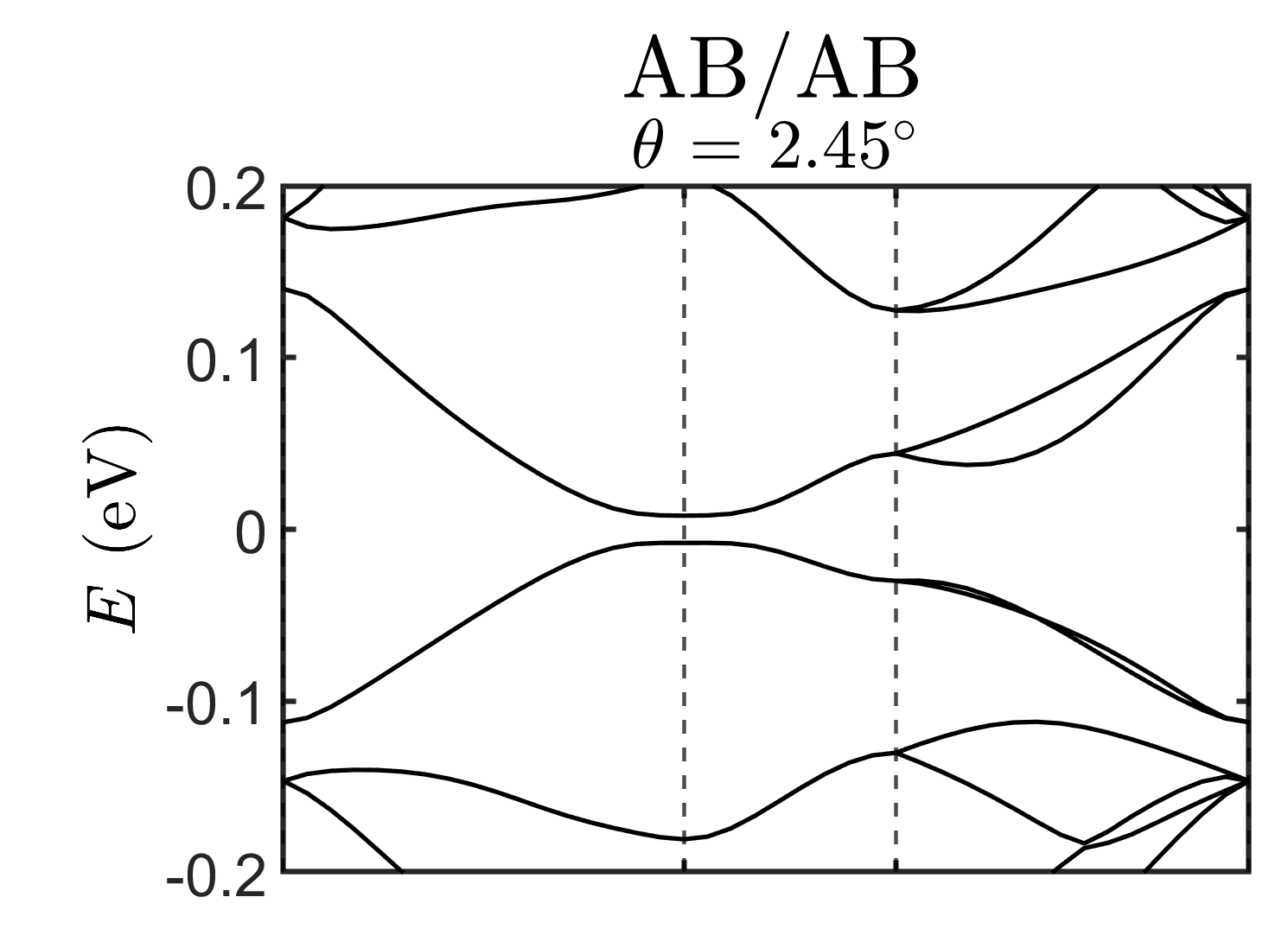}
\end{subfigure}
\begin{subfigure}{0.3060957\textwidth}
  \centering
  \includegraphics[width=1\linewidth]{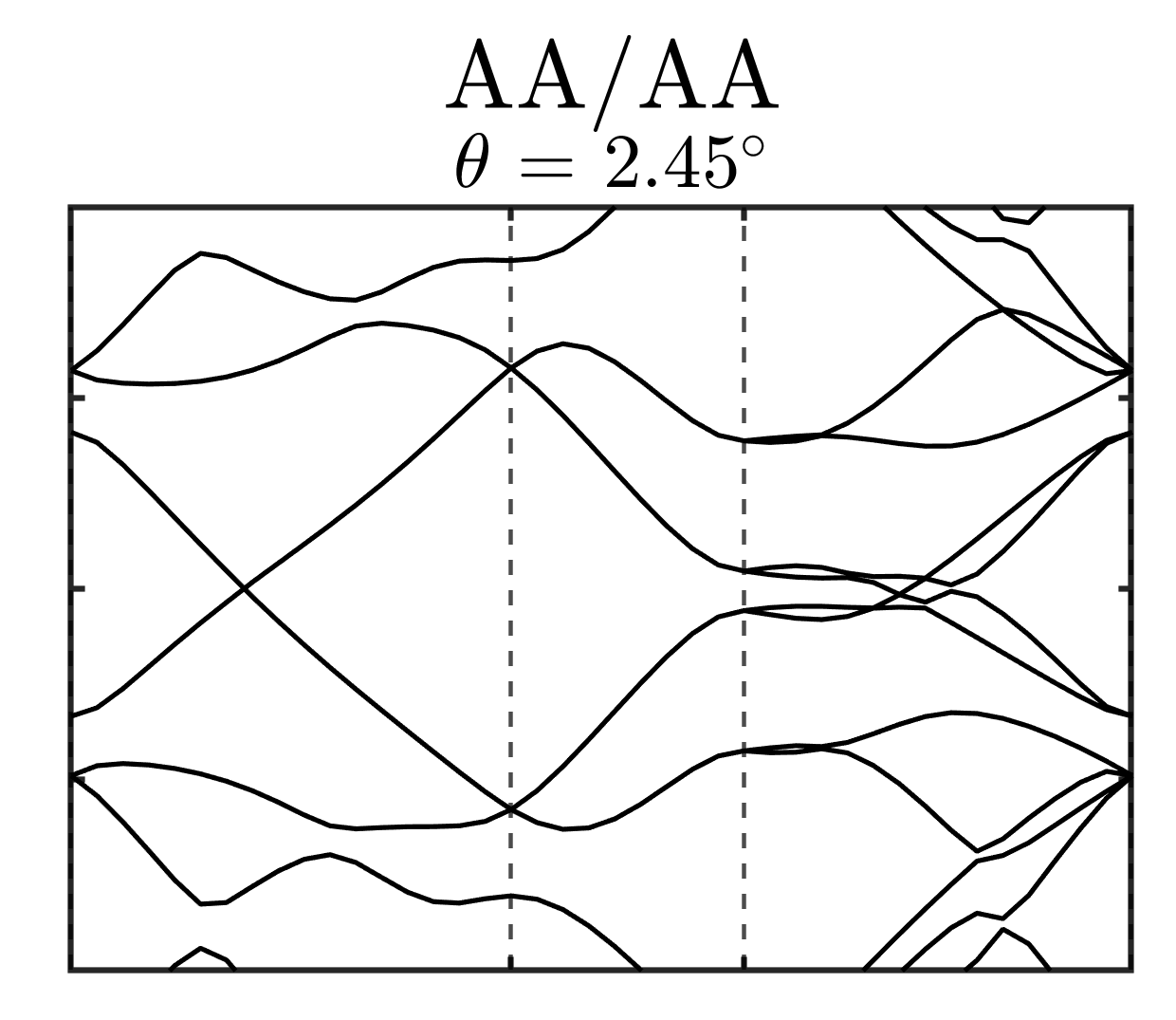}
\end{subfigure}
\begin{subfigure}{0.3060957\textwidth}
  \centering
  \includegraphics[width=1\linewidth]{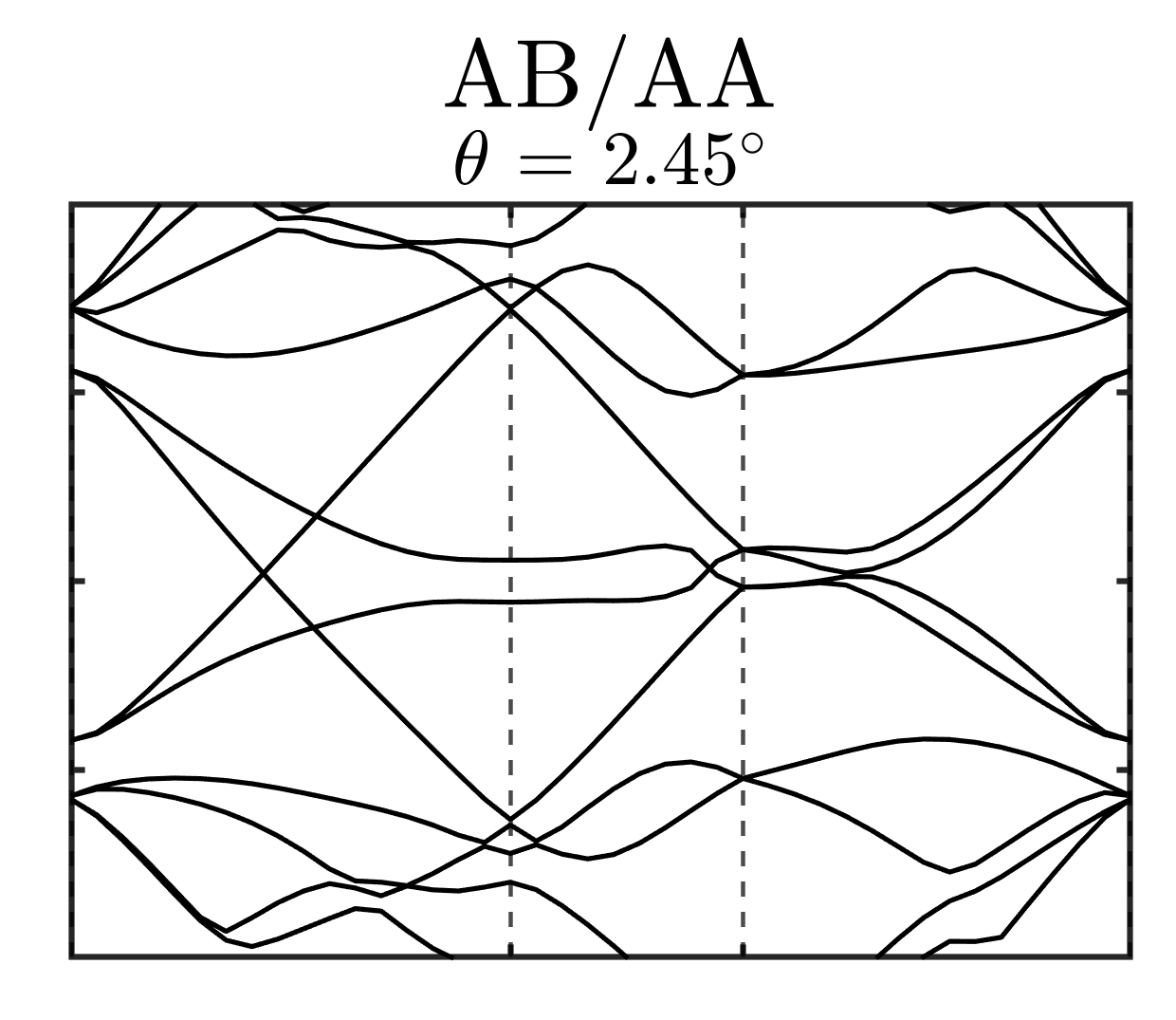}
\end{subfigure}
\begin{subfigure}{0.36780856\textwidth}
  \centering
  \includegraphics[width=1\linewidth]{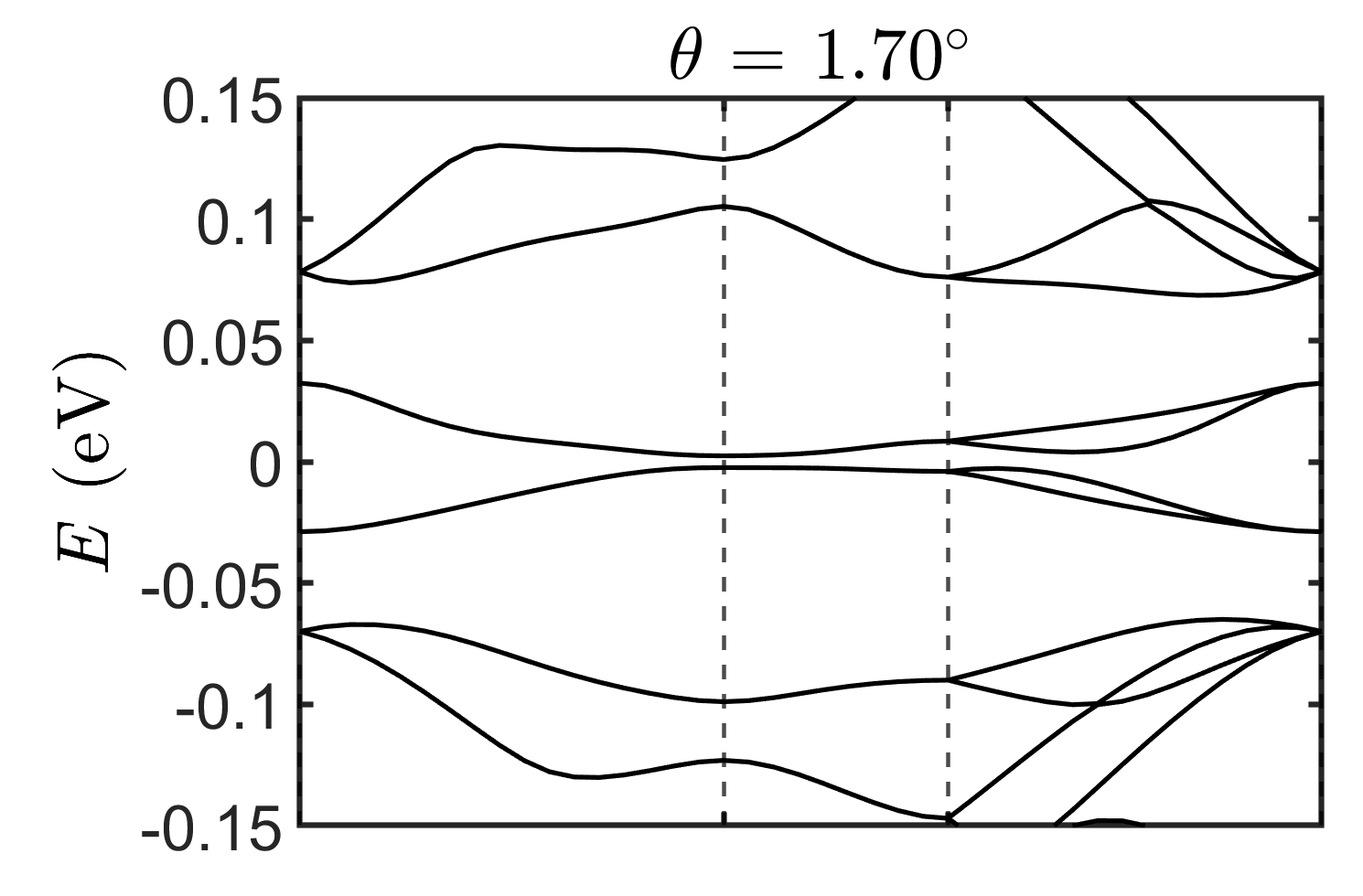}
\end{subfigure}
\begin{subfigure}{0.3060957\textwidth}
  \centering
  \includegraphics[width=1\linewidth]{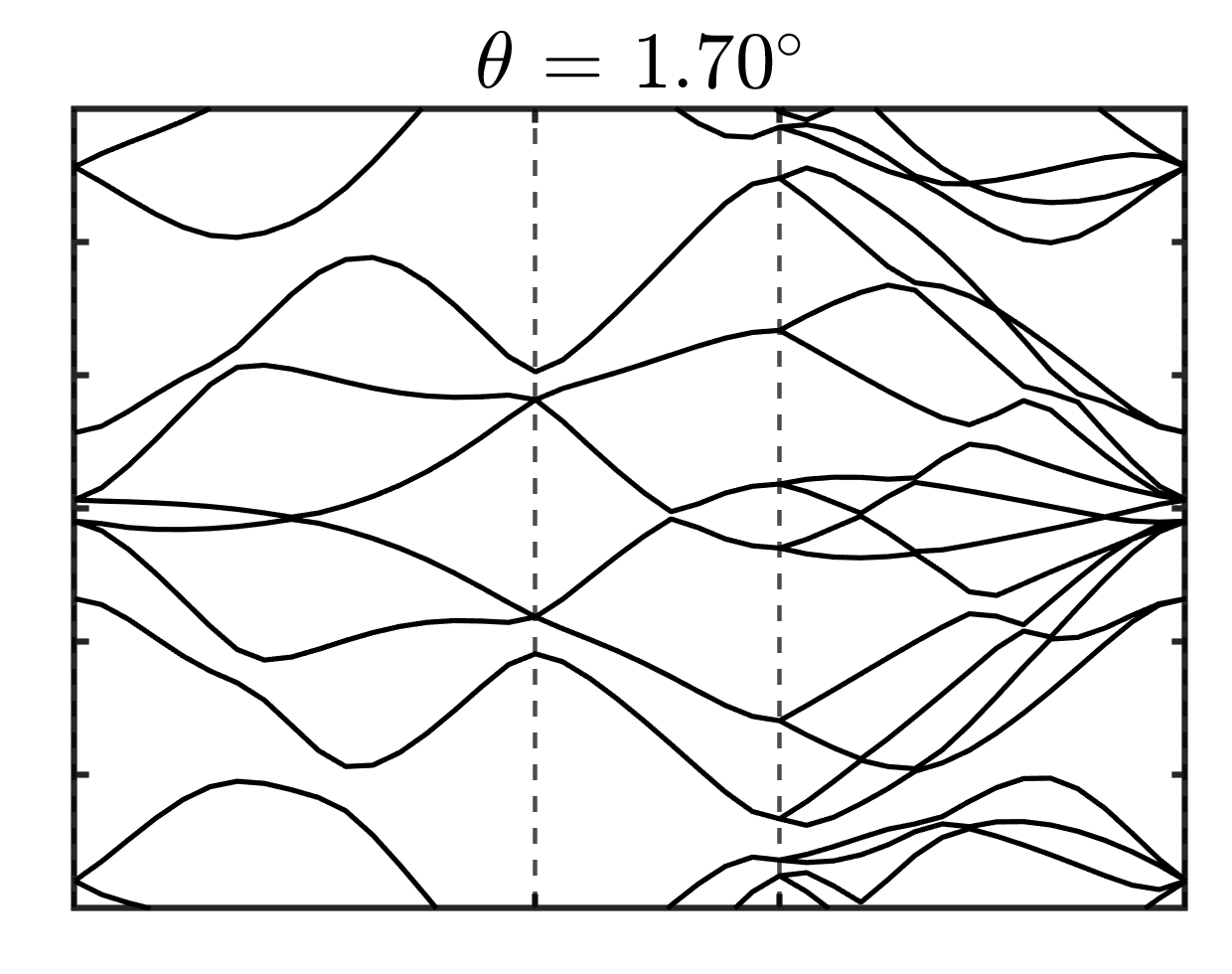}
\end{subfigure}
\begin{subfigure}{0.3060957\textwidth}
  \centering
  \includegraphics[width=1\linewidth]{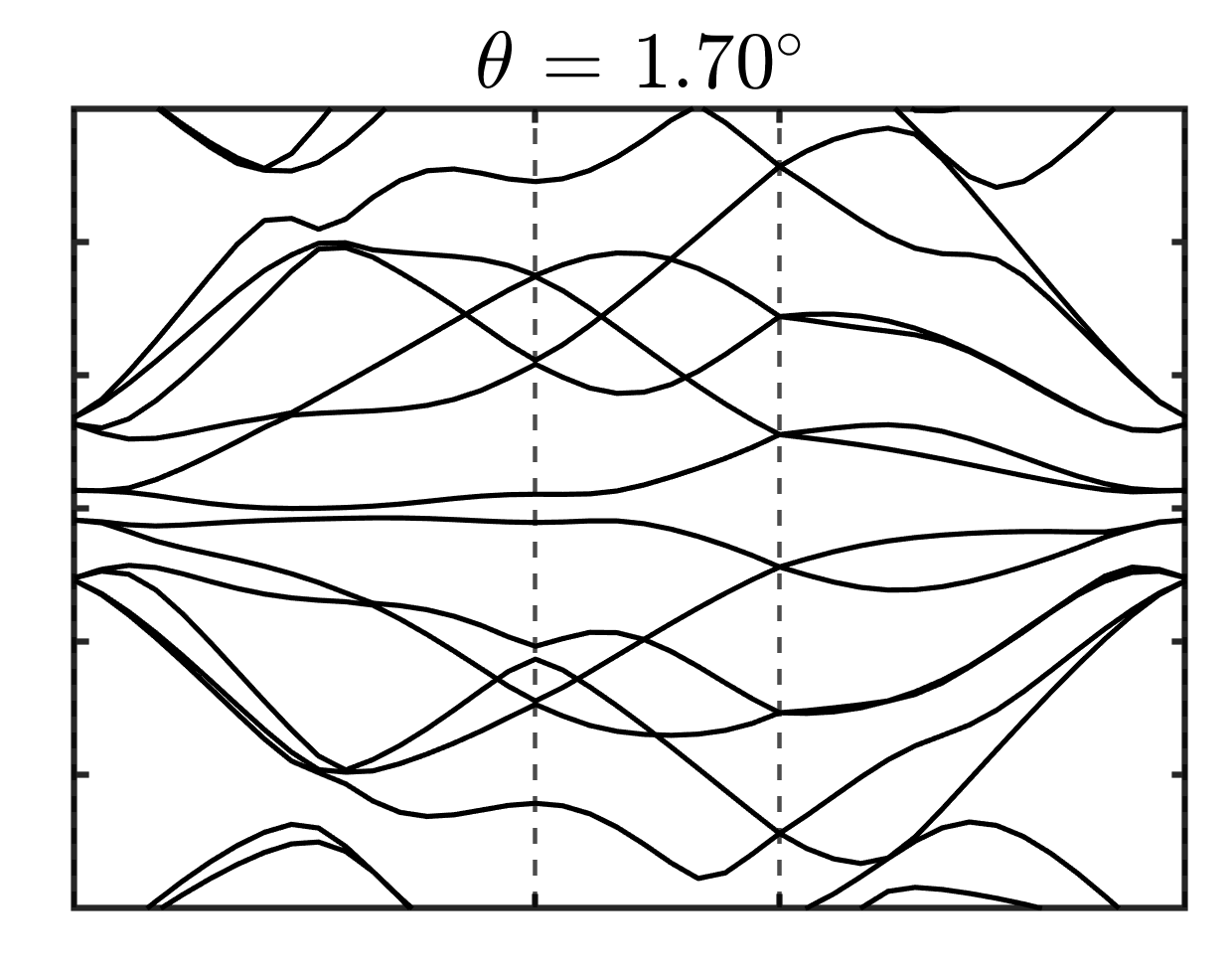}
\end{subfigure}
\begin{subfigure}{0.36780856\textwidth}
  \centering
  \includegraphics[width=1\linewidth]{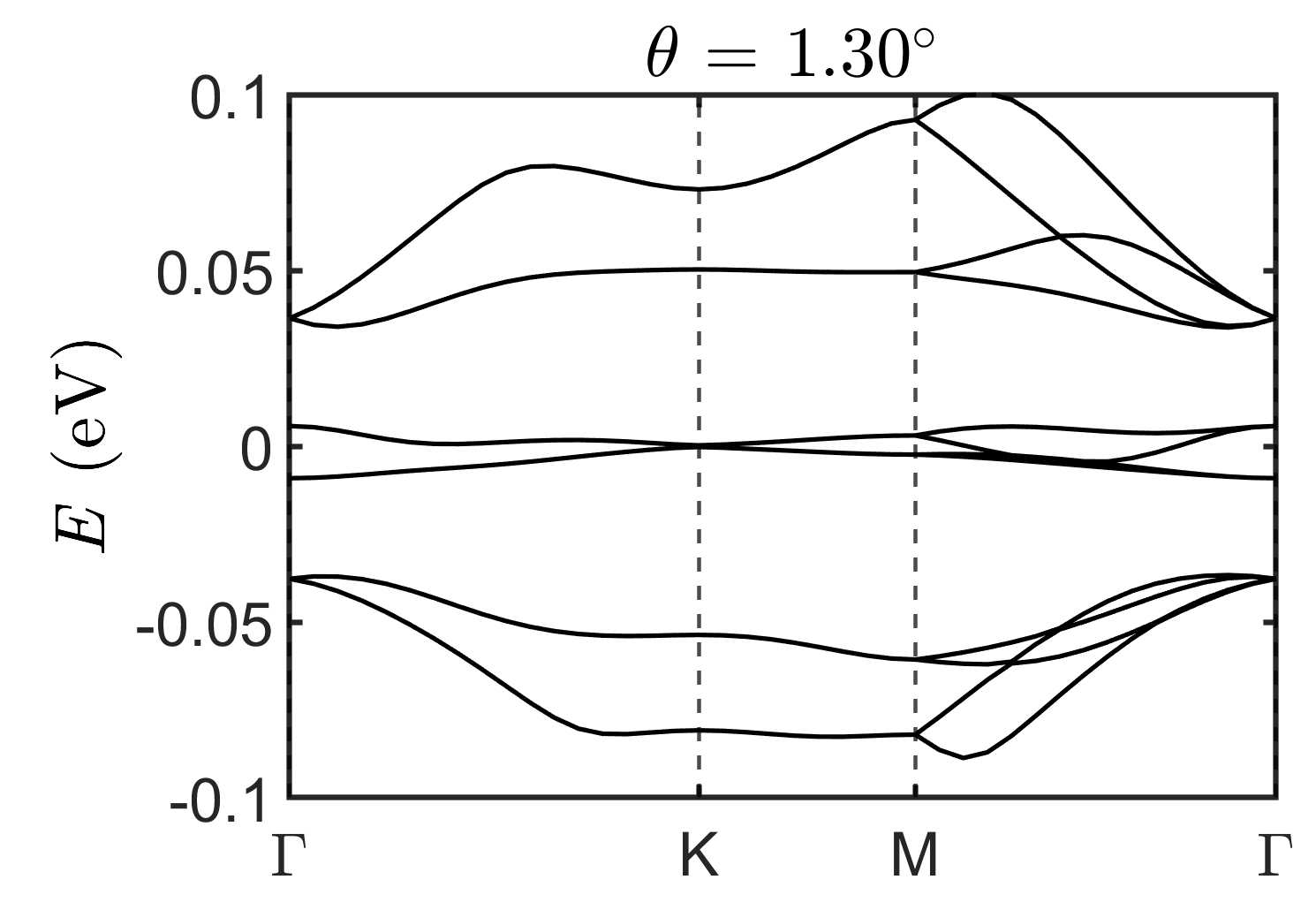}
\end{subfigure}
\begin{subfigure}{0.3060957\textwidth}
  \centering
  \includegraphics[width=1\linewidth]{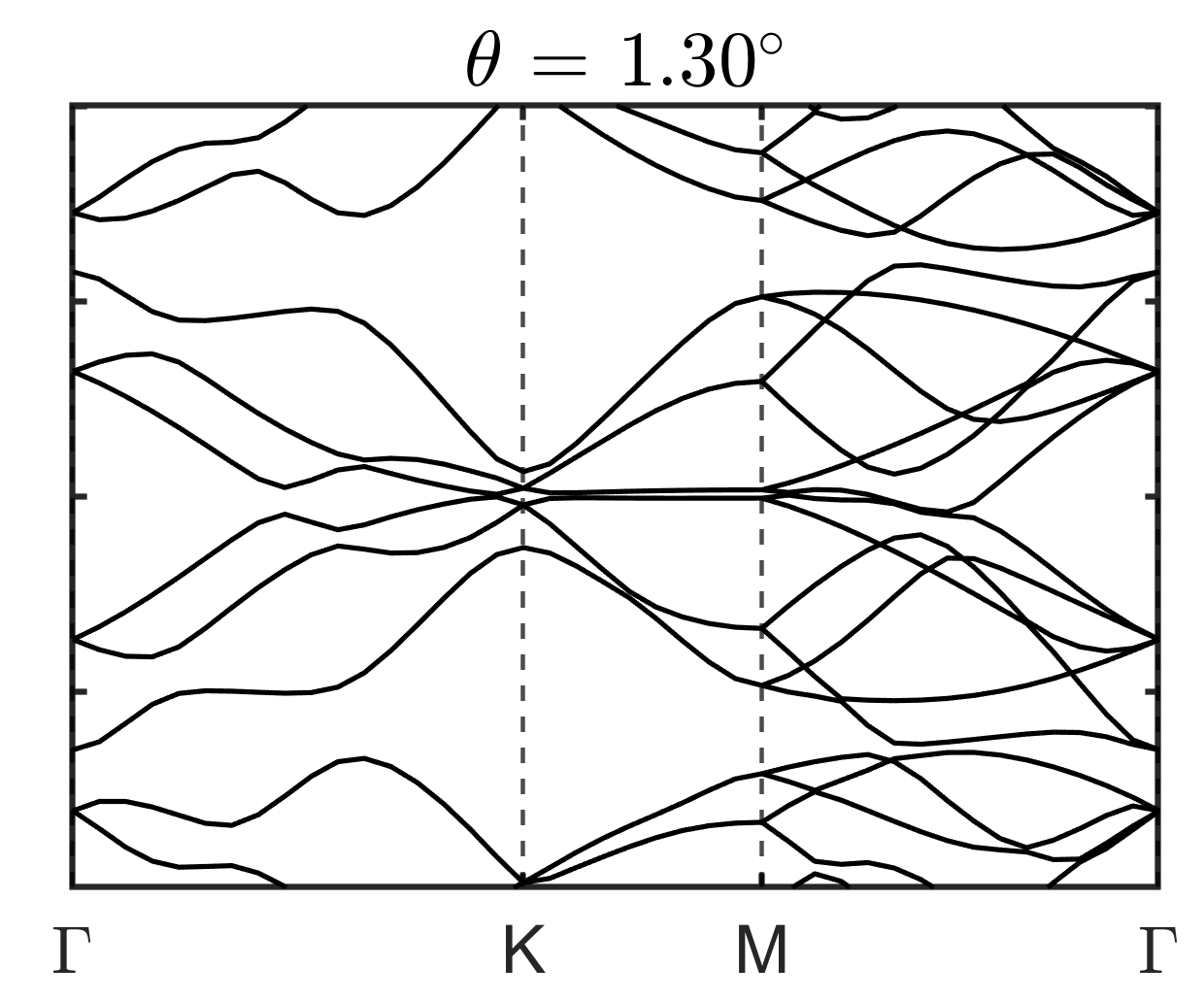}
\end{subfigure}
\begin{subfigure}{0.3060957\textwidth}
  \centering
  \includegraphics[width=1\linewidth]{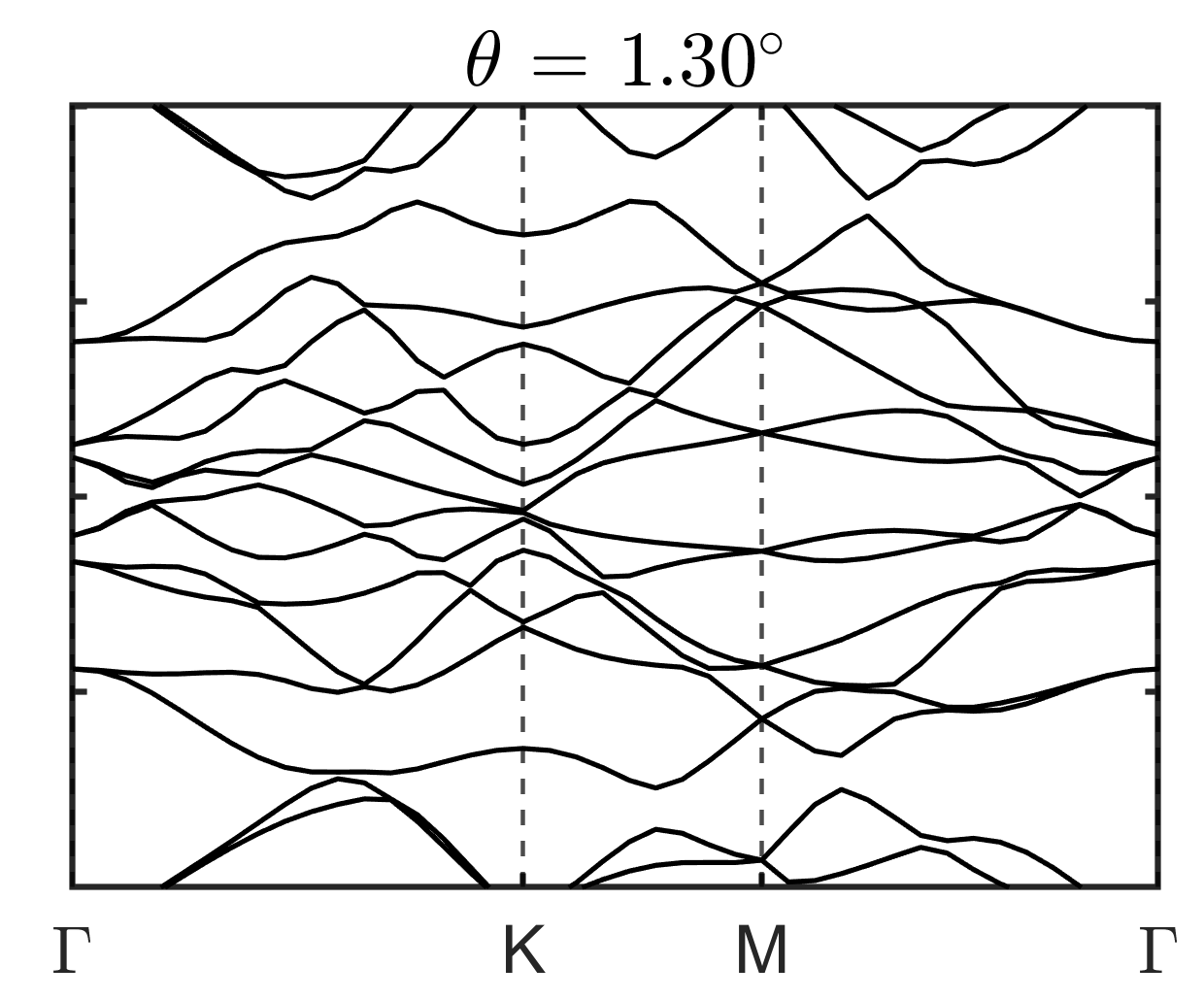}
\end{subfigure}
\caption{Comparison of band structures of AB/AB, AA/AA and AB/AA tDBLG. Results are shown for three twist angles: 2.45$\degree$, 1.70$\degree$, and 1.30$\degree$. The Fermi energy of the undoped systems is set to 0~eV.}
\label{BS}
\end{figure*}

\section{Results and Discussion}

\subsection{Relaxed atomic structure}

Similarly to tBLG, the relaxed tDBLG structures exhibit significant out-of-plane and in-plane atomic displacements for small twist angles~\cite{LREBM,ChoiY.W.2019Ibga,LinXianqing2020Pgma}. Figure~\ref{Z-surface} shows the out-of-plane displacements of AA/AA and AB/AB tDBLG for $\theta=0.73\degree$ (corresponding to $n=45$ and $m=46$). The atomic structure of the AB/AA tDBLG is shown in Appendix B; in this system, the structure of the AB bilayer is very similar to that of the AB/AB tDBLG and the structure of the AA bilayer is similar to that of AA/AA tDBLG. For all tDBLG structures, we find that the atomic structure of the inner two layers resembles closely that of isolated tBLG, see Figs.~\ref{Z-surface}(b), (e), (c) and (f), for example. In particular, the interlayer distances in the AA regions are larger than those in the AB regions, reflecting the larger interlayer separation of the untwisted AA bilayer (3.6~$\text{\AA}$ compared to 3.35~$\text{\AA}$ for the AB bilayer)~\cite{LREBM}. Interestingly, we find that the size of the AA regions in AB/AB tDBLG is somewhat larger than that of AA/AA tDBLG. Each AA region is surrounded by six triangular shaped AB and BA regions which are separated by thin lines where the bilayer has so-called saddle-point (sometimes denoted AA$^{\prime}$) stacking~\cite{META_AAp}. The size of the AB and BA regions is significantly larger than the AA regions. This is a consequence of atomic relaxations which favor low-energy AB stacking and shrink high-energy AA stacking regions. In AA/AA tDBLG we observe that the six AB and BA regions have the same out-of-plane displacement, while this is not the case in AB/AB tDBLG which only exhibits a three-fold symmetry with three of the six AB and BA regions surrounding the central AA spot having a larger displacement than the other three, see Fig.~\ref{Z-surface}(b). This can also be seen in the lower subpanels of Figs.~\ref{Z-surface}(c) and (f), which compare the out-of-plane displacements along the diagonal of the moir\'e unit cell to that of tBLG. The corrugation of tBLG and AA/AA tDBLG is symmetric with respect to the center of the AA region while that of AB/AB tDBLG is not. Figures~\ref{Z-surface}(c) and (f) also show that the magnitude of the inner layer out-of-plane displacements of AA/AA and AB/AB tDBLG is similar to that of tBLG.   

The out-of-plane displacements of the outer layers is shown in Figs.~\ref{Z-surface}(a) and (d). For AB/AB tDBLG, the outer layer displacements follow closely that of the inner layer. In contrast, there are significant differences between the inner and outer layer displacements for AA/AA tDBLG. The outer layers of AA/AA tDBLG also exhibit large out-of-plane displacements in the AA regions, but these neither connected by lines of saddle-point stacking nor are they surrounded by triangle shaped AB and BA regions. Instead, the AA regions are surrounded by a six-fold flower-shaped dip in the out-of-plane displacements followed by a roughly hexagon-shaped region of increased displacements, see also the upper subpanel in Fig.~\ref{Z-surface}(f).

The in-plane displacements of AA/AA and AB/AB tDBLG at $\theta=0.73\degree$ are shown in Fig.~\ref{inplane-disp}. Both inner and outer layers exhibit significant in-plane displacements with a vortex like shape around the AA regions. Interestingly, the vortices in layers 1 and 2 both revolve counter-clockwise in AB/AB tDBLG, while in AA/AA tDBLG the vortex in layer 1 revolves clockwise and the vortex of layer 2 counter clockwise. The magnitude of the in-plane displacements is always larger in the inner layers, and AA/AA tDBLG exhibits larger displacements than AB/AB tDBLG.

The observed in-plane and out-of-plane displacements of tDBLG can be understood by considering the energies of the various stacking arrangements. AA stacking has a high energy and therefore the relaxations reduce the size of these regions and increase the size of lower-energy AB regions. This is achieved by the vortex-shaped in-plane displacements which bring the atoms of the inner two layers closer to an AB stacking configuration. In AB/AB tDBLG, the atoms in the outer layers follow the in-plane displacements of the inner layers to preserve approximate AB stacking. Instead, atoms in the outer layers of AA/AA tDBLG move in the opposite direction to those in the inner layer to reduce AA stacking. This reduces the steric repulsion between atoms in inner and outer layers and creates to regions with smaller out-of-plane displacement around the AA centers, see Fig.~\ref{Z-surface}(d).

Finally, we explain why AB/AB tDBLG has a lower symmetry than AA/AA tDBLG and tBLG. While in tBLG all AB and BA stacked regions are equivalent, the presence of the outer layers in AB/AB tBLG results in two inequivalent stacking configurations: in one configuration, layers 1 and 3 (and also layers 2 and 4) are in an AA configuration (similar to the ABA trilayer stacking). In the other configuration, layers 1 and 3 are in a different AB configuration than layers 2 and 3 (similarly to the ABC trilayer stacking). The steric repulsion in the ABC configuration is smaller than in the ABA configuration resulting in a smaller out-of-plane displacement in three of the six triangle-shaped regions in Figs.~\ref{Z-surface} (a) and (b).

Figure~\ref{Theta} shows the twist-angle dependence of atomic relaxations of tDBLG. For both AA/AA and AB/AB tDBLG, the maximum out-of-plane displacement of the inner layers increases with decreasing twist angle, while minimum out-of-plane displacement decreases in qualitative agreement with tBLG~\cite{LREBM,ShaolongZheng2018ASaM,AC,JainSandeepK2017Sota,SETLA}, see Figs.~\ref{Theta}(a) and (b). Interestingly, the maximum and minimum displacements of tDBLG are always smaller than those of tBLG. This is a consequence of the presence of the outer layers which result in a van der Waals pressure that pushes the inner two layers closer together. In contrast to the inner layers, the maximum and minimum out-of-plane displacements of the outer layers do not sensitively depend on the twist angle and are close to the value of untwisted AA and AB stacked bilayers. 

The average in-plane displacements as function of twist angle are shown in Fig.~\ref{Theta}(c). At small twist angles the in-plane displacements increase significantly, reaching 10\% of the carbon-carbon bond length at 1$\degree$ for the inner layers. This is a consequence of the competition between the energy gain achieved by maximizing the size of AB regions and the energy cost of the required strain to achieve this. At small angles, the size of the AB and BA regions grows, significantly compensating the energy cost of large atomic displacements. We find that the in-plane displacements of AA/AA tDBLG are somewhat larger than those of AB/AB tDBLG. Moreover, the displacements of the outer layers are approximately a factor of three smaller than the inner layer displacements.  

Finally, Fig.~\ref{Theta}(d) shows the difference between the maximum and minimum carbon-carbon bond lengths as function of twist angle. We find that the changes in the bond length are quite small~\cite{JainSandeepK2017Sota,SETLA}, but increase significantly for small twist angles. Again, the changes in AA/AA tDBLG are larger than those in the AB/AB system with bonds length changes in the inner layers being larger than those in the outer layers.

\begin{figure*}[ht]
\centering
\begin{subfigure}[b]{0.45\textwidth}   
\includegraphics[width=1\textwidth]{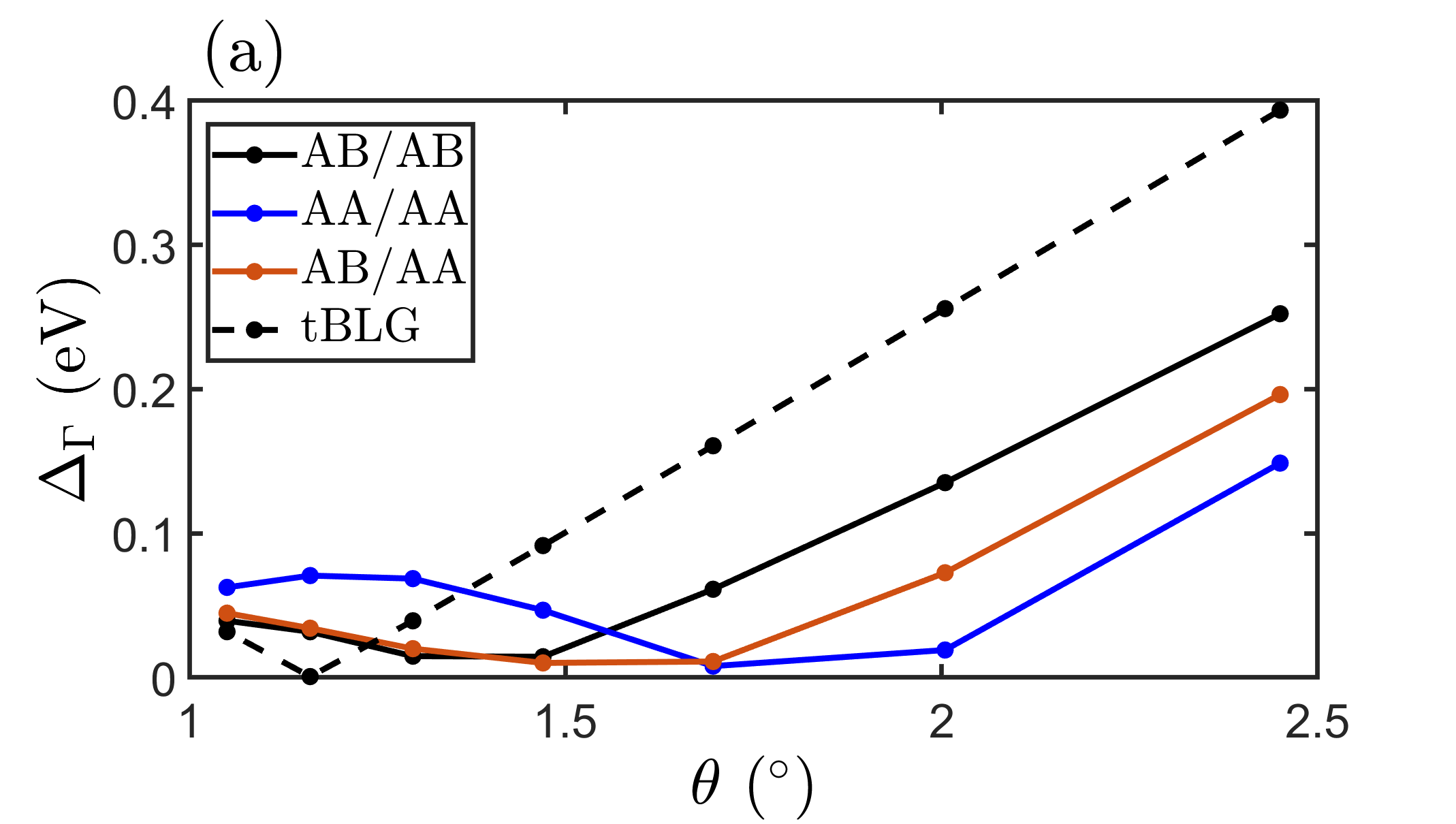}
\end{subfigure}
\begin{subfigure}[b]{0.45\textwidth}   
\includegraphics[width=1\textwidth]{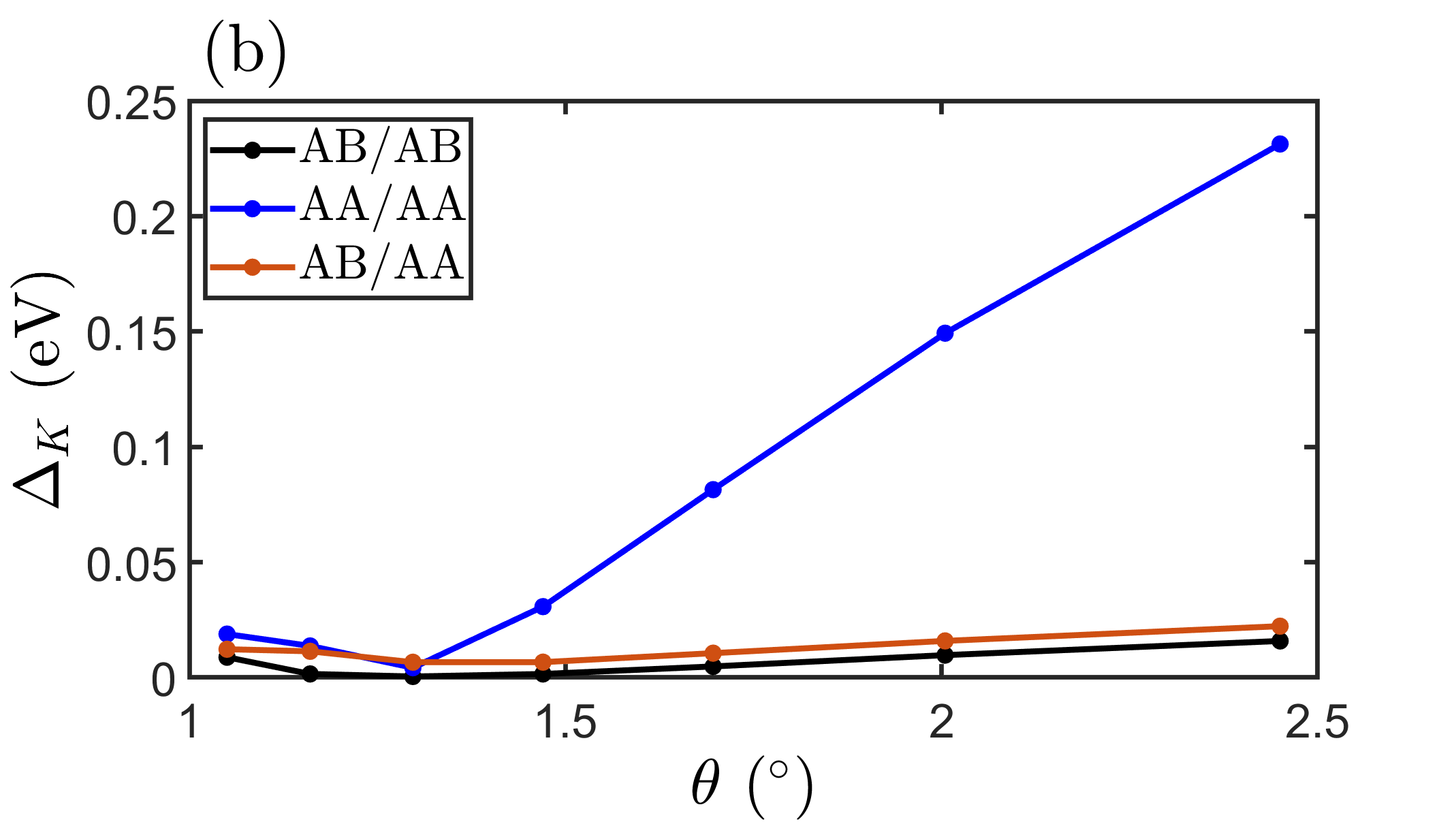}
\end{subfigure}
\caption{(a) Energy difference $\Delta_{\Gamma}$ between flat conduction and valence bands at the $\Gamma$ point in tDLBG as function of twist angle. We also show tBLG results for comparison. (b) Energy difference $\Delta_{K}$ between flat conduction and valence bands at the $K$ point in tDLBG as function of twist angle.}
\label{bandwidth}
\end{figure*}

\begin{figure*}[ht]
\begin{subfigure}{0.3513\textwidth}
  \centering
  \includegraphics[width=1\linewidth]{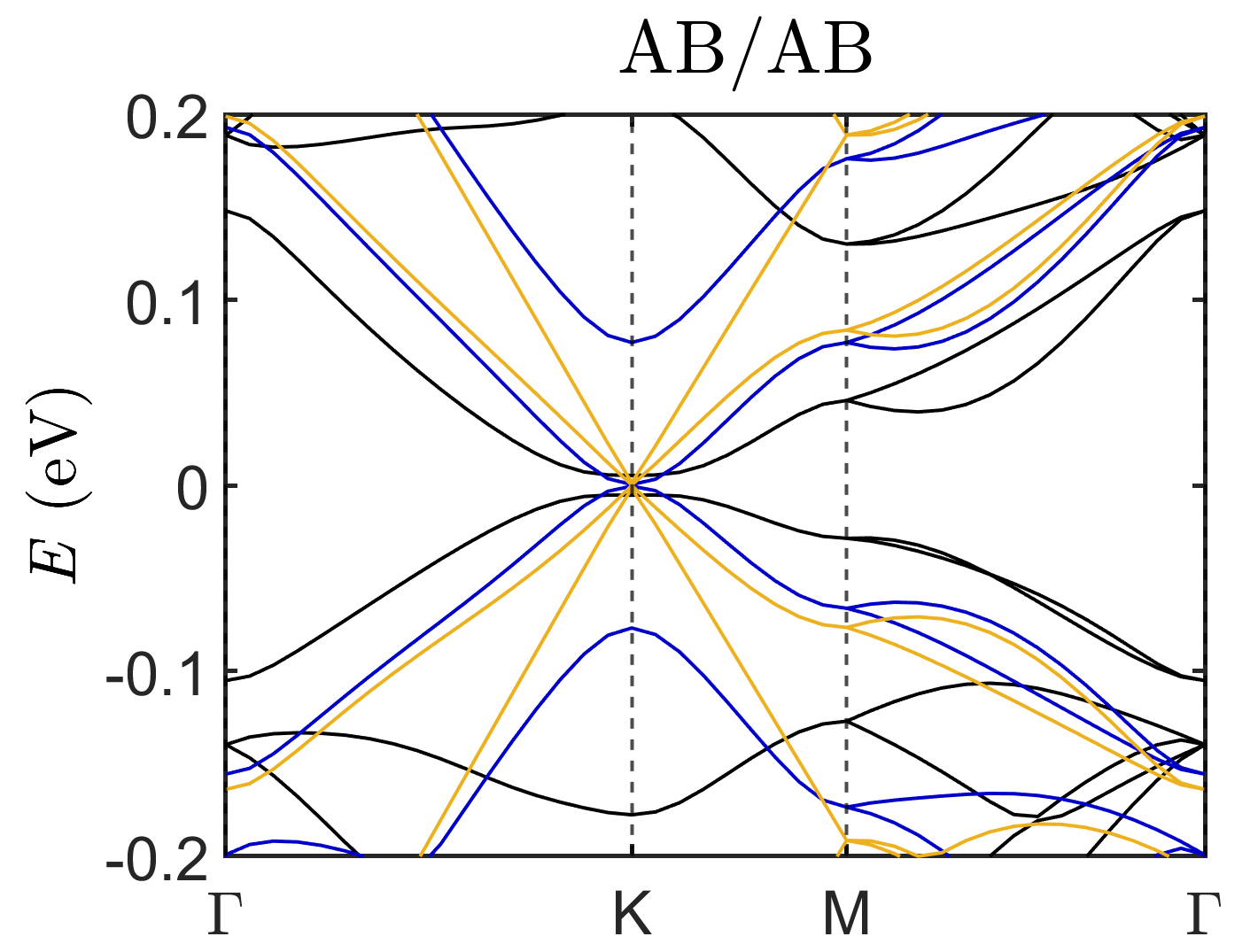}
\end{subfigure}
\begin{subfigure}{0.29935\textwidth}
  \centering
  \includegraphics[width=1\linewidth]{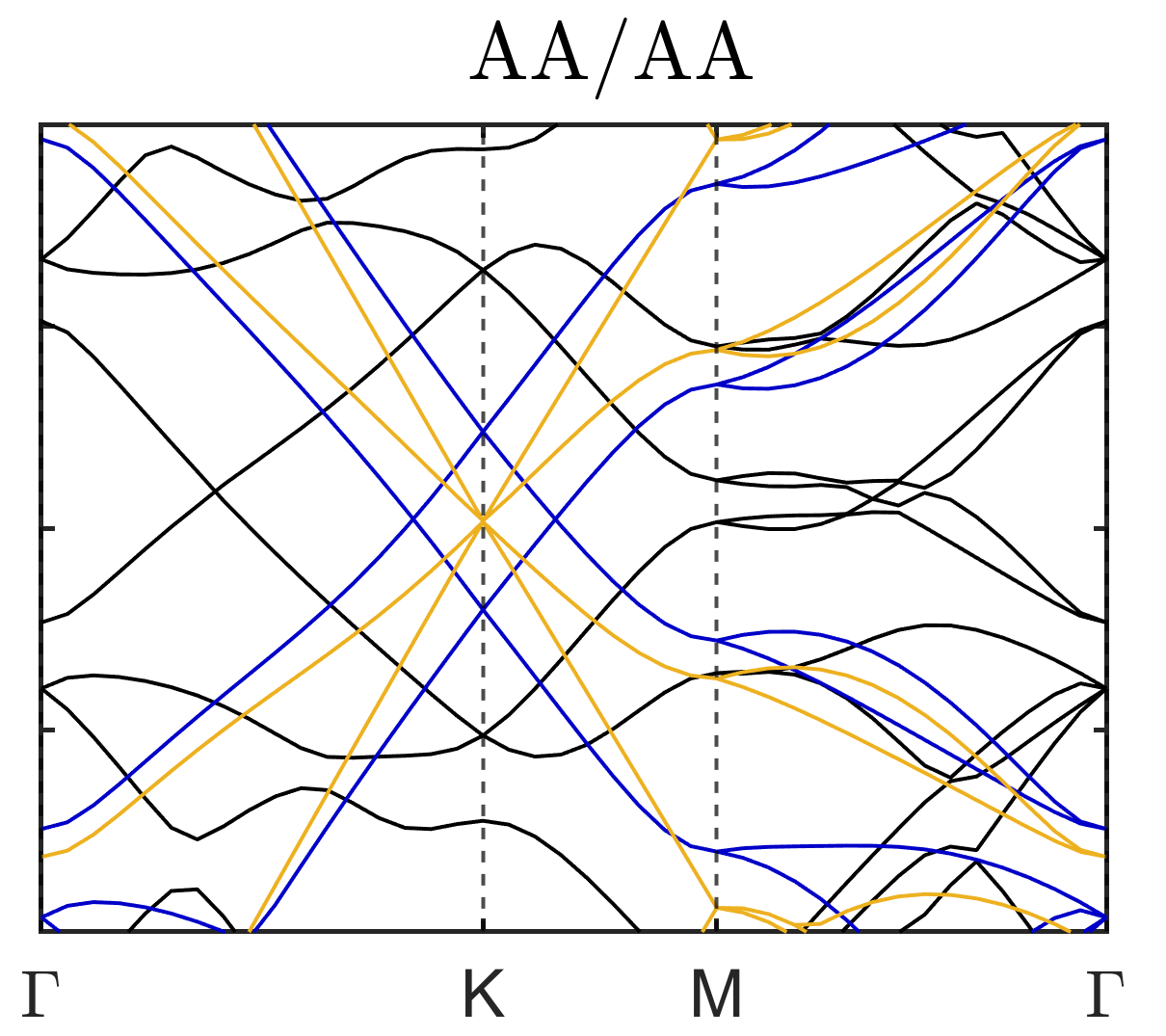}
\end{subfigure}
\begin{subfigure}{0.29935\textwidth}
  \centering
  \includegraphics[width=1\linewidth]{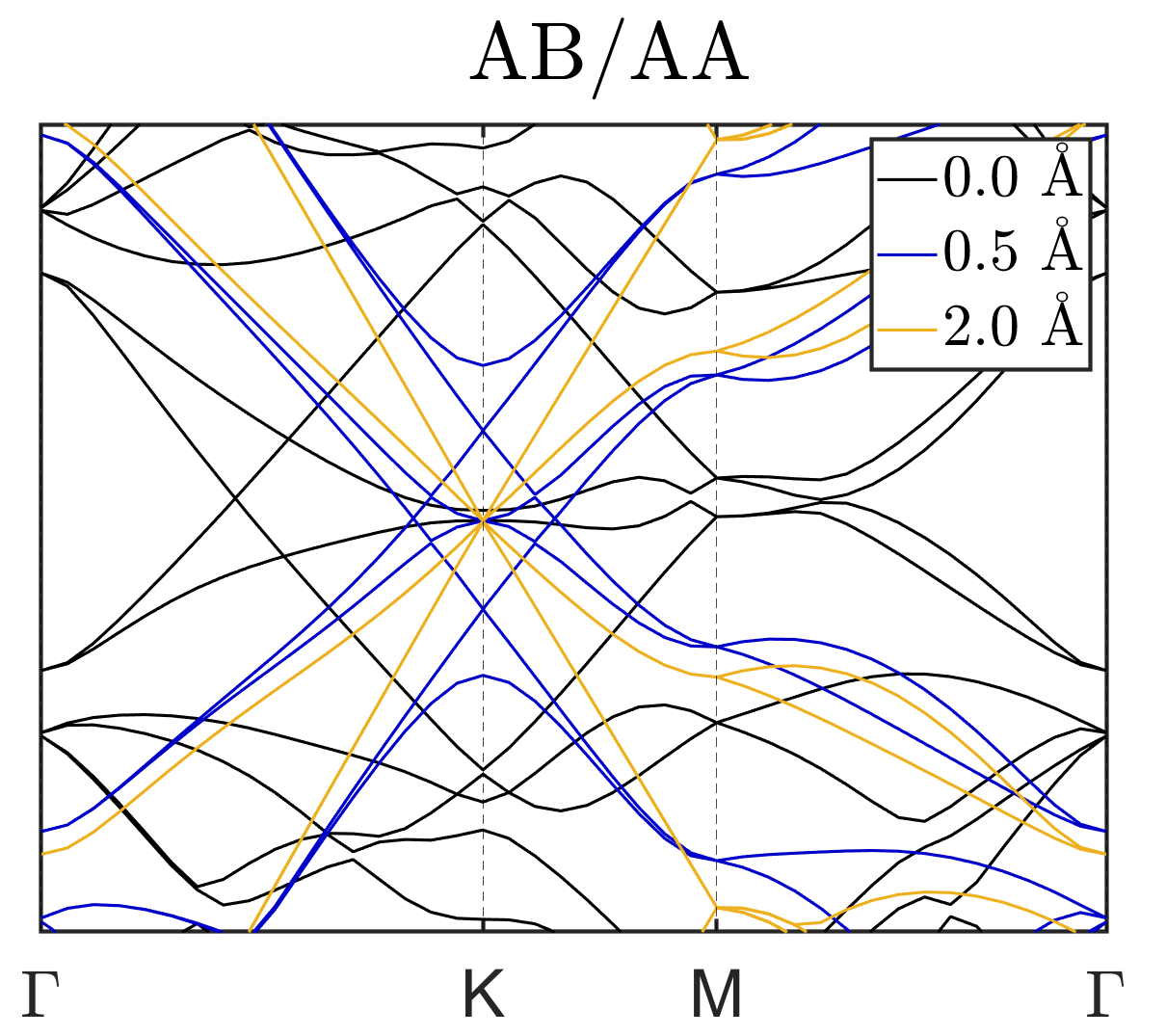}
\end{subfigure}
\caption{Evolution of the band structure of AB/AB, AA/AA and AB/AA tDBLG as the outer layers are rigidly shifted away from the two inner layers by $0.5~\text{\AA}$ and $2.0~\text{\AA}$ at a twist angle of 2.45$\degree$ and comparison to the band structure at the relaxed geometry. Note that for the separated structures, no on-site potential is added.}
\label{layer_separation}
\end{figure*}

\begin{figure*}[ht]
\begin{subfigure}{0.45\textwidth}
  \centering
  \includegraphics[width=1\linewidth]{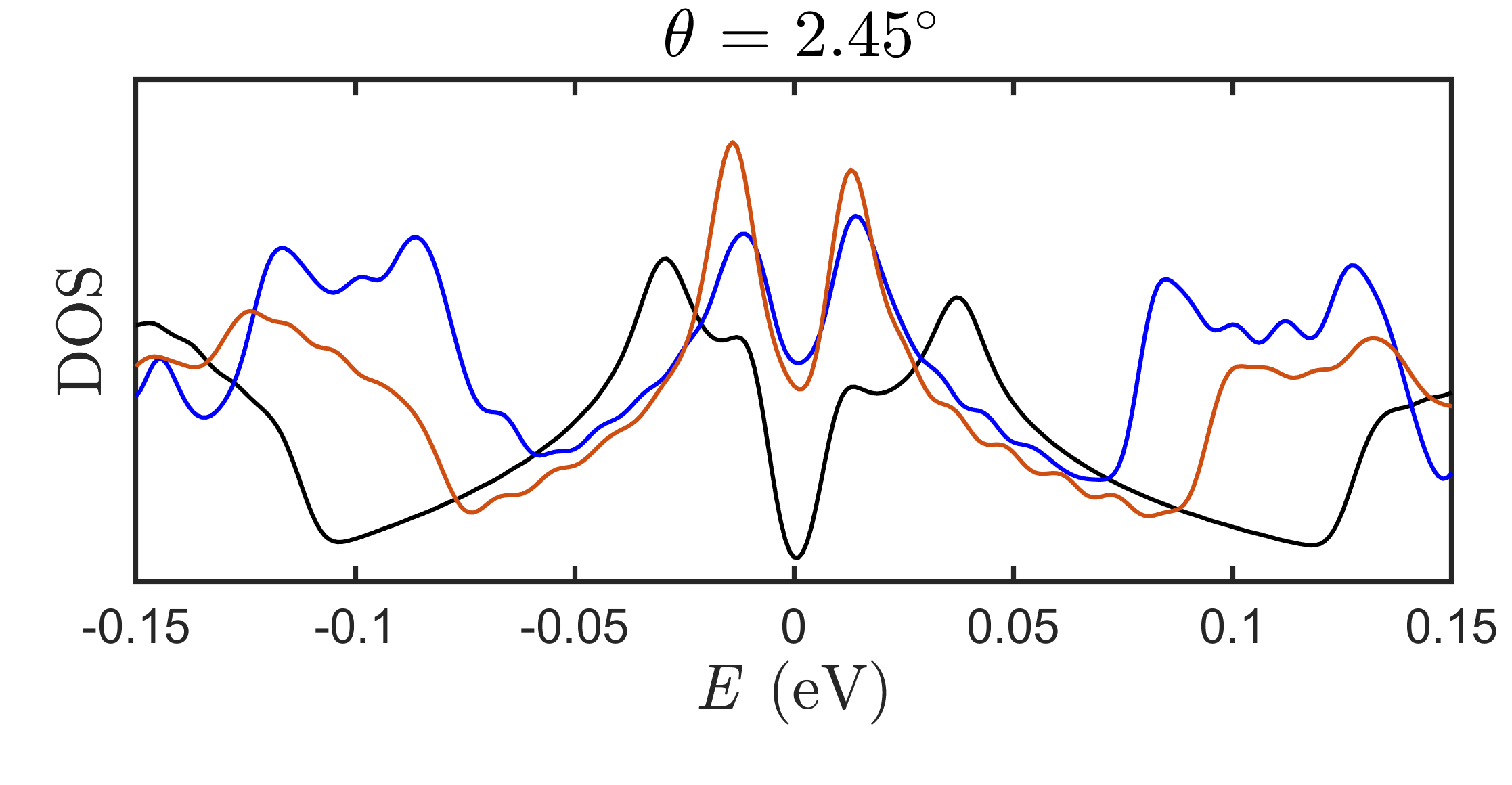}
\end{subfigure}
\begin{subfigure}{0.45\textwidth}
  \centering
  \includegraphics[width=1\linewidth]{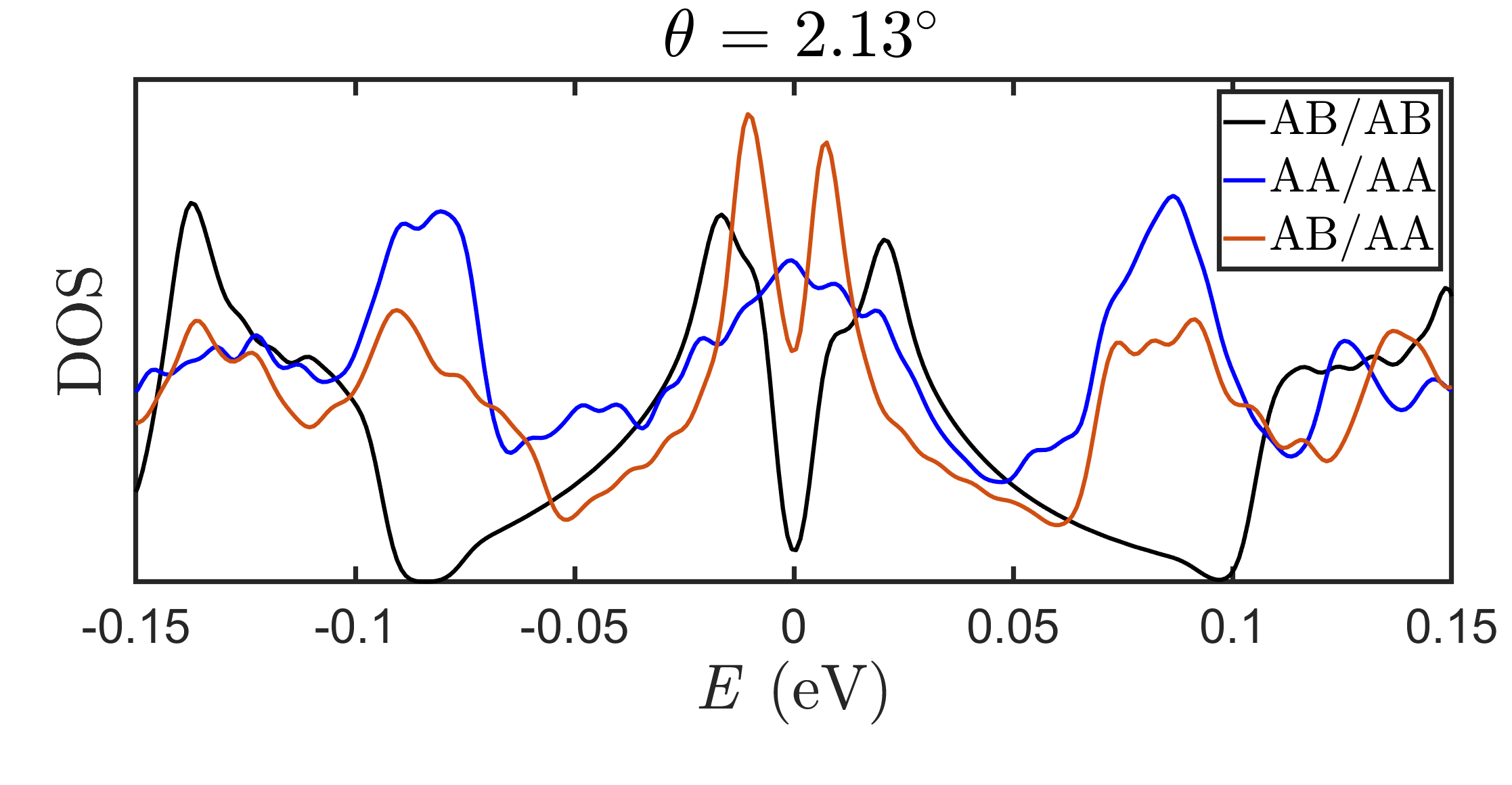}
\end{subfigure}
\begin{subfigure}{0.45\textwidth}
  \centering
  \includegraphics[width=1\linewidth]{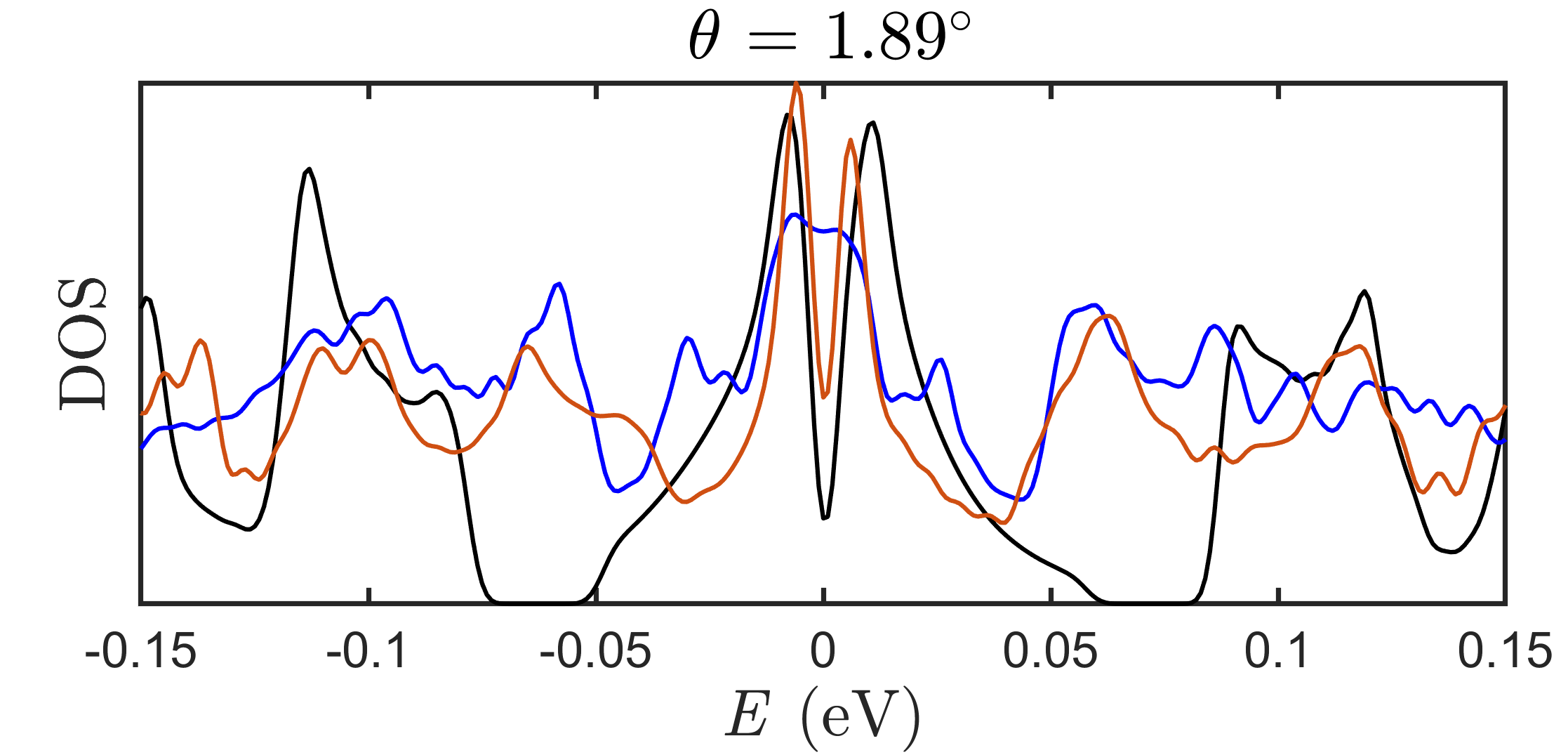}
\end{subfigure}
\begin{subfigure}{0.45\textwidth}
  \centering
  \includegraphics[width=1\linewidth]{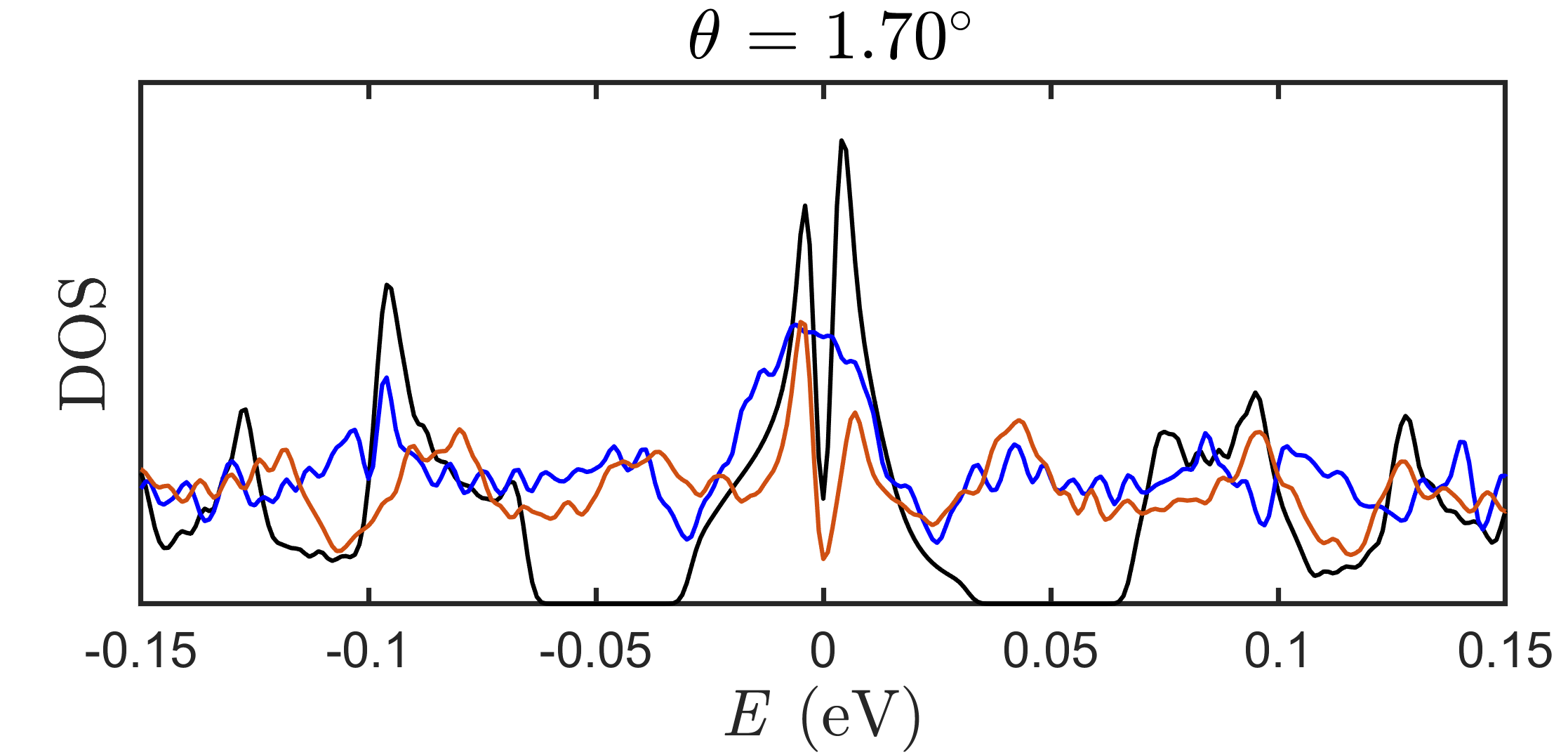}
\end{subfigure}
\caption{Density of states of AB/AB, AA/AA, and AB/AA tDBLG at four different twist angles. All energies are referenced to the Fermi energy of the undoped system.}
\label{DOS}
\end{figure*}

\subsection{Electronic structure} 

Figure~\ref{BS} shows the band structure of AB/AB, AA/AA and AB/AA tDBLG at three twist angles: 2.45$\degree$, 1.70$\degree$ and 1.30$\degree$ (results at other twist angles are shown in Appendix C). In the Appendices C and D, we show the how relaxations (relaxed vs. pristine structures in Appendix C) affect the band structure and also how only including out-of-plane relaxations is reasonable at large angles but not small ones (shown in Appendix D). For AB/AB tDBLG, we obtain a set of four bands near the Fermi level which are separated from all other bands by energy gaps~\cite{haddadi2019moir,STSCTDB,ChebroluNarasimha2019Fbit,KoshinoMikito2019Bsat}. The width of this band manifold decreases as the twist angle is reduced. The dispersion of the flat bands is qualitatively similar to that of tBLG. However, at larger twist angles (see, for example, the band structures for 2.45$\degree$ or 1.70$\degree$), the undoped material is not a semi-metal, but a semiconductor with small direct band gaps at the $K$ and $K^{\prime}$ points of the moir\'e Brillouin zone. Interestingly, the band gaps become smaller as the twist angle is reduced and vanish near the magic angle which is defined as the twist angle with the smallest width of the flat band manifold~\cite{RickhausPeter2019GOiT}. Figure~\ref{bandwidth}(a) shows the width of the flat bands (approximated by $\Delta_{\Gamma}$, the energy difference of the highest valence band and the lowest conduction band at the $\Gamma$ point) and demonstrates that the magic angle lies near 1.3$\degree$, somewhat larger than the value for tBLG (1.1$\degree$). The shape of $\Delta_{\Gamma}$ near the magic angle is flatter than that of tBLG which exhibits a clear V shape. This suggests that strong correlation phenomena in AB/AB tDBLG should be less sensitive to the precise value of the twist angle in the vicinity of the magic angle. Figure~\ref{bandwidth}(b) shows that energy gap $\Delta_{K}$ at the $K$ and $K^{\prime}$ points as function of twist angle. It can be seen that at twist angles smaller than the magic angle, the system exhibits again a non zero band gap. These findings for AB/AB tDBLG are in good agreement with previous studies~\cite{haddadi2019moir,ChoiY.W.2019Ibga,KoshinoMikito2019Bsat,ChebroluNarasimha2019Fbit,STSCTDB,BIBI}. 

The band structure of AA/AA tDBLG near the $K$ point is similar to that of the untwisted AA bilayer with two Dirac cones that are shifted up and down in energy. In contrast to AB/AB tDBLG, the low energy bands are not separated from the higher energy bands by energy gaps. As the twist angle is reduced, the energy splitting between the Dirac cones is reduced significantly. Figure~\ref{bandwidth} (b) shows that the energy splitting of the Dirac cones reaches a minimum at a twisted angle near 1.3$\degree$ and then increases again. In addition, it can be seen that the band structure at 2.45$\degree$ features extremely flat bands between $M$ and $\Gamma$. Similarly, we find ultraflat bands between $K$ and $M$ at $\theta=1.3\degree$. These findings suggest that AA/AA tDBLG is a promising candidate for observing strong correlation phenomena that can be tuned via the twist angle.

The band structure of AB/AA tDBLG contains elements from both AA/AA and AB/AB tDBLG. Specifically, we observe shifted Dirac cones at $K$ (see band structure in Fig.~\ref{BS} at 2.45$\degree$, for example) in addition to a set of bands that look similar to the tBLG bands, but with a gap at $K$. The bands flatten as the twist angle is reduced to 1.7$\degree$. At $\theta=1.3\degree$, the low-energy band structure features many entangled bands.

Interestingly, the band structures of all twisted double bilayers feature elements that are reminiscent of tBLG. This suggests that it might be instructive view the double bilayer as a central tBLG unit whose band structure is perturbed by the addition of the outer layers. To explore this viewpoint, we have studied the evolution of the double bilayer band structure as the distance of the outer layers is increased from the inner tBLG component. Figure~\ref{layer_separation} compares the band structures of AB/AB, AA/AA and AB/AA tDBLG at the relaxed atomic configuration with the result for configurations where the outer layers are rigidly shifted away from the inner layers by 0.5$~\text{\AA}$ and 2.0$~\text{\AA}$ (note that no on-site potential was added for these calculations). For the largest separation between inner and outer layers, we can clearly see a tBLG-like set of bands in addition to monolayer graphene bands for all tDBLG systems. As the distance between inner and outer layers is reduced, we find that the tBLG states and the monolayer graphene states hybridize. This pushes the tBLG states to lower energies resulting in an additional hybridization-induced band flattening, while the graphene states are pushed to higher energies. Of course, the detailed structure of the hybridized flat bands depends on the stacking of the outer layers. These findings suggests that it is possible to control flat bands properties via the hybridization of the inner tBLG with the outer layers. This hybridization can be modified by changing the outer layer stacking, shifting the on-site energies of the outer layers through application of a gate voltage or by changing the chemistry of the outer layers (for example, by using transition-metal dichalcogenides instead of graphene).

Finally, we present results for the density of states (DOS) of AA/AA, AB/AB and AB/AA tDBLG at four different twist angles in Fig.~\ref{DOS}. The (local) density of states can be directly measured in scanning tunneling experiments and the value of the DOS at the Fermi level is a key parameter in weak coupling approaches of electronic phase transitions. At a twist angle of 2.45$\degree$, the DOS of all tDLBG structures exhibits peaks originating from van Hove singularities (vHS) near the Fermi level of the undoped system. While the DOS of AA/AA and AB/AA tDBLG exhibit two peaks, the DOS of AB/AB tDBLG exhibits four peaks: the two smaller ones are closer to the Fermi level and their energies coincide with the peak positions of the AA/AA and AB/AB systems, while the two larger peaks are slightly farther away from the Fermi energy. Interestingly, AB/AA tDBLG exhibits the largest DOS values among the three tDBLG systems at this twist angle. The four peak structure in the DOS of AB/AB tDBLG can be traced back to its semiconducting band structure with the two smaller peaks arising from states near the valence and conduction band edges, while the larger peaks derive from bands near the $M$ point, similar to tBLG~\cite{KL}. Both AA/AA and AB/AA tDBLG are metallic and hybridization with the outer layers pushes the $M$ point states closer to the Fermi level giving rise to two large vHS peaks in the DOS. Reducing the twist angle to 2.13$\degree$ shifts the vHS peaks of the AB/AA and AB/AB systems closer to the Fermi level as a consequence of band flattening. For AA/AA tDBLG, the sharp vHS peaks disappear leaving only a single broad peak at the Fermi level.

At even smaller twist angles (bottom panels of Fig.~\ref{DOS}), the DOS of AB/AB tDBLG exhibits only two sharp vHS peaks because the band gap closes near the magic angle. In contrast, there are no qualitative changes in the DOS of AA/AA tDBLG (which exhibits a single broad peak near the Fermi level) and that of AB/AA tDBLG (which exhibits two sharp vHS peaks). At $\theta < 1.8\degree$, AB/AB tDBLG exhibits the highest vHS peaks among the three tDBLG structures and this likely the reason why this system is particularly susceptible to interaction-induced electronic phase transitions.

\section{Conclusions}

We have studied the atomic and electronic structure of AB/AB, AA/AA, and AB/AA tDBLG. In all systems, the atomic structure of the inner layers is similar to that of tBLG. In contrast, the structure of the outer layers depends on the stacking: Outer layers of AB bilayers follow the structure of the inner layers preserving the energetically favorable AB stacking, while atoms in the outer layers of AA bilayers attempt to avoid the unfavorable AA stacking resulting in a vortex-shaped in-plane displacement with an opposite sense of rotation than that of the inner layers. The electronic band structure of all tDBLG systems exhibits flat bands at small twist angles, but the shape of the bands depends sensitively on the stacking of the outer layers. To gain further insight, we analyze the evolution of the band structure as the outer layers are rigidly shifted away from the central tBLG unit, while retaining the atomic relaxations of each layer. This reveals that the hybridization between the flat bands of the tBLG and the graphene states of the outer layers leads to an additional band flattening and suggests the possibility of engineering flat band properties via the ``functionalization" of tBLG by additional layers of two-dimensional materials. We also study the density of states of the different tBLG systems and find that AB/AB and AB/AA tDBLG exhibits two sharp van Hove singularities near the Fermi level, while the AA/AA system only exhibits a single broad peak at small twist angles. Our findings suggest that the outer layer stacking results in qualitatively different flat band physics and introduces AA/AA and AB/AA tDBLG as promising moir\'e materials for studying strong electron correlations. 

\section{Acknowledgements}

We thank M. Kahk, K. Atalar, M. Newns and N. Gray Desai for helpful discussions. Z.A.H.G. was supported through a studentship in the Centre for Doctoral Training on Theory and Simulation of Materials at Imperial College London funded by the EPSRC (EP/L015579/1). We acknowledge funding from EPSRC Grant No. EP/S025324/1 and the Thomas Young Centre under Grant No. TYC-101. Via our membership of the UK's HEC Materials Chemistry Consortium, which is funded by EPSRC (EP/L000202 and EP/R029431), this work used the ARCHER UK National Supercomputing Service.

\bibliographystyle{apsrev4-1}
\bibliography{REF}

\begin{thebibliography}{80}%
\makeatletter
\providecommand \@ifxundefined [1]{%
 \@ifx{#1\undefined}
}%
\providecommand \@ifnum [1]{%
 \ifnum #1\expandafter \@firstoftwo
 \else \expandafter \@secondoftwo
 \fi
}%
\providecommand \@ifx [1]{%
 \ifx #1\expandafter \@firstoftwo
 \else \expandafter \@secondoftwo
 \fi
}%
\providecommand \natexlab [1]{#1}%
\providecommand \enquote  [1]{``#1''}%
\providecommand \bibnamefont  [1]{#1}%
\providecommand \bibfnamefont [1]{#1}%
\providecommand \citenamefont [1]{#1}%
\providecommand \href@noop [0]{\@secondoftwo}%
\providecommand \href [0]{\begingroup \@sanitize@url \@href}%
\providecommand \@href[1]{\@@startlink{#1}\@@href}%
\providecommand \@@href[1]{\endgroup#1\@@endlink}%
\providecommand \@sanitize@url [0]{\catcode `\\12\catcode `\$12\catcode
  `\&12\catcode `\#12\catcode `\^12\catcode `\_12\catcode `\%12\relax}%
\providecommand \@@startlink[1]{}%
\providecommand \@@endlink[0]{}%
\providecommand \url  [0]{\begingroup\@sanitize@url \@url }%
\providecommand \@url [1]{\endgroup\@href {#1}{\urlprefix }}%
\providecommand \urlprefix  [0]{URL }%
\providecommand \Eprint [0]{\href }%
\providecommand \doibase [0]{http://dx.doi.org/}%
\providecommand \selectlanguage [0]{\@gobble}%
\providecommand \bibinfo  [0]{\@secondoftwo}%
\providecommand \bibfield  [0]{\@secondoftwo}%
\providecommand \translation [1]{[#1]}%
\providecommand \BibitemOpen [0]{}%
\providecommand \bibitemStop [0]{}%
\providecommand \bibitemNoStop [0]{.\EOS\space}%
\providecommand \EOS [0]{\spacefactor3000\relax}%
\providecommand \BibitemShut  [1]{\csname bibitem#1\endcsname}%
\let\auto@bib@innerbib\@empty
\bibitem [{\citenamefont {dos Santos}\ \emph {et~al.}(2007)\citenamefont {dos
  Santos}, \citenamefont {Peres},\ and\ \citenamefont {Neto}}]{GBWT}%
  \BibitemOpen
  \bibfield  {author} {\bibinfo {author} {\bibfnamefont {J.~M. B.~L.}\
  \bibnamefont {dos Santos}}, \bibinfo {author} {\bibfnamefont {N.~M.~R.}\
  \bibnamefont {Peres}}, \ and\ \bibinfo {author} {\bibfnamefont {A.~H.~C.}\
  \bibnamefont {Neto}},\ }\href@noop {} {\bibfield  {journal} {\bibinfo
  {journal} {Phys. Rev. Lett.}\ }\textbf {\bibinfo {volume} {99}},\ \bibinfo
  {pages} {256802} (\bibinfo {year} {2007})}\BibitemShut {NoStop}%
\bibitem [{\citenamefont {Bistritzer}\ and\ \citenamefont
  {MacDonald}(2011)}]{Bistritzer12233}%
  \BibitemOpen
  \bibfield  {author} {\bibinfo {author} {\bibfnamefont {R.}~\bibnamefont
  {Bistritzer}}\ and\ \bibinfo {author} {\bibfnamefont {A.~H.}\ \bibnamefont
  {MacDonald}},\ }\href@noop {} {\bibfield  {journal} {\bibinfo  {journal}
  {PNAS}\ }\textbf {\bibinfo {volume} {108}},\ \bibinfo {pages} {12233}
  (\bibinfo {year} {2011})}\BibitemShut {NoStop}%
\bibitem [{\citenamefont {de~Laissardi\`ere}\ \emph {et~al.}(2010)\citenamefont
  {de~Laissardi\`ere}, \citenamefont {Mayou},\ and\ \citenamefont
  {Magaud}}]{LDE}%
  \BibitemOpen
  \bibfield  {author} {\bibinfo {author} {\bibfnamefont {G.~T.}\ \bibnamefont
  {de~Laissardi\`ere}}, \bibinfo {author} {\bibfnamefont {D.}~\bibnamefont
  {Mayou}}, \ and\ \bibinfo {author} {\bibfnamefont {L.}~\bibnamefont
  {Magaud}},\ }\href@noop {} {\bibfield  {journal} {\bibinfo  {journal} {Nano
  Lett.}\ }\textbf {\bibinfo {volume} {10}},\ \bibinfo {pages} {804} (\bibinfo
  {year} {2010})}\BibitemShut {NoStop}%
\bibitem [{\citenamefont {de~Laissardi\`ere}\ \emph {et~al.}(2012)\citenamefont
  {de~Laissardi\`ere}, \citenamefont {Mayou},\ and\ \citenamefont
  {Magaud}}]{NSCS}%
  \BibitemOpen
  \bibfield  {author} {\bibinfo {author} {\bibfnamefont {G.~T.}\ \bibnamefont
  {de~Laissardi\`ere}}, \bibinfo {author} {\bibfnamefont {D.}~\bibnamefont
  {Mayou}}, \ and\ \bibinfo {author} {\bibfnamefont {L.}~\bibnamefont
  {Magaud}},\ }\href@noop {} {\bibfield  {journal} {\bibinfo  {journal} {Phys.
  Rev. B}\ }\textbf {\bibinfo {volume} {86}},\ \bibinfo {pages} {125413}
  (\bibinfo {year} {2012})}\BibitemShut {NoStop}%
\bibitem [{\citenamefont {Su\'arez~Morell}\ \emph {et~al.}(2010)\citenamefont
  {Su\'arez~Morell}, \citenamefont {Correa}, \citenamefont {Vargas},
  \citenamefont {Pacheco},\ and\ \citenamefont
  {Barticevic}}]{PhysRevB.82.121407}%
  \BibitemOpen
  \bibfield  {author} {\bibinfo {author} {\bibfnamefont {E.}~\bibnamefont
  {Su\'arez~Morell}}, \bibinfo {author} {\bibfnamefont {J.~D.}\ \bibnamefont
  {Correa}}, \bibinfo {author} {\bibfnamefont {P.}~\bibnamefont {Vargas}},
  \bibinfo {author} {\bibfnamefont {M.}~\bibnamefont {Pacheco}}, \ and\
  \bibinfo {author} {\bibfnamefont {Z.}~\bibnamefont {Barticevic}},\
  }\href@noop {} {\bibfield  {journal} {\bibinfo  {journal} {Phys. Rev. B}\
  }\textbf {\bibinfo {volume} {82}},\ \bibinfo {pages} {121407} (\bibinfo
  {year} {2010})}\BibitemShut {NoStop}%
\bibitem [{\citenamefont {Carr}\ \emph {et~al.}(2017)\citenamefont {Carr},
  \citenamefont {Massatt}, \citenamefont {Fang}, \citenamefont {Cazeaux},
  \citenamefont {Luskin},\ and\ \citenamefont {Kaxiras}}]{Carr_2017}%
  \BibitemOpen
  \bibfield  {author} {\bibinfo {author} {\bibfnamefont {S.}~\bibnamefont
  {Carr}}, \bibinfo {author} {\bibfnamefont {D.}~\bibnamefont {Massatt}},
  \bibinfo {author} {\bibfnamefont {S.}~\bibnamefont {Fang}}, \bibinfo {author}
  {\bibfnamefont {P.}~\bibnamefont {Cazeaux}}, \bibinfo {author} {\bibfnamefont
  {M.}~\bibnamefont {Luskin}}, \ and\ \bibinfo {author} {\bibfnamefont
  {E.}~\bibnamefont {Kaxiras}},\ }\href@noop {} {\bibfield  {journal} {\bibinfo
   {journal} {Phys. Rev. B}\ }\textbf {\bibinfo {volume} {95}},\ \bibinfo
  {pages} {075420} (\bibinfo {year} {2017})}\BibitemShut {NoStop}%
\bibitem [{\citenamefont {Tritsaris}\ \emph {et~al.}(2020)\citenamefont
  {Tritsaris}, \citenamefont {Carr}, \citenamefont {Zhu}, \citenamefont {Xie},
  \citenamefont {Torrisi}, \citenamefont {Tang}, \citenamefont {Mattheakis},
  \citenamefont {Larson},\ and\ \citenamefont {Kaxiras}}]{Tritsaris_2020}%
  \BibitemOpen
  \bibfield  {author} {\bibinfo {author} {\bibfnamefont {G.~A.}\ \bibnamefont
  {Tritsaris}}, \bibinfo {author} {\bibfnamefont {S.}~\bibnamefont {Carr}},
  \bibinfo {author} {\bibfnamefont {Z.}~\bibnamefont {Zhu}}, \bibinfo {author}
  {\bibfnamefont {Y.}~\bibnamefont {Xie}}, \bibinfo {author} {\bibfnamefont
  {S.~B.}\ \bibnamefont {Torrisi}}, \bibinfo {author} {\bibfnamefont
  {J.}~\bibnamefont {Tang}}, \bibinfo {author} {\bibfnamefont {M.}~\bibnamefont
  {Mattheakis}}, \bibinfo {author} {\bibfnamefont {D.~T.}\ \bibnamefont
  {Larson}}, \ and\ \bibinfo {author} {\bibfnamefont {E.}~\bibnamefont
  {Kaxiras}},\ }\href {\doibase 10.1088/2053-1583/ab8f62} {\bibfield  {journal}
  {\bibinfo  {journal} {2D Materials}\ }\textbf {\bibinfo {volume} {7}},\
  \bibinfo {pages} {035028} (\bibinfo {year} {2020})}\BibitemShut {NoStop}%
\bibitem [{\citenamefont {Uchida}\ \emph {et~al.}(2014)\citenamefont {Uchida},
  \citenamefont {Furuya}, \citenamefont {Iwata},\ and\ \citenamefont
  {Oshiyama}}]{AC}%
  \BibitemOpen
  \bibfield  {author} {\bibinfo {author} {\bibfnamefont {K.}~\bibnamefont
  {Uchida}}, \bibinfo {author} {\bibfnamefont {S.}~\bibnamefont {Furuya}},
  \bibinfo {author} {\bibfnamefont {J.-I.}\ \bibnamefont {Iwata}}, \ and\
  \bibinfo {author} {\bibfnamefont {A.}~\bibnamefont {Oshiyama}},\ }\href@noop
  {} {\bibfield  {journal} {\bibinfo  {journal} {Phys. Rev. B}\ }\textbf
  {\bibinfo {volume} {90}},\ \bibinfo {pages} {155451} (\bibinfo {year}
  {2014})}\BibitemShut {NoStop}%
\bibitem [{\citenamefont {Jain}\ \emph {et~al.}(2017)\citenamefont {Jain},
  \citenamefont {Juri\u{c}i\'c},\ and\ \citenamefont
  {Barkema}}]{JainSandeepK2017Sota}%
  \BibitemOpen
  \bibfield  {author} {\bibinfo {author} {\bibfnamefont {S.~K.}\ \bibnamefont
  {Jain}}, \bibinfo {author} {\bibfnamefont {V.}~\bibnamefont {Juri\u{c}i\'c}},
  \ and\ \bibinfo {author} {\bibfnamefont {G.~T.}\ \bibnamefont {Barkema}},\
  }\href@noop {} {\bibfield  {journal} {\bibinfo  {journal} {2D Mater.}\
  }\textbf {\bibinfo {volume} {4}},\ \bibinfo {pages} {015018} (\bibinfo {year}
  {2017})}\BibitemShut {NoStop}%
\bibitem [{\citenamefont {Gargiulo}\ and\ \citenamefont
  {Yazyev}(2018)}]{SETLA}%
  \BibitemOpen
  \bibfield  {author} {\bibinfo {author} {\bibfnamefont {F.}~\bibnamefont
  {Gargiulo}}\ and\ \bibinfo {author} {\bibfnamefont {O.~V.}\ \bibnamefont
  {Yazyev}},\ }\href@noop {} {\bibfield  {journal} {\bibinfo  {journal} {2D
  Mater.}\ }\textbf {\bibinfo {volume} {5}},\ \bibinfo {pages} {015019}
  (\bibinfo {year} {2018})}\BibitemShut {NoStop}%
\bibitem [{\citenamefont {Carr}\ \emph {et~al.}(2018)\citenamefont {Carr},
  \citenamefont {Fang}, \citenamefont {Jarillo-Herrero},\ and\ \citenamefont
  {Kaxiras}}]{PDTBLG}%
  \BibitemOpen
  \bibfield  {author} {\bibinfo {author} {\bibfnamefont {S.}~\bibnamefont
  {Carr}}, \bibinfo {author} {\bibfnamefont {S.}~\bibnamefont {Fang}}, \bibinfo
  {author} {\bibfnamefont {P.}~\bibnamefont {Jarillo-Herrero}}, \ and\ \bibinfo
  {author} {\bibfnamefont {E.}~\bibnamefont {Kaxiras}},\ }\href@noop {}
  {\bibfield  {journal} {\bibinfo  {journal} {Phys. Rev. B}\ }\textbf {\bibinfo
  {volume} {98}},\ \bibinfo {pages} {085144} (\bibinfo {year}
  {2018})}\BibitemShut {NoStop}%
\bibitem [{\citenamefont {Carr}\ \emph {et~al.}(2019)\citenamefont {Carr},
  \citenamefont {Fang}, \citenamefont {Zhu},\ and\ \citenamefont
  {Kaxiras}}]{KDP}%
  \BibitemOpen
  \bibfield  {author} {\bibinfo {author} {\bibfnamefont {S.}~\bibnamefont
  {Carr}}, \bibinfo {author} {\bibfnamefont {S.}~\bibnamefont {Fang}}, \bibinfo
  {author} {\bibfnamefont {Z.}~\bibnamefont {Zhu}}, \ and\ \bibinfo {author}
  {\bibfnamefont {E.}~\bibnamefont {Kaxiras}},\ }\href@noop {} {\bibfield
  {journal} {\bibinfo  {journal} {Phys. Rev. Research}\ }\textbf {\bibinfo
  {volume} {1}},\ \bibinfo {pages} {013001} (\bibinfo {year}
  {2019})}\BibitemShut {NoStop}%
\bibitem [{\citenamefont {Gonz\'alez}\ and\ \citenamefont
  {Stauber}(2019)}]{KL}%
  \BibitemOpen
  \bibfield  {author} {\bibinfo {author} {\bibfnamefont {J.}~\bibnamefont
  {Gonz\'alez}}\ and\ \bibinfo {author} {\bibfnamefont {T.}~\bibnamefont
  {Stauber}},\ }\href@noop {} {\bibfield  {journal} {\bibinfo  {journal} {Phys.
  Rev. Lett.}\ }\textbf {\bibinfo {volume} {122}},\ \bibinfo {pages} {026801}
  (\bibinfo {year} {2019})}\BibitemShut {NoStop}%
\bibitem [{\citenamefont {Goodwin}\ \emph
  {et~al.}(2020{\natexlab{a}})\citenamefont {Goodwin}, \citenamefont {Vitale},
  \citenamefont {Corsetti}, \citenamefont {Efetov}, \citenamefont {Mostofi},\
  and\ \citenamefont {Lischner}}]{PHD_3}%
  \BibitemOpen
  \bibfield  {author} {\bibinfo {author} {\bibfnamefont {Z.~A.~H.}\
  \bibnamefont {Goodwin}}, \bibinfo {author} {\bibfnamefont {V.}~\bibnamefont
  {Vitale}}, \bibinfo {author} {\bibfnamefont {F.}~\bibnamefont {Corsetti}},
  \bibinfo {author} {\bibfnamefont {D.}~\bibnamefont {Efetov}}, \bibinfo
  {author} {\bibfnamefont {A.~A.}\ \bibnamefont {Mostofi}}, \ and\ \bibinfo
  {author} {\bibfnamefont {J.}~\bibnamefont {Lischner}},\ }\href@noop {}
  {\bibfield  {journal} {\bibinfo  {journal} {Phys. Rev. B}\ }\textbf {\bibinfo
  {volume} {101}},\ \bibinfo {pages} {165110} (\bibinfo {year}
  {2020}{\natexlab{a}})}\BibitemShut {NoStop}%
\bibitem [{\citenamefont {Stauber}\ and\ \citenamefont {Kohler}(2016)}]{PTBLG}%
  \BibitemOpen
  \bibfield  {author} {\bibinfo {author} {\bibfnamefont {T.}~\bibnamefont
  {Stauber}}\ and\ \bibinfo {author} {\bibfnamefont {H.}~\bibnamefont
  {Kohler}},\ }\href@noop {} {\bibfield  {journal} {\bibinfo  {journal} {Nano
  Lett.}\ }\textbf {\bibinfo {volume} {16}},\ \bibinfo {pages} {6844} (\bibinfo
  {year} {2016})}\BibitemShut {NoStop}%
\bibitem [{\citenamefont {Pizarro}\ \emph {et~al.}(2019)\citenamefont
  {Pizarro}, \citenamefont {Rosner}, \citenamefont {Thomale}, \citenamefont
  {Valent},\ and\ \citenamefont {Wehling}}]{CCRPA}%
  \BibitemOpen
  \bibfield  {author} {\bibinfo {author} {\bibfnamefont {J.~M.}\ \bibnamefont
  {Pizarro}}, \bibinfo {author} {\bibfnamefont {M.}~\bibnamefont {Rosner}},
  \bibinfo {author} {\bibfnamefont {R.}~\bibnamefont {Thomale}}, \bibinfo
  {author} {\bibfnamefont {R.}~\bibnamefont {Valent}}, \ and\ \bibinfo {author}
  {\bibfnamefont {T.~O.}\ \bibnamefont {Wehling}},\ }\href@noop {} {\bibfield
  {journal} {\bibinfo  {journal} {Phys. Rev. B}\ }\textbf {\bibinfo {volume}
  {100}},\ \bibinfo {pages} {161102(R)} (\bibinfo {year} {2019})}\BibitemShut
  {NoStop}%
\bibitem [{\citenamefont {Goodwin}\ \emph
  {et~al.}(2019{\natexlab{a}})\citenamefont {Goodwin}, \citenamefont
  {Corsetti}, \citenamefont {Mostofi},\ and\ \citenamefont {Lischner}}]{PHD_2}%
  \BibitemOpen
  \bibfield  {author} {\bibinfo {author} {\bibfnamefont {Z.~A.~H.}\
  \bibnamefont {Goodwin}}, \bibinfo {author} {\bibfnamefont {F.}~\bibnamefont
  {Corsetti}}, \bibinfo {author} {\bibfnamefont {A.~A.}\ \bibnamefont
  {Mostofi}}, \ and\ \bibinfo {author} {\bibfnamefont {J.}~\bibnamefont
  {Lischner}},\ }\href@noop {} {\bibfield  {journal} {\bibinfo  {journal}
  {Phys. Rev. B}\ }\textbf {\bibinfo {volume} {100}},\ \bibinfo {pages}
  {235424} (\bibinfo {year} {2019}{\natexlab{a}})}\BibitemShut {NoStop}%
\bibitem [{\citenamefont {Padhi}\ \emph {et~al.}(2018)\citenamefont {Padhi},
  \citenamefont {Setty},\ and\ \citenamefont {Phillips}}]{WC}%
  \BibitemOpen
  \bibfield  {author} {\bibinfo {author} {\bibfnamefont {B.}~\bibnamefont
  {Padhi}}, \bibinfo {author} {\bibfnamefont {C.}~\bibnamefont {Setty}}, \ and\
  \bibinfo {author} {\bibfnamefont {P.~W.}\ \bibnamefont {Phillips}},\
  }\href@noop {} {\bibfield  {journal} {\bibinfo  {journal} {Nano Lett.}\
  }\textbf {\bibinfo {volume} {18}},\ \bibinfo {pages} {6175} (\bibinfo {year}
  {2018})}\BibitemShut {NoStop}%
\bibitem [{\citenamefont {Gonzalez-Arraga}\ \emph {et~al.}(2017)\citenamefont
  {Gonzalez-Arraga}, \citenamefont {Lado}, \citenamefont {Guinea},\ and\
  \citenamefont {San-Jose}}]{ECM}%
  \BibitemOpen
  \bibfield  {author} {\bibinfo {author} {\bibfnamefont {L.~A.}\ \bibnamefont
  {Gonzalez-Arraga}}, \bibinfo {author} {\bibfnamefont {J.~L.}\ \bibnamefont
  {Lado}}, \bibinfo {author} {\bibfnamefont {F.}~\bibnamefont {Guinea}}, \ and\
  \bibinfo {author} {\bibfnamefont {P.}~\bibnamefont {San-Jose}},\ }\href@noop
  {} {\bibfield  {journal} {\bibinfo  {journal} {Phys. Rev. Lett.}\ }\textbf
  {\bibinfo {volume} {119}},\ \bibinfo {pages} {107201} (\bibinfo {year}
  {2017})}\BibitemShut {NoStop}%
\bibitem [{\citenamefont {Guinea}\ and\ \citenamefont {Walet}(2018)}]{EE}%
  \BibitemOpen
  \bibfield  {author} {\bibinfo {author} {\bibfnamefont {F.}~\bibnamefont
  {Guinea}}\ and\ \bibinfo {author} {\bibfnamefont {N.~R.}\ \bibnamefont
  {Walet}},\ }\href@noop {} {\bibfield  {journal} {\bibinfo  {journal} {PNAS}\
  }\textbf {\bibinfo {volume} {115}},\ \bibinfo {pages} {13174–13179}
  (\bibinfo {year} {2018})}\BibitemShut {NoStop}%
\bibitem [{\citenamefont {Goodwin}\ \emph
  {et~al.}(2020{\natexlab{b}})\citenamefont {Goodwin}, \citenamefont {Vitale},
  \citenamefont {Liang}, \citenamefont {Mostofi},\ and\ \citenamefont
  {Lischner}}]{PHD_4}%
  \BibitemOpen
  \bibfield  {author} {\bibinfo {author} {\bibfnamefont {Z.~A.~H.}\
  \bibnamefont {Goodwin}}, \bibinfo {author} {\bibfnamefont {V.}~\bibnamefont
  {Vitale}}, \bibinfo {author} {\bibfnamefont {X.}~\bibnamefont {Liang}},
  \bibinfo {author} {\bibfnamefont {A.~A.}\ \bibnamefont {Mostofi}}, \ and\
  \bibinfo {author} {\bibfnamefont {J.}~\bibnamefont {Lischner}},\ }\href@noop
  {} {\bibfield  {journal} {\bibinfo  {journal} {Electron. Struct.}\ }\textbf
  {\bibinfo {volume} {2}},\ \bibinfo {pages} {034001} (\bibinfo {year}
  {2020}{\natexlab{b}})}\BibitemShut {NoStop}%
\bibitem [{\citenamefont {Tarnopolsky}\ \emph {et~al.}(2019)\citenamefont
  {Tarnopolsky}, \citenamefont {Kruchkov},\ and\ \citenamefont
  {Vishwanath}}]{OMACM}%
  \BibitemOpen
  \bibfield  {author} {\bibinfo {author} {\bibfnamefont {G.}~\bibnamefont
  {Tarnopolsky}}, \bibinfo {author} {\bibfnamefont {A.~J.}\ \bibnamefont
  {Kruchkov}}, \ and\ \bibinfo {author} {\bibfnamefont {A.}~\bibnamefont
  {Vishwanath}},\ }\href@noop {} {\bibfield  {journal} {\bibinfo  {journal}
  {Phys. Rev. Lett.}\ }\textbf {\bibinfo {volume} {122}},\ \bibinfo {pages}
  {106405} (\bibinfo {year} {2019})}\BibitemShut {NoStop}%
\bibitem [{\citenamefont {Goodwin}\ \emph
  {et~al.}(2019{\natexlab{b}})\citenamefont {Goodwin}, \citenamefont
  {Corsetti}, \citenamefont {Mostofi},\ and\ \citenamefont {Lischner}}]{PHD_1}%
  \BibitemOpen
  \bibfield  {author} {\bibinfo {author} {\bibfnamefont {Z.~A.~H.}\
  \bibnamefont {Goodwin}}, \bibinfo {author} {\bibfnamefont {F.}~\bibnamefont
  {Corsetti}}, \bibinfo {author} {\bibfnamefont {A.~A.}\ \bibnamefont
  {Mostofi}}, \ and\ \bibinfo {author} {\bibfnamefont {J.}~\bibnamefont
  {Lischner}},\ }\href@noop {} {\bibfield  {journal} {\bibinfo  {journal}
  {Phys. Rev. B}\ }\textbf {\bibinfo {volume} {100}},\ \bibinfo {pages}
  {121106(R)} (\bibinfo {year} {2019}{\natexlab{b}})}\BibitemShut {NoStop}%
\bibitem [{\citenamefont {Cao}\ \emph {et~al.}(2018{\natexlab{a}})\citenamefont
  {Cao}, \citenamefont {Fatemi}, \citenamefont {Fang}, \citenamefont
  {Watanabe}, \citenamefont {Taniguchi}, \citenamefont {Kaxiras},\ and\
  \citenamefont {Jarillo-Herrero}}]{YuanCao2018Usim}%
  \BibitemOpen
  \bibfield  {author} {\bibinfo {author} {\bibfnamefont {Y.}~\bibnamefont
  {Cao}}, \bibinfo {author} {\bibfnamefont {V.}~\bibnamefont {Fatemi}},
  \bibinfo {author} {\bibfnamefont {S.}~\bibnamefont {Fang}}, \bibinfo {author}
  {\bibfnamefont {K.}~\bibnamefont {Watanabe}}, \bibinfo {author}
  {\bibfnamefont {T.}~\bibnamefont {Taniguchi}}, \bibinfo {author}
  {\bibfnamefont {E.}~\bibnamefont {Kaxiras}}, \ and\ \bibinfo {author}
  {\bibfnamefont {P.}~\bibnamefont {Jarillo-Herrero}},\ }\href@noop {}
  {\bibfield  {journal} {\bibinfo  {journal} {Nature}\ }\textbf {\bibinfo
  {volume} {556}},\ \bibinfo {pages} {43} (\bibinfo {year}
  {2018}{\natexlab{a}})}\BibitemShut {NoStop}%
\bibitem [{\citenamefont {Cao}\ \emph {et~al.}(2018{\natexlab{b}})\citenamefont
  {Cao}, \citenamefont {Fatemi}, \citenamefont {Demir}, \citenamefont {Fang},
  \citenamefont {Tomarken}, \citenamefont {Luo}, \citenamefont
  {Sanchez-Yamagishi}, \citenamefont {Watanabe}, \citenamefont {Taniguchi},
  \citenamefont {Kaxiras}, \citenamefont {Ashoori},\ and\ \citenamefont
  {Jarillo-Herrero}}]{YuanCao2018Ciba}%
  \BibitemOpen
  \bibfield  {author} {\bibinfo {author} {\bibfnamefont {Y.}~\bibnamefont
  {Cao}}, \bibinfo {author} {\bibfnamefont {V.}~\bibnamefont {Fatemi}},
  \bibinfo {author} {\bibfnamefont {A.}~\bibnamefont {Demir}}, \bibinfo
  {author} {\bibfnamefont {S.}~\bibnamefont {Fang}}, \bibinfo {author}
  {\bibfnamefont {S.~L.}\ \bibnamefont {Tomarken}}, \bibinfo {author}
  {\bibfnamefont {J.~Y.}\ \bibnamefont {Luo}}, \bibinfo {author} {\bibfnamefont
  {J.~D.}\ \bibnamefont {Sanchez-Yamagishi}}, \bibinfo {author} {\bibfnamefont
  {K.}~\bibnamefont {Watanabe}}, \bibinfo {author} {\bibfnamefont
  {T.}~\bibnamefont {Taniguchi}}, \bibinfo {author} {\bibfnamefont
  {E.}~\bibnamefont {Kaxiras}}, \bibinfo {author} {\bibfnamefont {R.~C.}\
  \bibnamefont {Ashoori}}, \ and\ \bibinfo {author} {\bibfnamefont
  {P.}~\bibnamefont {Jarillo-Herrero}},\ }\href@noop {} {\bibfield  {journal}
  {\bibinfo  {journal} {Nature}\ }\textbf {\bibinfo {volume} {556}},\ \bibinfo
  {pages} {80} (\bibinfo {year} {2018}{\natexlab{b}})}\BibitemShut {NoStop}%
\bibitem [{\citenamefont {Yankowitz}\ \emph {et~al.}(2019)\citenamefont
  {Yankowitz}, \citenamefont {Chen}, \citenamefont {Polshyn}, \citenamefont
  {Zhang}, \citenamefont {Watanabe}, \citenamefont {Taniguchi}, \citenamefont
  {Graf}, \citenamefont {Young},\ and\ \citenamefont {Dean}}]{TSTBLG}%
  \BibitemOpen
  \bibfield  {author} {\bibinfo {author} {\bibfnamefont {M.}~\bibnamefont
  {Yankowitz}}, \bibinfo {author} {\bibfnamefont {S.}~\bibnamefont {Chen}},
  \bibinfo {author} {\bibfnamefont {H.}~\bibnamefont {Polshyn}}, \bibinfo
  {author} {\bibfnamefont {Y.}~\bibnamefont {Zhang}}, \bibinfo {author}
  {\bibfnamefont {K.}~\bibnamefont {Watanabe}}, \bibinfo {author}
  {\bibfnamefont {T.}~\bibnamefont {Taniguchi}}, \bibinfo {author}
  {\bibfnamefont {D.}~\bibnamefont {Graf}}, \bibinfo {author} {\bibfnamefont
  {A.~F.}\ \bibnamefont {Young}}, \ and\ \bibinfo {author} {\bibfnamefont
  {C.~R.}\ \bibnamefont {Dean}},\ }\href@noop {} {\bibfield  {journal}
  {\bibinfo  {journal} {Science}\ }\textbf {\bibinfo {volume} {363}},\ \bibinfo
  {pages} {1059} (\bibinfo {year} {2019})}\BibitemShut {NoStop}%
\bibitem [{\citenamefont {Lu}\ \emph {et~al.}(2019)\citenamefont {Lu},
  \citenamefont {Stepanov}, \citenamefont {Yang}, \citenamefont {Xie},
  \citenamefont {Aamir}, \citenamefont {Das}, \citenamefont {Urgell},
  \citenamefont {Watanabe}, \citenamefont {Taniguchi}, \citenamefont {Zhang},
  \citenamefont {Bachtold}, \citenamefont {MacDonald},\ and\ \citenamefont
  {Efetov}}]{SOM}%
  \BibitemOpen
  \bibfield  {author} {\bibinfo {author} {\bibfnamefont {X.}~\bibnamefont
  {Lu}}, \bibinfo {author} {\bibfnamefont {P.}~\bibnamefont {Stepanov}},
  \bibinfo {author} {\bibfnamefont {W.}~\bibnamefont {Yang}}, \bibinfo {author}
  {\bibfnamefont {M.}~\bibnamefont {Xie}}, \bibinfo {author} {\bibfnamefont
  {M.~A.}\ \bibnamefont {Aamir}}, \bibinfo {author} {\bibfnamefont
  {I.}~\bibnamefont {Das}}, \bibinfo {author} {\bibfnamefont {C.}~\bibnamefont
  {Urgell}}, \bibinfo {author} {\bibfnamefont {K.}~\bibnamefont {Watanabe}},
  \bibinfo {author} {\bibfnamefont {T.}~\bibnamefont {Taniguchi}}, \bibinfo
  {author} {\bibfnamefont {G.}~\bibnamefont {Zhang}}, \bibinfo {author}
  {\bibfnamefont {A.}~\bibnamefont {Bachtold}}, \bibinfo {author}
  {\bibfnamefont {A.~H.}\ \bibnamefont {MacDonald}}, \ and\ \bibinfo {author}
  {\bibfnamefont {D.~K.}\ \bibnamefont {Efetov}},\ }\href@noop {} {\bibfield
  {journal} {\bibinfo  {journal} {Nature}\ }\textbf {\bibinfo {volume} {574}},\
  \bibinfo {pages} {653–657} (\bibinfo {year} {2019})}\BibitemShut {NoStop}%
\bibitem [{\citenamefont {Xie}\ \emph {et~al.}(2019)\citenamefont {Xie},
  \citenamefont {Lian}, \citenamefont {J\"{a}ck}, \citenamefont {Liu},
  \citenamefont {Chiu}, \citenamefont {Watanabe}, \citenamefont {Taniguchi},
  \citenamefont {Bernevig},\ and\ \citenamefont {Yazdani}}]{NAT_SS}%
  \BibitemOpen
  \bibfield  {author} {\bibinfo {author} {\bibfnamefont {Y.}~\bibnamefont
  {Xie}}, \bibinfo {author} {\bibfnamefont {B.}~\bibnamefont {Lian}}, \bibinfo
  {author} {\bibfnamefont {B.}~\bibnamefont {J\"{a}ck}}, \bibinfo {author}
  {\bibfnamefont {X.}~\bibnamefont {Liu}}, \bibinfo {author} {\bibfnamefont
  {C.-L.}\ \bibnamefont {Chiu}}, \bibinfo {author} {\bibfnamefont
  {K.}~\bibnamefont {Watanabe}}, \bibinfo {author} {\bibfnamefont
  {T.}~\bibnamefont {Taniguchi}}, \bibinfo {author} {\bibfnamefont {B.~A.}\
  \bibnamefont {Bernevig}}, \ and\ \bibinfo {author} {\bibfnamefont
  {A.}~\bibnamefont {Yazdani}},\ }\href@noop {} {\bibfield  {journal} {\bibinfo
   {journal} {Nature}\ }\textbf {\bibinfo {volume} {572}},\ \bibinfo {pages}
  {101} (\bibinfo {year} {2019})}\BibitemShut {NoStop}%
\bibitem [{\citenamefont {Kerelsky}\ \emph {et~al.}(2019)\citenamefont
  {Kerelsky}, \citenamefont {McGilly}, \citenamefont {Kennes}, \citenamefont
  {Xian}, \citenamefont {Yankowitz}, \citenamefont {Chen}, \citenamefont
  {Watanabe}, \citenamefont {Taniguchi}, \citenamefont {Hone}, \citenamefont
  {Dean}, \citenamefont {Rubio},\ and\ \citenamefont {Pasupathy}}]{NAT_MEI}%
  \BibitemOpen
  \bibfield  {author} {\bibinfo {author} {\bibfnamefont {A.}~\bibnamefont
  {Kerelsky}}, \bibinfo {author} {\bibfnamefont {L.~J.}\ \bibnamefont
  {McGilly}}, \bibinfo {author} {\bibfnamefont {D.~M.}\ \bibnamefont {Kennes}},
  \bibinfo {author} {\bibfnamefont {L.}~\bibnamefont {Xian}}, \bibinfo {author}
  {\bibfnamefont {M.}~\bibnamefont {Yankowitz}}, \bibinfo {author}
  {\bibfnamefont {S.}~\bibnamefont {Chen}}, \bibinfo {author} {\bibfnamefont
  {K.}~\bibnamefont {Watanabe}}, \bibinfo {author} {\bibfnamefont
  {T.}~\bibnamefont {Taniguchi}}, \bibinfo {author} {\bibfnamefont
  {J.}~\bibnamefont {Hone}}, \bibinfo {author} {\bibfnamefont {C.}~\bibnamefont
  {Dean}}, \bibinfo {author} {\bibfnamefont {A.}~\bibnamefont {Rubio}}, \ and\
  \bibinfo {author} {\bibfnamefont {A.~N.}\ \bibnamefont {Pasupathy}},\
  }\href@noop {} {\bibfield  {journal} {\bibinfo  {journal} {Nature}\ }\textbf
  {\bibinfo {volume} {572}},\ \bibinfo {pages} {95} (\bibinfo {year}
  {2019})}\BibitemShut {NoStop}%
\bibitem [{\citenamefont {Jiang}\ \emph {et~al.}(2019)\citenamefont {Jiang},
  \citenamefont {Lai}, \citenamefont {Watanabe}, \citenamefont {Taniguchi},
  \citenamefont {Haule}, \citenamefont {Mao},\ and\ \citenamefont
  {Andrei}}]{NAT_CO}%
  \BibitemOpen
  \bibfield  {author} {\bibinfo {author} {\bibfnamefont {Y.}~\bibnamefont
  {Jiang}}, \bibinfo {author} {\bibfnamefont {X.}~\bibnamefont {Lai}}, \bibinfo
  {author} {\bibfnamefont {K.}~\bibnamefont {Watanabe}}, \bibinfo {author}
  {\bibfnamefont {T.}~\bibnamefont {Taniguchi}}, \bibinfo {author}
  {\bibfnamefont {K.}~\bibnamefont {Haule}}, \bibinfo {author} {\bibfnamefont
  {J.}~\bibnamefont {Mao}}, \ and\ \bibinfo {author} {\bibfnamefont {E.~Y.}\
  \bibnamefont {Andrei}},\ }\href@noop {} {\bibfield  {journal} {\bibinfo
  {journal} {Nature}\ }\textbf {\bibinfo {volume} {573}},\ \bibinfo {pages}
  {91} (\bibinfo {year} {2019})}\BibitemShut {NoStop}%
\bibitem [{\citenamefont {Xie}\ and\ \citenamefont {MacDonald}(2020)}]{SCHFC}%
  \BibitemOpen
  \bibfield  {author} {\bibinfo {author} {\bibfnamefont {M.}~\bibnamefont
  {Xie}}\ and\ \bibinfo {author} {\bibfnamefont {A.~H.}\ \bibnamefont
  {MacDonald}},\ }\href@noop {} {\bibfield  {journal} {\bibinfo  {journal}
  {Phys. Rev. Lett.}\ }\textbf {\bibinfo {volume} {124}},\ \bibinfo {pages}
  {097601} (\bibinfo {year} {2020})}\BibitemShut {NoStop}%
\bibitem [{\citenamefont {Ochi}\ \emph {et~al.}(2018)\citenamefont {Ochi},
  \citenamefont {Koshino},\ and\ \citenamefont {Kuroki}}]{PCIS}%
  \BibitemOpen
  \bibfield  {author} {\bibinfo {author} {\bibfnamefont {M.}~\bibnamefont
  {Ochi}}, \bibinfo {author} {\bibfnamefont {M.}~\bibnamefont {Koshino}}, \
  and\ \bibinfo {author} {\bibfnamefont {K.}~\bibnamefont {Kuroki}},\
  }\href@noop {} {\bibfield  {journal} {\bibinfo  {journal} {Phys. Rev. B}\
  }\textbf {\bibinfo {volume} {98}},\ \bibinfo {pages} {081102(R)} (\bibinfo
  {year} {2018})}\BibitemShut {NoStop}%
\bibitem [{\citenamefont {Kennes}\ \emph {et~al.}(2018)\citenamefont {Kennes},
  \citenamefont {Lischner},\ and\ \citenamefont {Karrasch}}]{SCDID}%
  \BibitemOpen
  \bibfield  {author} {\bibinfo {author} {\bibfnamefont {D.~M.}\ \bibnamefont
  {Kennes}}, \bibinfo {author} {\bibfnamefont {J.}~\bibnamefont {Lischner}}, \
  and\ \bibinfo {author} {\bibfnamefont {C.}~\bibnamefont {Karrasch}},\
  }\href@noop {} {\bibfield  {journal} {\bibinfo  {journal} {Phys. Rev. B}\
  }\textbf {\bibinfo {volume} {98}},\ \bibinfo {pages} {241407(R)} (\bibinfo
  {year} {2018})}\BibitemShut {NoStop}%
\bibitem [{\citenamefont {Naik}\ and\ \citenamefont {Jain}(2018)}]{UFSS}%
  \BibitemOpen
  \bibfield  {author} {\bibinfo {author} {\bibfnamefont {M.~H.}\ \bibnamefont
  {Naik}}\ and\ \bibinfo {author} {\bibfnamefont {M.}~\bibnamefont {Jain}},\
  }\href@noop {} {\bibfield  {journal} {\bibinfo  {journal} {Phys. Rev. Lett.}\
  }\textbf {\bibinfo {volume} {121}},\ \bibinfo {pages} {266401} (\bibinfo
  {year} {2018})}\BibitemShut {NoStop}%
\bibitem [{\citenamefont {Wu}\ \emph {et~al.}(2019)\citenamefont {Wu},
  \citenamefont {Lovorn}, \citenamefont {Tutuc}, \citenamefont {Martin},\ and\
  \citenamefont {MacDonald}}]{TITMD}%
  \BibitemOpen
  \bibfield  {author} {\bibinfo {author} {\bibfnamefont {F.}~\bibnamefont
  {Wu}}, \bibinfo {author} {\bibfnamefont {T.}~\bibnamefont {Lovorn}}, \bibinfo
  {author} {\bibfnamefont {E.}~\bibnamefont {Tutuc}}, \bibinfo {author}
  {\bibfnamefont {I.}~\bibnamefont {Martin}}, \ and\ \bibinfo {author}
  {\bibfnamefont {A.~H.}\ \bibnamefont {MacDonald}},\ }\href@noop {} {\bibfield
   {journal} {\bibinfo  {journal} {Phys. Rev. Lett.}\ }\textbf {\bibinfo
  {volume} {122}},\ \bibinfo {pages} {086402} (\bibinfo {year}
  {2019})}\BibitemShut {NoStop}%
\bibitem [{\citenamefont {Wang}\ \emph {et~al.}(2012)\citenamefont {Wang},
  \citenamefont {Kalantar-Zadeh}, \citenamefont {Kis}, \citenamefont
  {Coleman},\ and\ \citenamefont {Strano}}]{WangQing2012Eaoo}%
  \BibitemOpen
  \bibfield  {author} {\bibinfo {author} {\bibfnamefont {Q.}~\bibnamefont
  {Wang}}, \bibinfo {author} {\bibfnamefont {K.}~\bibnamefont
  {Kalantar-Zadeh}}, \bibinfo {author} {\bibfnamefont {A.}~\bibnamefont {Kis}},
  \bibinfo {author} {\bibfnamefont {J.}~\bibnamefont {Coleman}}, \ and\
  \bibinfo {author} {\bibfnamefont {M.}~\bibnamefont {Strano}},\ }\href@noop {}
  {\bibfield  {journal} {\bibinfo  {journal} {Nat. Nanotechnol.}\ }\textbf
  {\bibinfo {volume} {7}},\ \bibinfo {pages} {699} (\bibinfo {year}
  {2012})}\BibitemShut {NoStop}%
\bibitem [{\citenamefont {Jariwala}\ \emph {et~al.}(2014)\citenamefont
  {Jariwala}, \citenamefont {Sangwan}, \citenamefont {Lauhon}, \citenamefont
  {Marks},\ and\ \citenamefont {Hersam}}]{JariwalaDeep2014Edaf}%
  \BibitemOpen
  \bibfield  {author} {\bibinfo {author} {\bibfnamefont {D.}~\bibnamefont
  {Jariwala}}, \bibinfo {author} {\bibfnamefont {V.~K.}\ \bibnamefont
  {Sangwan}}, \bibinfo {author} {\bibfnamefont {L.~J.}\ \bibnamefont {Lauhon}},
  \bibinfo {author} {\bibfnamefont {T.~J.}\ \bibnamefont {Marks}}, \ and\
  \bibinfo {author} {\bibfnamefont {M.~C.}\ \bibnamefont {Hersam}},\
  }\href@noop {} {\bibfield  {journal} {\bibinfo  {journal} {ACS nano}\
  }\textbf {\bibinfo {volume} {8}},\ \bibinfo {pages} {1102} (\bibinfo {year}
  {2014})}\BibitemShut {NoStop}%
\bibitem [{\citenamefont {Mak}\ and\ \citenamefont
  {Shan}(2016)}]{KinFaiMak2016Paoo}%
  \BibitemOpen
  \bibfield  {author} {\bibinfo {author} {\bibfnamefont {K.~F.}\ \bibnamefont
  {Mak}}\ and\ \bibinfo {author} {\bibfnamefont {J.}~\bibnamefont {Shan}},\
  }\href@noop {} {\bibfield  {journal} {\bibinfo  {journal} {Nat. Photon.}\
  }\textbf {\bibinfo {volume} {10}},\ \bibinfo {pages} {216} (\bibinfo {year}
  {2016})}\BibitemShut {NoStop}%
\bibitem [{\citenamefont {Koshino}(2019)}]{KoshinoMikito2019Bsat}%
  \BibitemOpen
  \bibfield  {author} {\bibinfo {author} {\bibfnamefont {M.}~\bibnamefont
  {Koshino}},\ }\href@noop {} {\bibfield  {journal} {\bibinfo  {journal} {Phys.
  Rev. B}\ }\textbf {\bibinfo {volume} {99}},\ \bibinfo {pages} {235406}
  (\bibinfo {year} {2019})}\BibitemShut {NoStop}%
\bibitem [{\citenamefont {Liu}\ \emph {et~al.}(2020)\citenamefont {Liu},
  \citenamefont {Hao}, \citenamefont {Khalaf}, \citenamefont {Lee},
  \citenamefont {Watanabe}, \citenamefont {Taniguchi}, \citenamefont
  {Vishwanath},\ and\ \citenamefont {Kim}}]{BIBI}%
  \BibitemOpen
  \bibfield  {author} {\bibinfo {author} {\bibfnamefont {X.}~\bibnamefont
  {Liu}}, \bibinfo {author} {\bibfnamefont {Z.}~\bibnamefont {Hao}}, \bibinfo
  {author} {\bibfnamefont {E.}~\bibnamefont {Khalaf}}, \bibinfo {author}
  {\bibfnamefont {J.~Y.}\ \bibnamefont {Lee}}, \bibinfo {author} {\bibfnamefont
  {K.}~\bibnamefont {Watanabe}}, \bibinfo {author} {\bibfnamefont
  {T.}~\bibnamefont {Taniguchi}}, \bibinfo {author} {\bibfnamefont
  {A.}~\bibnamefont {Vishwanath}}, \ and\ \bibinfo {author} {\bibfnamefont
  {P.}~\bibnamefont {Kim}},\ }\href@noop {} {\bibfield  {journal} {\bibinfo
  {journal} {Nature}\ }\textbf {\bibinfo {volume} {583}},\ \bibinfo {pages}
  {221} (\bibinfo {year} {2020})}\BibitemShut {NoStop}%
\bibitem [{\citenamefont {Haddadi}\ \emph {et~al.}(2020)\citenamefont
  {Haddadi}, \citenamefont {Wu}, \citenamefont {Kruchkov},\ and\ \citenamefont
  {Yazyev}}]{haddadi2019moir}%
  \BibitemOpen
  \bibfield  {author} {\bibinfo {author} {\bibfnamefont {F.}~\bibnamefont
  {Haddadi}}, \bibinfo {author} {\bibfnamefont {Q.}~\bibnamefont {Wu}},
  \bibinfo {author} {\bibfnamefont {A.~J.}\ \bibnamefont {Kruchkov}}, \ and\
  \bibinfo {author} {\bibfnamefont {O.~V.}\ \bibnamefont {Yazyev}},\
  }\href@noop {} {\bibfield  {journal} {\bibinfo  {journal} {Nano Lett.}\
  }\textbf {\bibinfo {volume} {20}},\ \bibinfo {pages} {2410–2415} (\bibinfo
  {year} {2020})}\BibitemShut {NoStop}%
\bibitem [{\citenamefont {Burg}\ \emph {et~al.}(2019)\citenamefont {Burg},
  \citenamefont {Zhu}, \citenamefont {Taniguchi}, \citenamefont {Watanabe},
  \citenamefont {MacDonald},\ and\ \citenamefont
  {Tutuc}}]{PhysRevLett.123.197702}%
  \BibitemOpen
  \bibfield  {author} {\bibinfo {author} {\bibfnamefont {G.~W.}\ \bibnamefont
  {Burg}}, \bibinfo {author} {\bibfnamefont {J.}~\bibnamefont {Zhu}}, \bibinfo
  {author} {\bibfnamefont {T.}~\bibnamefont {Taniguchi}}, \bibinfo {author}
  {\bibfnamefont {K.}~\bibnamefont {Watanabe}}, \bibinfo {author}
  {\bibfnamefont {A.~H.}\ \bibnamefont {MacDonald}}, \ and\ \bibinfo {author}
  {\bibfnamefont {E.}~\bibnamefont {Tutuc}},\ }\href@noop {} {\bibfield
  {journal} {\bibinfo  {journal} {Phys. Rev. Lett.}\ }\textbf {\bibinfo
  {volume} {123}},\ \bibinfo {pages} {197702} (\bibinfo {year}
  {2019})}\BibitemShut {NoStop}%
\bibitem [{\citenamefont {Leey}\ \emph {et~al.}(2019)\citenamefont {Leey},
  \citenamefont {Khalafy}, \citenamefont {Liu}, \citenamefont {Liu},
  \citenamefont {Hao}, \citenamefont {Kim},\ and\ \citenamefont
  {Vishwanath}}]{STSCTDB}%
  \BibitemOpen
  \bibfield  {author} {\bibinfo {author} {\bibfnamefont {J.~Y.}\ \bibnamefont
  {Leey}}, \bibinfo {author} {\bibfnamefont {E.}~\bibnamefont {Khalafy}},
  \bibinfo {author} {\bibfnamefont {S.}~\bibnamefont {Liu}}, \bibinfo {author}
  {\bibfnamefont {X.}~\bibnamefont {Liu}}, \bibinfo {author} {\bibfnamefont
  {Z.}~\bibnamefont {Hao}}, \bibinfo {author} {\bibfnamefont {P.}~\bibnamefont
  {Kim}}, \ and\ \bibinfo {author} {\bibfnamefont {A.}~\bibnamefont
  {Vishwanath}},\ }\href@noop {} {\bibfield  {journal} {\bibinfo  {journal}
  {Nat. Commun.}\ }\textbf {\bibinfo {volume} {10}},\ \bibinfo {pages} {5333}
  (\bibinfo {year} {2019})}\BibitemShut {NoStop}%
\bibitem [{\citenamefont {Samajdar}\ and\ \citenamefont
  {Scheurer}(2020)}]{Samajdar2020}%
  \BibitemOpen
  \bibfield  {author} {\bibinfo {author} {\bibfnamefont {R.}~\bibnamefont
  {Samajdar}}\ and\ \bibinfo {author} {\bibfnamefont {M.~S.}\ \bibnamefont
  {Scheurer}},\ }\href@noop {} {\bibfield  {journal} {\bibinfo  {journal}
  {Phys. Rev. B}\ }\textbf {\bibinfo {volume} {102}},\ \bibinfo {pages}
  {064501} (\bibinfo {year} {2020})}\BibitemShut {NoStop}%
\bibitem [{\citenamefont {Cao}\ \emph {et~al.}(2020)\citenamefont {Cao},
  \citenamefont {Rodan-Legrain}, \citenamefont {Rubies-Bigorda}, \citenamefont
  {Park}, \citenamefont {Watanabe}, \citenamefont {Taniguchi},\ and\
  \citenamefont {Jarillo-Herrero}}]{cao2019electric}%
  \BibitemOpen
  \bibfield  {author} {\bibinfo {author} {\bibfnamefont {Y.}~\bibnamefont
  {Cao}}, \bibinfo {author} {\bibfnamefont {D.}~\bibnamefont {Rodan-Legrain}},
  \bibinfo {author} {\bibfnamefont {O.}~\bibnamefont {Rubies-Bigorda}},
  \bibinfo {author} {\bibfnamefont {J.~M.}\ \bibnamefont {Park}}, \bibinfo
  {author} {\bibfnamefont {K.}~\bibnamefont {Watanabe}}, \bibinfo {author}
  {\bibfnamefont {T.}~\bibnamefont {Taniguchi}}, \ and\ \bibinfo {author}
  {\bibfnamefont {P.}~\bibnamefont {Jarillo-Herrero}},\ }\href@noop {}
  {\bibfield  {journal} {\bibinfo  {journal} {Nature}\ }\textbf {\bibinfo
  {volume} {583}},\ \bibinfo {pages} {215} (\bibinfo {year}
  {2020})}\BibitemShut {NoStop}%
\bibitem [{\citenamefont {Shen}\ \emph {et~al.}(2020)\citenamefont {Shen},
  \citenamefont {Chu}, \citenamefont {Wu}, \citenamefont {Li}, \citenamefont
  {Wang}, \citenamefont {Zhao}, \citenamefont {Tang}, \citenamefont {Liu},
  \citenamefont {Tian}, \citenamefont {Watanabe}, \citenamefont {Taniguchi},
  \citenamefont {Yang}, \citenamefont {Meng}, \citenamefont {Shi},
  \citenamefont {Yazyev},\ and\ \citenamefont {Zhang}}]{TCT}%
  \BibitemOpen
  \bibfield  {author} {\bibinfo {author} {\bibfnamefont {C.}~\bibnamefont
  {Shen}}, \bibinfo {author} {\bibfnamefont {Y.}~\bibnamefont {Chu}}, \bibinfo
  {author} {\bibfnamefont {Q.}~\bibnamefont {Wu}}, \bibinfo {author}
  {\bibfnamefont {N.}~\bibnamefont {Li}}, \bibinfo {author} {\bibfnamefont
  {S.}~\bibnamefont {Wang}}, \bibinfo {author} {\bibfnamefont {Y.}~\bibnamefont
  {Zhao}}, \bibinfo {author} {\bibfnamefont {J.}~\bibnamefont {Tang}}, \bibinfo
  {author} {\bibfnamefont {J.}~\bibnamefont {Liu}}, \bibinfo {author}
  {\bibfnamefont {J.}~\bibnamefont {Tian}}, \bibinfo {author} {\bibfnamefont
  {K.}~\bibnamefont {Watanabe}}, \bibinfo {author} {\bibfnamefont
  {T.}~\bibnamefont {Taniguchi}}, \bibinfo {author} {\bibfnamefont
  {R.}~\bibnamefont {Yang}}, \bibinfo {author} {\bibfnamefont {Z.~Y.}\
  \bibnamefont {Meng}}, \bibinfo {author} {\bibfnamefont {D.}~\bibnamefont
  {Shi}}, \bibinfo {author} {\bibfnamefont {O.~V.}\ \bibnamefont {Yazyev}}, \
  and\ \bibinfo {author} {\bibfnamefont {G.}~\bibnamefont {Zhang}},\
  }\href@noop {} {\bibfield  {journal} {\bibinfo  {journal} {Nat. Phys.}\
  }\textbf {\bibinfo {volume} {16}},\ \bibinfo {pages} {520–525} (\bibinfo
  {year} {2020})}\BibitemShut {NoStop}%
\bibitem [{\citenamefont {Rickhaus}\ \emph {et~al.}(2019)\citenamefont
  {Rickhaus}, \citenamefont {Zheng}, \citenamefont {Lado}, \citenamefont {Lee},
  \citenamefont {Kurzmann}, \citenamefont {Eich}, \citenamefont {Pisoni},
  \citenamefont {Tong}, \citenamefont {Garreis}, \citenamefont {Gold},
  \citenamefont {Masseroni}, \citenamefont {Taniguchi}, \citenamefont
  {Wantanabe}, \citenamefont {Ihn},\ and\ \citenamefont
  {Ensslin}}]{RickhausPeter2019GOiT}%
  \BibitemOpen
  \bibfield  {author} {\bibinfo {author} {\bibfnamefont {P.}~\bibnamefont
  {Rickhaus}}, \bibinfo {author} {\bibfnamefont {G.}~\bibnamefont {Zheng}},
  \bibinfo {author} {\bibfnamefont {J.~L.}\ \bibnamefont {Lado}}, \bibinfo
  {author} {\bibfnamefont {Y.}~\bibnamefont {Lee}}, \bibinfo {author}
  {\bibfnamefont {A.}~\bibnamefont {Kurzmann}}, \bibinfo {author}
  {\bibfnamefont {M.}~\bibnamefont {Eich}}, \bibinfo {author} {\bibfnamefont
  {R.}~\bibnamefont {Pisoni}}, \bibinfo {author} {\bibfnamefont
  {C.}~\bibnamefont {Tong}}, \bibinfo {author} {\bibfnamefont {R.}~\bibnamefont
  {Garreis}}, \bibinfo {author} {\bibfnamefont {C.}~\bibnamefont {Gold}},
  \bibinfo {author} {\bibfnamefont {M.}~\bibnamefont {Masseroni}}, \bibinfo
  {author} {\bibfnamefont {T.}~\bibnamefont {Taniguchi}}, \bibinfo {author}
  {\bibfnamefont {K.}~\bibnamefont {Wantanabe}}, \bibinfo {author}
  {\bibfnamefont {T.}~\bibnamefont {Ihn}}, \ and\ \bibinfo {author}
  {\bibfnamefont {K.}~\bibnamefont {Ensslin}},\ }\href@noop {} {\bibfield
  {journal} {\bibinfo  {journal} {Nano lett.}\ }\textbf {\bibinfo {volume}
  {19}},\ \bibinfo {pages} {8821–8828} (\bibinfo {year} {2019})}\BibitemShut
  {NoStop}%
\bibitem [{\citenamefont {Choi}\ and\ \citenamefont
  {Choi}(2019)}]{ChoiY.W.2019Ibga}%
  \BibitemOpen
  \bibfield  {author} {\bibinfo {author} {\bibfnamefont {Y.}~\bibnamefont
  {Choi}}\ and\ \bibinfo {author} {\bibfnamefont {H.}~\bibnamefont {Choi}},\
  }\href@noop {} {\bibfield  {journal} {\bibinfo  {journal} {Phys. Rev. B}\
  }\textbf {\bibinfo {volume} {100}},\ \bibinfo {pages} {201402(R)} (\bibinfo
  {year} {2019})}\BibitemShut {NoStop}%
\bibitem [{\citenamefont {Chebrolu}\ \emph {et~al.}(2019)\citenamefont
  {Chebrolu}, \citenamefont {Chittari},\ and\ \citenamefont
  {Jung}}]{ChebroluNarasimha2019Fbit}%
  \BibitemOpen
  \bibfield  {author} {\bibinfo {author} {\bibfnamefont {N.}~\bibnamefont
  {Chebrolu}}, \bibinfo {author} {\bibfnamefont {B.}~\bibnamefont {Chittari}},
  \ and\ \bibinfo {author} {\bibfnamefont {J.}~\bibnamefont {Jung}},\
  }\href@noop {} {\bibfield  {journal} {\bibinfo  {journal} {Phys. Rev. B}\
  }\textbf {\bibinfo {volume} {99}},\ \bibinfo {pages} {235417} (\bibinfo
  {year} {2019})}\BibitemShut {NoStop}%
\bibitem [{\citenamefont {Culchac}\ \emph {et~al.}(2020)\citenamefont
  {Culchac}, \citenamefont {Del~Grande}, \citenamefont {Capaz}, \citenamefont
  {Chico},\ and\ \citenamefont {Morell}}]{CulchacF.J.2020Fbag}%
  \BibitemOpen
  \bibfield  {author} {\bibinfo {author} {\bibfnamefont {F.~J.}\ \bibnamefont
  {Culchac}}, \bibinfo {author} {\bibfnamefont {R.~R.}\ \bibnamefont
  {Del~Grande}}, \bibinfo {author} {\bibfnamefont {R.~B.}\ \bibnamefont
  {Capaz}}, \bibinfo {author} {\bibfnamefont {L.}~\bibnamefont {Chico}}, \ and\
  \bibinfo {author} {\bibfnamefont {E.~S.}\ \bibnamefont {Morell}},\
  }\href@noop {} {\bibfield  {journal} {\bibinfo  {journal} {Nanoscale}\
  }\textbf {\bibinfo {volume} {12}},\ \bibinfo {pages} {5014} (\bibinfo {year}
  {2020})}\BibitemShut {NoStop}%
\bibitem [{\citenamefont {Adak}\ \emph {et~al.}(2020)\citenamefont {Adak},
  \citenamefont {Sinha}, \citenamefont {Ghorai}, \citenamefont {Sangani},
  \citenamefont {Watanabe}, \citenamefont {Taniguchi}, \citenamefont
  {Sensarma},\ and\ \citenamefont {Deshmukh}}]{AdakPratap2020Tbag}%
  \BibitemOpen
  \bibfield  {author} {\bibinfo {author} {\bibfnamefont {P.}~\bibnamefont
  {Adak}}, \bibinfo {author} {\bibfnamefont {S.}~\bibnamefont {Sinha}},
  \bibinfo {author} {\bibfnamefont {U.}~\bibnamefont {Ghorai}}, \bibinfo
  {author} {\bibfnamefont {L.}~\bibnamefont {Sangani}}, \bibinfo {author}
  {\bibfnamefont {K.}~\bibnamefont {Watanabe}}, \bibinfo {author}
  {\bibfnamefont {T.}~\bibnamefont {Taniguchi}}, \bibinfo {author}
  {\bibfnamefont {R.}~\bibnamefont {Sensarma}}, \ and\ \bibinfo {author}
  {\bibfnamefont {M.}~\bibnamefont {Deshmukh}},\ }\href@noop {} {\bibfield
  {journal} {\bibinfo  {journal} {Phys. Rev. B}\ }\textbf {\bibinfo {volume}
  {101}},\ \bibinfo {pages} {125428} (\bibinfo {year} {2020})}\BibitemShut
  {NoStop}%
\bibitem [{\citenamefont {Wu}\ and\ \citenamefont
  {Sarma}(2020)}]{WuFengcheng2020Fasi}%
  \BibitemOpen
  \bibfield  {author} {\bibinfo {author} {\bibfnamefont {F.}~\bibnamefont
  {Wu}}\ and\ \bibinfo {author} {\bibfnamefont {S.}~\bibnamefont {Sarma}},\
  }\href@noop {} {\bibfield  {journal} {\bibinfo  {journal} {Phys. Rev. B}\
  }\textbf {\bibinfo {volume} {101}},\ \bibinfo {pages} {155149} (\bibinfo
  {year} {2020})}\BibitemShut {NoStop}%
\bibitem [{\citenamefont {Lee}\ \emph {et~al.}(2016)\citenamefont {Lee},
  \citenamefont {Kim}, \citenamefont {Hembram}, \citenamefont {Kim},
  \citenamefont {Min}, \citenamefont {Park}, \citenamefont {Lee}, \citenamefont
  {Moon}, \citenamefont {Lee}, \citenamefont {Lee},\ and\ \citenamefont
  {John}}]{META_AAp}%
  \BibitemOpen
  \bibfield  {author} {\bibinfo {author} {\bibfnamefont {J.-K.}\ \bibnamefont
  {Lee}}, \bibinfo {author} {\bibfnamefont {J.-G.}\ \bibnamefont {Kim}},
  \bibinfo {author} {\bibfnamefont {K.~P. S.~S.}\ \bibnamefont {Hembram}},
  \bibinfo {author} {\bibfnamefont {Y.-I.}\ \bibnamefont {Kim}}, \bibinfo
  {author} {\bibfnamefont {B.-K.}\ \bibnamefont {Min}}, \bibinfo {author}
  {\bibfnamefont {Y.}~\bibnamefont {Park}}, \bibinfo {author} {\bibfnamefont
  {J.-K.}\ \bibnamefont {Lee}}, \bibinfo {author} {\bibfnamefont {D.~J.}\
  \bibnamefont {Moon}}, \bibinfo {author} {\bibfnamefont {W.}~\bibnamefont
  {Lee}}, \bibinfo {author} {\bibfnamefont {S.-G.}\ \bibnamefont {Lee}}, \ and\
  \bibinfo {author} {\bibfnamefont {P.}~\bibnamefont {John}},\ }\href@noop {}
  {\bibfield  {journal} {\bibinfo  {journal} {Sci. Rep.}\ }\textbf {\bibinfo
  {volume} {6}},\ \bibinfo {pages} {39624} (\bibinfo {year}
  {2016})}\BibitemShut {NoStop}%
\bibitem [{\citenamefont {Rakhmanov}\ \emph {et~al.}(2012)\citenamefont
  {Rakhmanov}, \citenamefont {Rozhkov}, \citenamefont {Sboychakov},\ and\
  \citenamefont {Nori}}]{RakhmanovAL2012IotA}%
  \BibitemOpen
  \bibfield  {author} {\bibinfo {author} {\bibfnamefont {A.~L.}\ \bibnamefont
  {Rakhmanov}}, \bibinfo {author} {\bibfnamefont {A.~V.}\ \bibnamefont
  {Rozhkov}}, \bibinfo {author} {\bibfnamefont {A.~O.}\ \bibnamefont
  {Sboychakov}}, \ and\ \bibinfo {author} {\bibfnamefont {F.}~\bibnamefont
  {Nori}},\ }\href@noop {} {\bibfield  {journal} {\bibinfo  {journal} {Phys.
  Rev. Lett.}\ }\textbf {\bibinfo {volume} {109}},\ \bibinfo {pages} {206801}
  (\bibinfo {year} {2012})}\BibitemShut {NoStop}%
\bibitem [{\citenamefont {Rozhkov}\ \emph {et~al.}(2016)\citenamefont
  {Rozhkov}, \citenamefont {Sboychakov}, \citenamefont {Rakhmanov},\ and\
  \citenamefont {Nori}}]{RozhkovA.V2016Epog}%
  \BibitemOpen
  \bibfield  {author} {\bibinfo {author} {\bibfnamefont {A.}~\bibnamefont
  {Rozhkov}}, \bibinfo {author} {\bibfnamefont {A.}~\bibnamefont {Sboychakov}},
  \bibinfo {author} {\bibfnamefont {A.}~\bibnamefont {Rakhmanov}}, \ and\
  \bibinfo {author} {\bibfnamefont {F.}~\bibnamefont {Nori}},\ }\href@noop {}
  {\bibfield  {journal} {\bibinfo  {journal} {Phys. Rep.}\ }\textbf {\bibinfo
  {volume} {648}},\ \bibinfo {pages} {1} (\bibinfo {year} {2016})}\BibitemShut
  {NoStop}%
\bibitem [{\citenamefont {Wang}\ and\ \citenamefont
  {Jin}(2012)}]{WangDali2012Tetc}%
  \BibitemOpen
  \bibfield  {author} {\bibinfo {author} {\bibfnamefont {D.}~\bibnamefont
  {Wang}}\ and\ \bibinfo {author} {\bibfnamefont {G.}~\bibnamefont {Jin}},\
  }\href@noop {} {\bibfield  {journal} {\bibinfo  {journal} {J. Appl. Phys.}\
  }\textbf {\bibinfo {volume} {112}},\ \bibinfo {pages} {053714} (\bibinfo
  {year} {2012})}\BibitemShut {NoStop}%
\bibitem [{\citenamefont {Tsai}\ \emph {et~al.}(2012)\citenamefont {Tsai},
  \citenamefont {Chiu}, \citenamefont {Ho},\ and\ \citenamefont
  {Lin}}]{TsaiSing-Jyun2012GLli}%
  \BibitemOpen
  \bibfield  {author} {\bibinfo {author} {\bibfnamefont {S.-J.}\ \bibnamefont
  {Tsai}}, \bibinfo {author} {\bibfnamefont {Y.-H.}\ \bibnamefont {Chiu}},
  \bibinfo {author} {\bibfnamefont {Y.-H.}\ \bibnamefont {Ho}}, \ and\ \bibinfo
  {author} {\bibfnamefont {M.-F.}\ \bibnamefont {Lin}},\ }\href@noop {}
  {\bibfield  {journal} {\bibinfo  {journal} {Chem. Phys. Lett.}\ }\textbf
  {\bibinfo {volume} {550}},\ \bibinfo {pages} {104} (\bibinfo {year}
  {2012})}\BibitemShut {NoStop}%
\bibitem [{\citenamefont {Abdullah}\ \emph {et~al.}(2018)\citenamefont
  {Abdullah}, \citenamefont {Al~Ezzi},\ and\ \citenamefont
  {Bahlouli}}]{AbdullahHasanM.2018EtaK}%
  \BibitemOpen
  \bibfield  {author} {\bibinfo {author} {\bibfnamefont {H.~M.}\ \bibnamefont
  {Abdullah}}, \bibinfo {author} {\bibfnamefont {M.}~\bibnamefont {Al~Ezzi}}, \
  and\ \bibinfo {author} {\bibfnamefont {H.}~\bibnamefont {Bahlouli}},\
  }\href@noop {} {\bibfield  {journal} {\bibinfo  {journal} {J. Appl. Phys.}\
  }\textbf {\bibinfo {volume} {124}},\ \bibinfo {pages} {204303} (\bibinfo
  {year} {2018})}\BibitemShut {NoStop}%
\bibitem [{\citenamefont {Neto}\ \emph {et~al.}(2009)\citenamefont {Neto},
  \citenamefont {Guinea}, \citenamefont {Peres}, \citenamefont {Novoselov},\
  and\ \citenamefont {Geim}}]{EPG}%
  \BibitemOpen
  \bibfield  {author} {\bibinfo {author} {\bibfnamefont {A.~H.~C.}\
  \bibnamefont {Neto}}, \bibinfo {author} {\bibfnamefont {F.}~\bibnamefont
  {Guinea}}, \bibinfo {author} {\bibfnamefont {N.~M.~R.}\ \bibnamefont
  {Peres}}, \bibinfo {author} {\bibfnamefont {K.~S.}\ \bibnamefont
  {Novoselov}}, \ and\ \bibinfo {author} {\bibfnamefont {A.~K.}\ \bibnamefont
  {Geim}},\ }\href@noop {} {\bibfield  {journal} {\bibinfo  {journal} {Rev.
  Mod. Phys.}\ }\textbf {\bibinfo {volume} {81}},\ \bibinfo {pages} {109}
  (\bibinfo {year} {2009})}\BibitemShut {NoStop}%
\bibitem [{\citenamefont {Savini}\ \emph {et~al.}(2011)\citenamefont {Savini},
  \citenamefont {Dappe}, \citenamefont {\"{O}berg}, \citenamefont {Charlier},
  \citenamefont {Katsnelson},\ and\ \citenamefont {Fasolino}}]{SAVINI201162}%
  \BibitemOpen
  \bibfield  {author} {\bibinfo {author} {\bibfnamefont {G.}~\bibnamefont
  {Savini}}, \bibinfo {author} {\bibfnamefont {Y.}~\bibnamefont {Dappe}},
  \bibinfo {author} {\bibfnamefont {S.}~\bibnamefont {\"{O}berg}}, \bibinfo
  {author} {\bibfnamefont {J.-C.}\ \bibnamefont {Charlier}}, \bibinfo {author}
  {\bibfnamefont {M.}~\bibnamefont {Katsnelson}}, \ and\ \bibinfo {author}
  {\bibfnamefont {A.}~\bibnamefont {Fasolino}},\ }\href@noop {} {\bibfield
  {journal} {\bibinfo  {journal} {Carbon}\ }\textbf {\bibinfo {volume} {49}},\
  \bibinfo {pages} {62 } (\bibinfo {year} {2011})}\BibitemShut {NoStop}%
\bibitem [{\citenamefont {Shallcross}\ \emph {et~al.}(2010)\citenamefont
  {Shallcross}, \citenamefont {Sharma}, \citenamefont {Kandelaki},\ and\
  \citenamefont {Pankratov}}]{PhysRevB.81.165105}%
  \BibitemOpen
  \bibfield  {author} {\bibinfo {author} {\bibfnamefont {S.}~\bibnamefont
  {Shallcross}}, \bibinfo {author} {\bibfnamefont {S.}~\bibnamefont {Sharma}},
  \bibinfo {author} {\bibfnamefont {E.}~\bibnamefont {Kandelaki}}, \ and\
  \bibinfo {author} {\bibfnamefont {O.~A.}\ \bibnamefont {Pankratov}},\
  }\href@noop {} {\bibfield  {journal} {\bibinfo  {journal} {Phys. Rev. B}\
  }\textbf {\bibinfo {volume} {81}},\ \bibinfo {pages} {165105} (\bibinfo
  {year} {2010})}\BibitemShut {NoStop}%
\bibitem [{\citenamefont {Plimpton}(1995)}]{LAMMPS}%
  \BibitemOpen
  \bibfield  {author} {\bibinfo {author} {\bibfnamefont {S.}~\bibnamefont
  {Plimpton}},\ }\href@noop {} {\bibfield  {journal} {\bibinfo  {journal} {J.
  Comp. Phys.}\ }\textbf {\bibinfo {volume} {117}},\ \bibinfo {pages} {1}
  (\bibinfo {year} {1995})}\BibitemShut {NoStop}%
\bibitem [{\citenamefont {O’Connor}\ \emph {et~al.}(2015)\citenamefont
  {O’Connor}, \citenamefont {Andzelm},\ and\ \citenamefont
  {Robbins}}]{AIREBO}%
  \BibitemOpen
  \bibfield  {author} {\bibinfo {author} {\bibfnamefont {T.~C.}\ \bibnamefont
  {O’Connor}}, \bibinfo {author} {\bibfnamefont {J.}~\bibnamefont {Andzelm}},
  \ and\ \bibinfo {author} {\bibfnamefont {M.~O.}\ \bibnamefont {Robbins}},\
  }\href@noop {} {\bibfield  {journal} {\bibinfo  {journal} {J. Chem. Phys.}\
  }\textbf {\bibinfo {volume} {142}},\ \bibinfo {pages} {024903} (\bibinfo
  {year} {2015})}\BibitemShut {NoStop}%
\bibitem [{\citenamefont {Guinea}\ and\ \citenamefont
  {Walet}(2019)}]{GuineaF.2019Cmft}%
  \BibitemOpen
  \bibfield  {author} {\bibinfo {author} {\bibfnamefont {F.}~\bibnamefont
  {Guinea}}\ and\ \bibinfo {author} {\bibfnamefont {N.}~\bibnamefont {Walet}},\
  }\href@noop {} {\bibfield  {journal} {\bibinfo  {journal} {Phys. Rev. B}\
  }\textbf {\bibinfo {volume} {99}},\ \bibinfo {pages} {205134} (\bibinfo
  {year} {2019})}\BibitemShut {NoStop}%
\bibitem [{\citenamefont {van Wijk}\ \emph {et~al.}(2015)\citenamefont {van
  Wijk}, \citenamefont {Schuring}, \citenamefont {Katsnelson},\ and\
  \citenamefont {Fasolino}}]{Wijk_2015}%
  \BibitemOpen
  \bibfield  {author} {\bibinfo {author} {\bibfnamefont {M.~M.}\ \bibnamefont
  {van Wijk}}, \bibinfo {author} {\bibfnamefont {A.}~\bibnamefont {Schuring}},
  \bibinfo {author} {\bibfnamefont {M.~I.}\ \bibnamefont {Katsnelson}}, \ and\
  \bibinfo {author} {\bibfnamefont {A.}~\bibnamefont {Fasolino}},\ }\href@noop
  {} {\bibfield  {journal} {\bibinfo  {journal} {2D Mater.}\ }\textbf {\bibinfo
  {volume} {2}},\ \bibinfo {pages} {034010} (\bibinfo {year}
  {2015})}\BibitemShut {NoStop}%
\bibitem [{\citenamefont {Kolmogorov}\ and\ \citenamefont {Crespi}(2005)}]{KC}%
  \BibitemOpen
  \bibfield  {author} {\bibinfo {author} {\bibfnamefont {A.~N.}\ \bibnamefont
  {Kolmogorov}}\ and\ \bibinfo {author} {\bibfnamefont {V.~H.}\ \bibnamefont
  {Crespi}},\ }\href@noop {} {\bibfield  {journal} {\bibinfo  {journal} {Phys.
  Rev. B}\ }\textbf {\bibinfo {volume} {71}},\ \bibinfo {pages} {235415}
  (\bibinfo {year} {2005})}\BibitemShut {NoStop}%
\bibitem [{\citenamefont {Angeli}\ \emph {et~al.}(2018)\citenamefont {Angeli},
  \citenamefont {Mandelli}, \citenamefont {Valli}, \citenamefont {Amaricci},
  \citenamefont {Capone}, \citenamefont {Tosatti},\ and\ \citenamefont
  {Fabrizio}}]{AngeliM2018ED6s}%
  \BibitemOpen
  \bibfield  {author} {\bibinfo {author} {\bibfnamefont {M.}~\bibnamefont
  {Angeli}}, \bibinfo {author} {\bibfnamefont {D.}~\bibnamefont {Mandelli}},
  \bibinfo {author} {\bibfnamefont {A.}~\bibnamefont {Valli}}, \bibinfo
  {author} {\bibfnamefont {A.}~\bibnamefont {Amaricci}}, \bibinfo {author}
  {\bibfnamefont {M.}~\bibnamefont {Capone}}, \bibinfo {author} {\bibfnamefont
  {E.}~\bibnamefont {Tosatti}}, \ and\ \bibinfo {author} {\bibfnamefont
  {M.}~\bibnamefont {Fabrizio}},\ }\href@noop {} {\bibfield  {journal}
  {\bibinfo  {journal} {Phys. Rev. B}\ }\textbf {\bibinfo {volume} {98}},\
  \bibinfo {pages} {235137} (\bibinfo {year} {2018})}\BibitemShut {NoStop}%
\bibitem [{\citenamefont {Bitzek}\ \emph {et~al.}(2006)\citenamefont {Bitzek},
  \citenamefont {Koskinen}, \citenamefont {G\"ahler}, \citenamefont {Moseler},\
  and\ \citenamefont {Gumbsch}}]{PhysRevLett.97.170201}%
  \BibitemOpen
  \bibfield  {author} {\bibinfo {author} {\bibfnamefont {E.}~\bibnamefont
  {Bitzek}}, \bibinfo {author} {\bibfnamefont {P.}~\bibnamefont {Koskinen}},
  \bibinfo {author} {\bibfnamefont {F.}~\bibnamefont {G\"ahler}}, \bibinfo
  {author} {\bibfnamefont {M.}~\bibnamefont {Moseler}}, \ and\ \bibinfo
  {author} {\bibfnamefont {P.}~\bibnamefont {Gumbsch}},\ }\href@noop {}
  {\bibfield  {journal} {\bibinfo  {journal} {Phys. Rev. Lett.}\ }\textbf
  {\bibinfo {volume} {97}},\ \bibinfo {pages} {170201} (\bibinfo {year}
  {2006})}\BibitemShut {NoStop}%
\bibitem [{\citenamefont {Slater}\ and\ \citenamefont {Koster}(1954)}]{SK}%
  \BibitemOpen
  \bibfield  {author} {\bibinfo {author} {\bibfnamefont {J.~C.}\ \bibnamefont
  {Slater}}\ and\ \bibinfo {author} {\bibfnamefont {G.~F.}\ \bibnamefont
  {Koster}},\ }\href@noop {} {\bibfield  {journal} {\bibinfo  {journal} {Phys.
  Rev.}\ }\textbf {\bibinfo {volume} {94}},\ \bibinfo {pages} {1498} (\bibinfo
  {year} {1954})}\BibitemShut {NoStop}%
\bibitem [{\citenamefont {Corsetti}\ \emph {et~al.}(2017)\citenamefont
  {Corsetti}, \citenamefont {Mostofi},\ and\ \citenamefont {Lischner}}]{FC}%
  \BibitemOpen
  \bibfield  {author} {\bibinfo {author} {\bibfnamefont {F.}~\bibnamefont
  {Corsetti}}, \bibinfo {author} {\bibfnamefont {A.~A.}\ \bibnamefont
  {Mostofi}}, \ and\ \bibinfo {author} {\bibfnamefont {J.}~\bibnamefont
  {Lischner}},\ }\href@noop {} {\bibfield  {journal} {\bibinfo  {journal} {2D
  Mater.}\ }\textbf {\bibinfo {volume} {4}},\ \bibinfo {pages} {025070}
  (\bibinfo {year} {2017})}\BibitemShut {NoStop}%
\bibitem [{\citenamefont {Nam}\ and\ \citenamefont {Koshino}(2017)}]{LREBM}%
  \BibitemOpen
  \bibfield  {author} {\bibinfo {author} {\bibfnamefont {N.~N.~T.}\
  \bibnamefont {Nam}}\ and\ \bibinfo {author} {\bibfnamefont {M.}~\bibnamefont
  {Koshino}},\ }\href@noop {} {\bibfield  {journal} {\bibinfo  {journal} {Phys.
  Rev. B}\ }\textbf {\bibinfo {volume} {96}},\ \bibinfo {pages} {075311}
  (\bibinfo {year} {2017})}\BibitemShut {NoStop}%
\bibitem [{\citenamefont {Lin}\ \emph {et~al.}(2020)\citenamefont {Lin},
  \citenamefont {Zhu},\ and\ \citenamefont {Ni}}]{LinXianqing2020Pgma}%
  \BibitemOpen
  \bibfield  {author} {\bibinfo {author} {\bibfnamefont {X.}~\bibnamefont
  {Lin}}, \bibinfo {author} {\bibfnamefont {H.}~\bibnamefont {Zhu}}, \ and\
  \bibinfo {author} {\bibfnamefont {J.}~\bibnamefont {Ni}},\ }\href@noop {}
  {\bibfield  {journal} {\bibinfo  {journal} {Phys. Rev. B}\ }\textbf {\bibinfo
  {volume} {101}},\ \bibinfo {pages} {155405} (\bibinfo {year}
  {2020})}\BibitemShut {NoStop}%
\bibitem [{\citenamefont {Zheng}\ \emph {et~al.}(2018)\citenamefont {Zheng},
  \citenamefont {Cao}, \citenamefont {Liu},\ and\ \citenamefont
  {Peng}}]{ShaolongZheng2018ASaM}%
  \BibitemOpen
  \bibfield  {author} {\bibinfo {author} {\bibfnamefont {S.}~\bibnamefont
  {Zheng}}, \bibinfo {author} {\bibfnamefont {Q.}~\bibnamefont {Cao}}, \bibinfo
  {author} {\bibfnamefont {S.}~\bibnamefont {Liu}}, \ and\ \bibinfo {author}
  {\bibfnamefont {Q.}~\bibnamefont {Peng}},\ }\href@noop {} {\bibfield
  {journal} {\bibinfo  {journal} {J. Compos. Sci.}\ }\textbf {\bibinfo {volume}
  {3}},\ \bibinfo {pages} {2} (\bibinfo {year} {2018})}\BibitemShut {NoStop}%
\bibitem [{\citenamefont {Prentice}(2020)}]{onetep_2020}%
  \BibitemOpen
  \bibfield  {author} {\bibinfo {author} {\bibfnamefont {J.~C. A.~t.}\
  \bibnamefont {Prentice}},\ }\href@noop {} {\bibfield  {journal} {\bibinfo
  {journal} {J. Chem. Phys.}\ }\textbf {\bibinfo {volume} {152}},\ \bibinfo
  {pages} {174111} (\bibinfo {year} {2020})}\BibitemShut {NoStop}%
\bibitem [{\citenamefont {Ratcliff}\ \emph {et~al.}(2018)\citenamefont
  {Ratcliff}, \citenamefont {Conduit}, \citenamefont {Hine},\ and\
  \citenamefont {Haynes}}]{Ratclif_PRB98_2018}%
  \BibitemOpen
  \bibfield  {author} {\bibinfo {author} {\bibfnamefont {L.~E.}\ \bibnamefont
  {Ratcliff}}, \bibinfo {author} {\bibfnamefont {G.~J.}\ \bibnamefont
  {Conduit}}, \bibinfo {author} {\bibfnamefont {N.~D.~M.}\ \bibnamefont
  {Hine}}, \ and\ \bibinfo {author} {\bibfnamefont {P.~D.}\ \bibnamefont
  {Haynes}},\ }\href@noop {} {\bibfield  {journal} {\bibinfo  {journal} {Phys.
  Rev. B}\ }\textbf {\bibinfo {volume} {98}},\ \bibinfo {pages} {125123}
  (\bibinfo {year} {2018})}\BibitemShut {NoStop}%
\bibitem [{\citenamefont {Perdew}\ \emph {et~al.}(1996)\citenamefont {Perdew},
  \citenamefont {Burke},\ and\ \citenamefont {Ernzerhof}}]{PBE_PRL77}%
  \BibitemOpen
  \bibfield  {author} {\bibinfo {author} {\bibfnamefont {J.~P.}\ \bibnamefont
  {Perdew}}, \bibinfo {author} {\bibfnamefont {K.}~\bibnamefont {Burke}}, \
  and\ \bibinfo {author} {\bibfnamefont {M.}~\bibnamefont {Ernzerhof}},\
  }\href@noop {} {\bibfield  {journal} {\bibinfo  {journal} {Phys. Rev. Lett.}\
  }\textbf {\bibinfo {volume} {77}},\ \bibinfo {pages} {3865} (\bibinfo {year}
  {1996})}\BibitemShut {NoStop}%
\bibitem [{\citenamefont {Bl\"ochl}(1994)}]{PAW_PRB50}%
  \BibitemOpen
  \bibfield  {author} {\bibinfo {author} {\bibfnamefont {P.~E.}\ \bibnamefont
  {Bl\"ochl}},\ }\href@noop {} {\bibfield  {journal} {\bibinfo  {journal}
  {Phys. Rev. B}\ }\textbf {\bibinfo {volume} {50}},\ \bibinfo {pages} {17953}
  (\bibinfo {year} {1994})}\BibitemShut {NoStop}%
\bibitem [{\citenamefont {Jollet}\ \emph {et~al.}(2014)\citenamefont {Jollet},
  \citenamefont {Torrent},\ and\ \citenamefont {Holzwarth}}]{JOLLET20141246}%
  \BibitemOpen
  \bibfield  {author} {\bibinfo {author} {\bibfnamefont {F.}~\bibnamefont
  {Jollet}}, \bibinfo {author} {\bibfnamefont {M.}~\bibnamefont {Torrent}}, \
  and\ \bibinfo {author} {\bibfnamefont {N.}~\bibnamefont {Holzwarth}},\
  }\href@noop {} {\bibfield  {journal} {\bibinfo  {journal} {Comput. Phys.
  Commun.}\ }\textbf {\bibinfo {volume} {185}},\ \bibinfo {pages} {1246 }
  (\bibinfo {year} {2014})}\BibitemShut {NoStop}%
\bibitem [{\citenamefont {Ruiz-Serrano}\ and\ \citenamefont
  {Skylaris}(2013)}]{Serrano_JCP139_2013}%
  \BibitemOpen
  \bibfield  {author} {\bibinfo {author} {\bibfnamefont {{\`A}.}~\bibnamefont
  {Ruiz-Serrano}}\ and\ \bibinfo {author} {\bibfnamefont {C.-K.}\ \bibnamefont
  {Skylaris}},\ }\href@noop {} {\bibfield  {journal} {\bibinfo  {journal} {J.
  Chem. Phys.}\ }\textbf {\bibinfo {volume} {139}},\ \bibinfo {pages} {054107}
  (\bibinfo {year} {2013})}\BibitemShut {NoStop}%
\bibitem [{\citenamefont {Marzari}\ \emph {et~al.}(1997)\citenamefont
  {Marzari}, \citenamefont {Vanderbilt},\ and\ \citenamefont
  {Payne}}]{Marzari_PRL79_1997}%
  \BibitemOpen
  \bibfield  {author} {\bibinfo {author} {\bibfnamefont {N.}~\bibnamefont
  {Marzari}}, \bibinfo {author} {\bibfnamefont {D.}~\bibnamefont {Vanderbilt}},
  \ and\ \bibinfo {author} {\bibfnamefont {M.~C.}\ \bibnamefont {Payne}},\
  }\href@noop {} {\bibfield  {journal} {\bibinfo  {journal} {Phys. Rev. Lett.}\
  }\textbf {\bibinfo {volume} {79}},\ \bibinfo {pages} {1337} (\bibinfo {year}
  {1997})}\BibitemShut {NoStop}%
\end{thebibliography}%

\newpage

\onecolumngrid

\appendix

\section{Additional results: DFT comparison}

\begin{figure*}[ht]
\begin{subfigure}{0.4369\textwidth}
  \centering
  \includegraphics[width=1\linewidth]{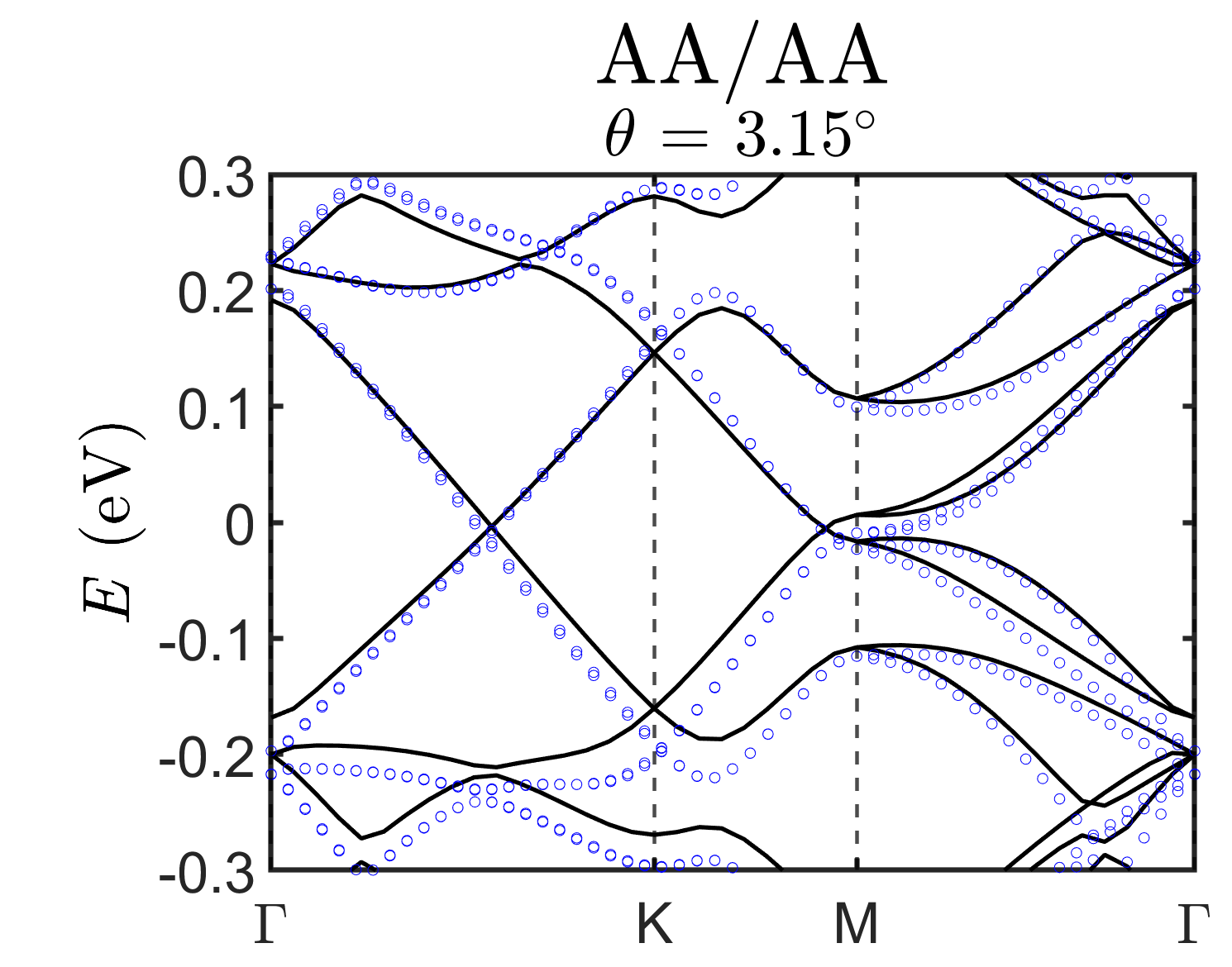}
\end{subfigure}
\begin{subfigure}{0.3631\textwidth}
  \centering
  \includegraphics[width=1\linewidth]{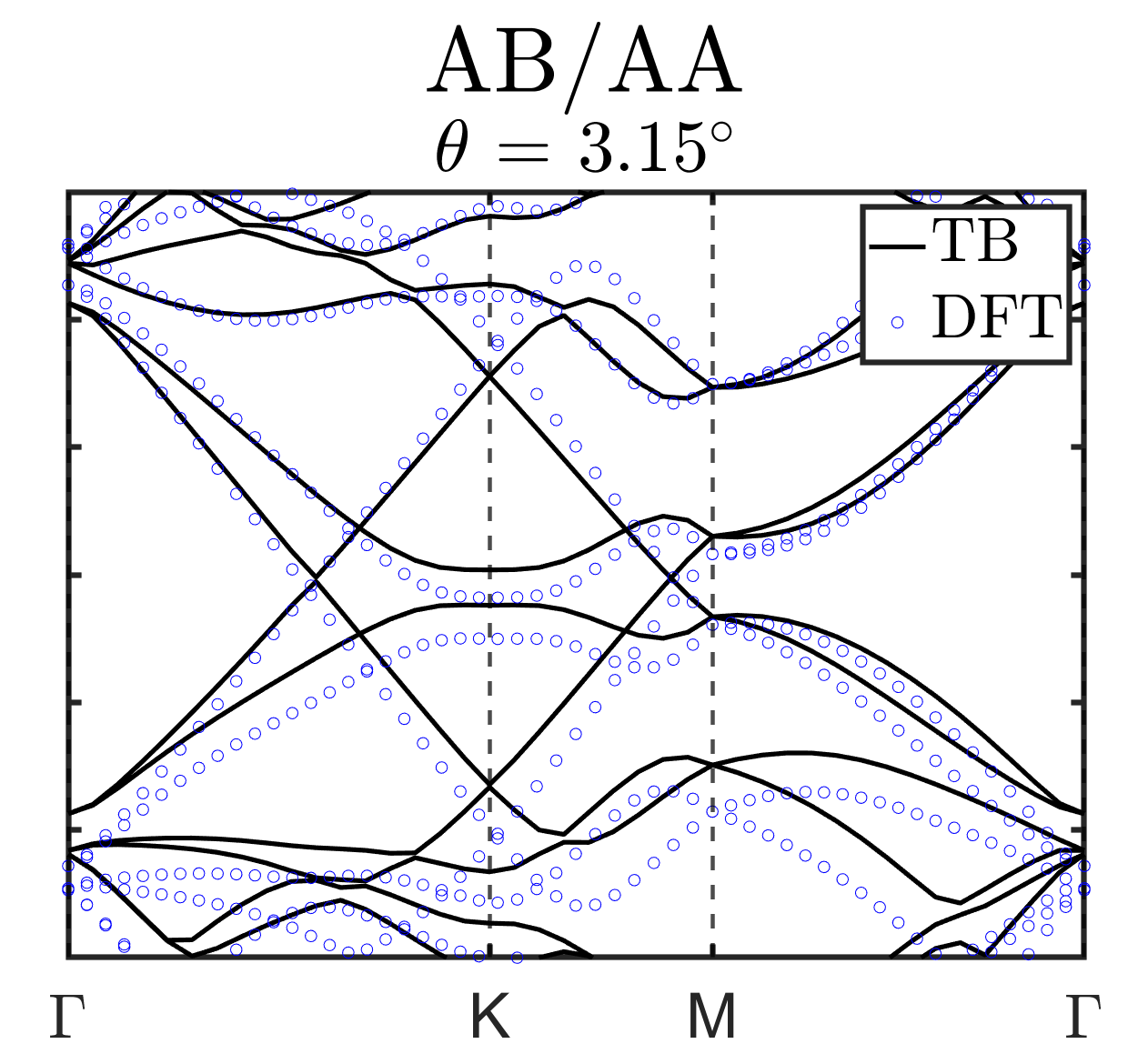}
\end{subfigure}
\caption{Comparison between tight-binding and \textit{ab initio} DFT band structures for AA/AA and AB/AA tDBLG. Note that an on-site potential is included in the tight-binding Hamiltonian, as described under Methods. DFT calculations were carried out using the \textsc{ONETEP} code~\cite{onetep_2020,Ratclif_PRB98_2018}. For these, the Perdew-Burke-Ernzerhof exchange-correlation functional~\cite{PBE_PRL77} with projector-augmented-wave pseudopotentials~\cite{PAW_PRB50,JOLLET20141246} was employed and the kinetic-energy cutoff was set to 800~eV. A minimal basis consisting of four nonorthogonal generalized Wannier functions per carbon atom was used. Because of the metallic nature of these systems, the ensemble-DFT approach was used~\cite{Serrano_JCP139_2013,Marzari_PRL79_1997}.}
\label{TB_DFT}
\end{figure*}

In Figure~\ref{TB_DFT}, we compare the atomistic tight-binding
calculations against ab initio DFT calculations (details of which can be found in the caption of Fig.~\ref{TB_DFT}). In both calculations, the classically relaxed structures were used. We find good agreement between both methods, and therefore, we
have confidence in our tight-binding model.

\section{Additional results: relaxations of AB/AA}

\begin{figure*}[ht]
    \centering
    \begin{subfigure}[b]{0.47\textwidth}
        \centering
        \includegraphics[width=\textwidth]{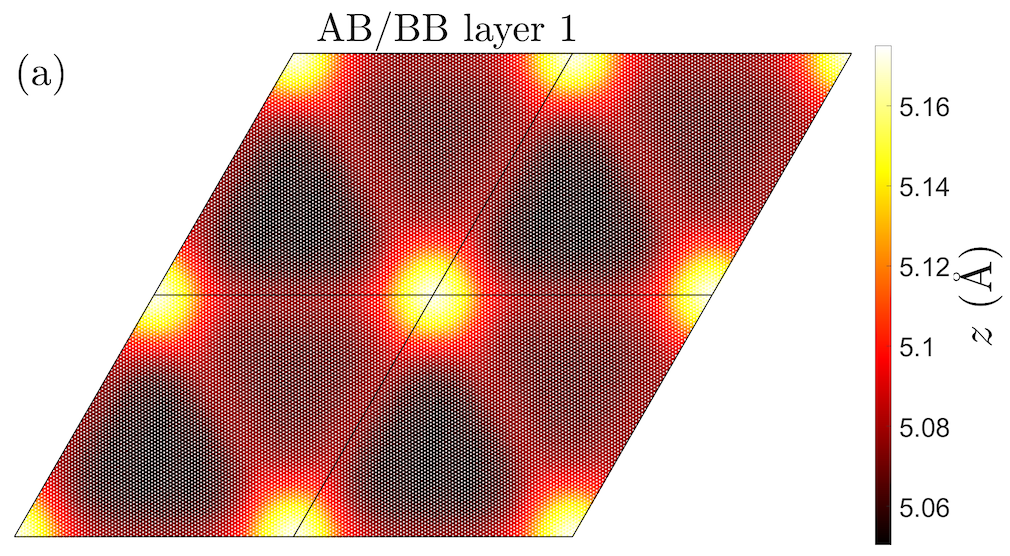}
    \end{subfigure}
    \begin{subfigure}[b]{0.47\textwidth}  
        \centering 
        \includegraphics[width=\textwidth]{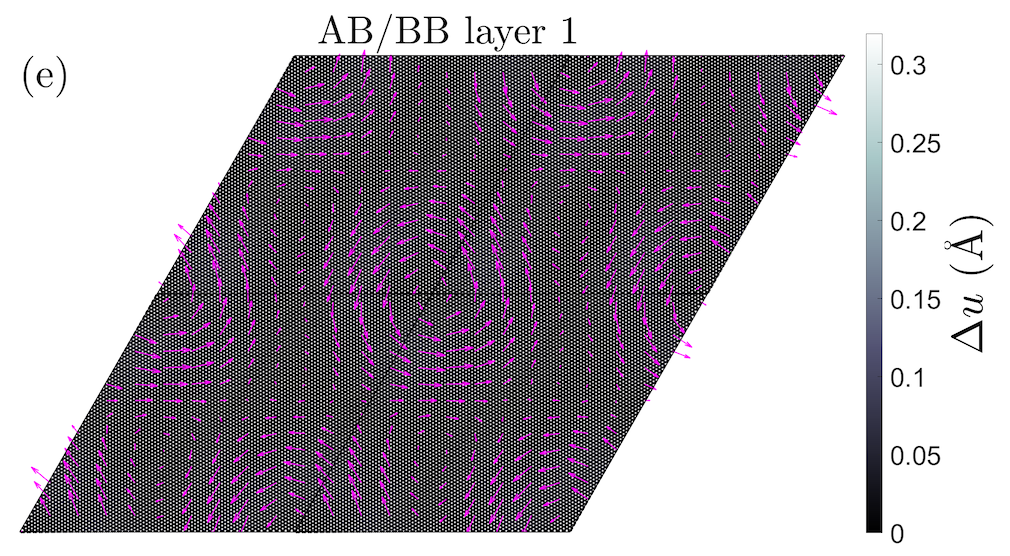}
    \end{subfigure}
    \hfill
    \vskip\baselineskip
    \begin{subfigure}[b]{0.47\textwidth}  
        \centering 
        \includegraphics[width=\textwidth]{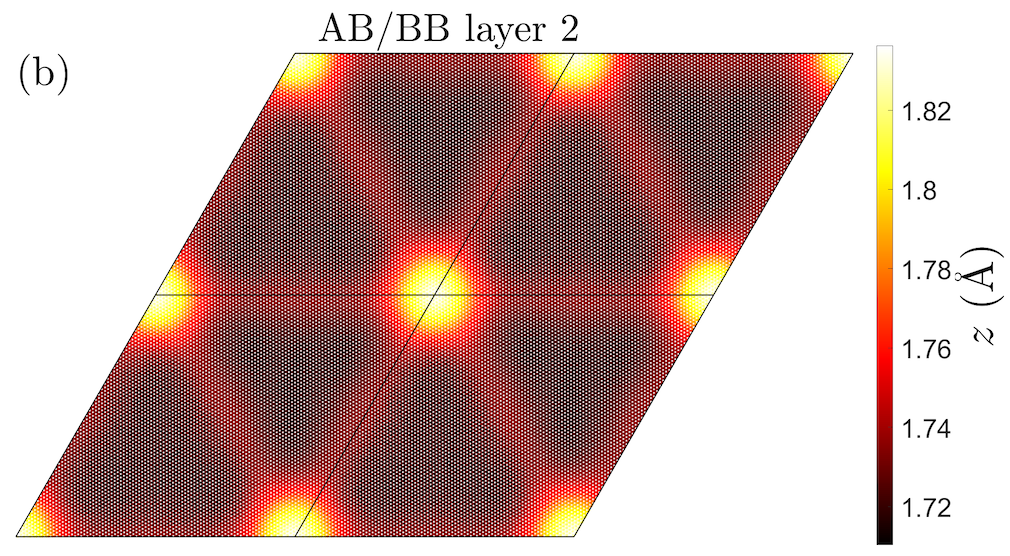}
    \end{subfigure}
    \begin{subfigure}[b]{0.47\textwidth}   
        \centering 
        \includegraphics[width=\textwidth]{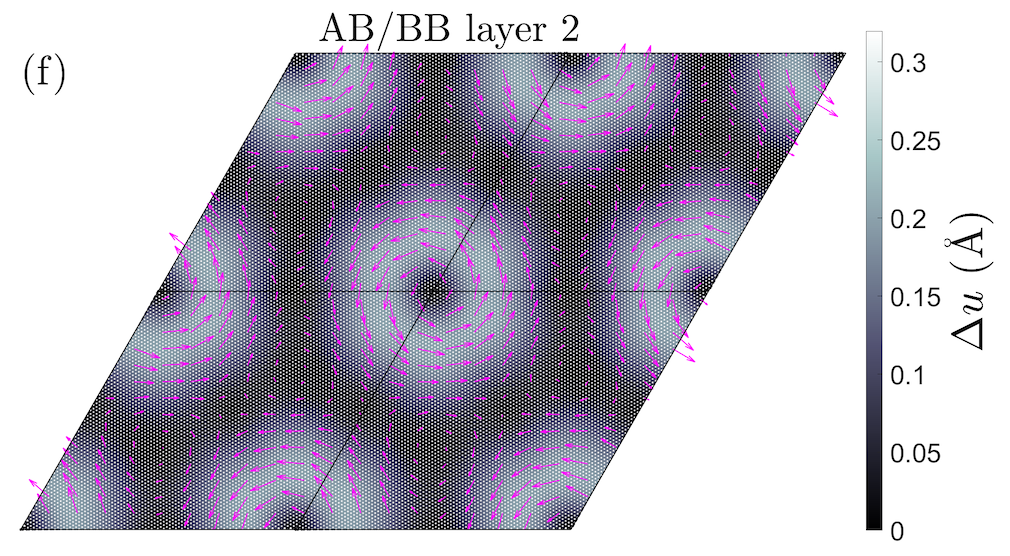}
    \end{subfigure}
    \vskip\baselineskip
    \begin{subfigure}[b]{0.47\textwidth}
        \centering
        \includegraphics[width=\textwidth]{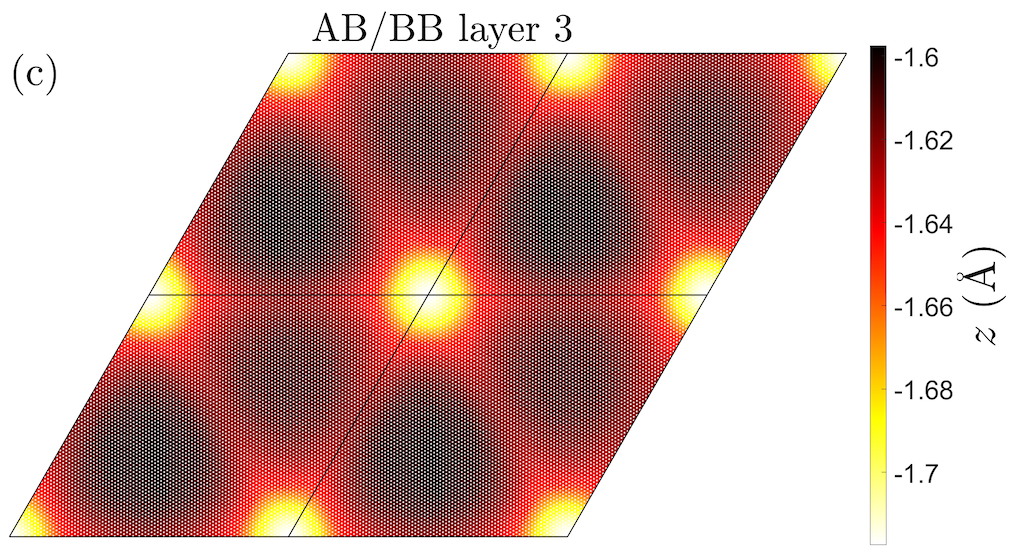}
    \end{subfigure}
    \begin{subfigure}[b]{0.47\textwidth}  
        \centering 
        \includegraphics[width=\textwidth]{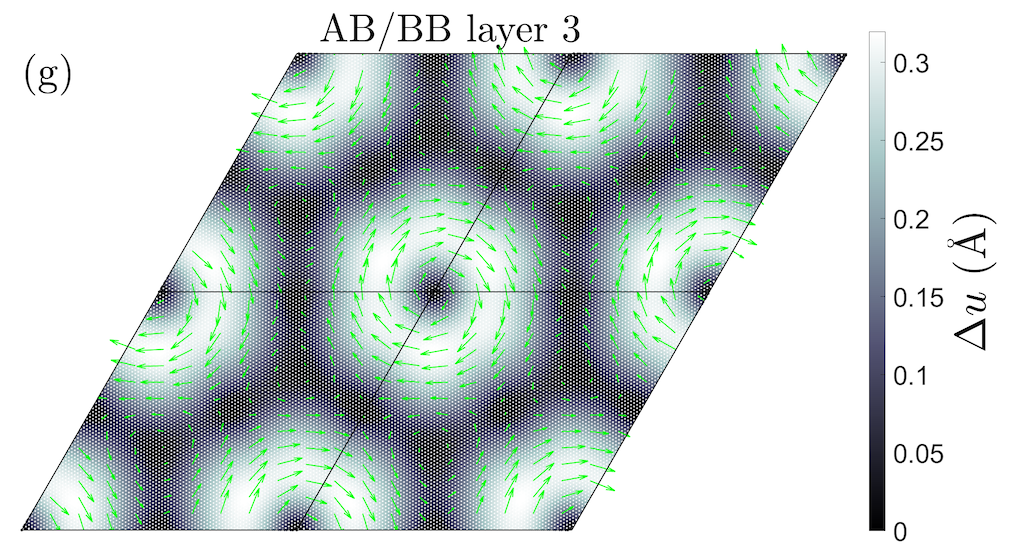}
    \end{subfigure}
    \hfill
    \vskip\baselineskip
    \begin{subfigure}[b]{0.47\textwidth}  
        \centering 
        \includegraphics[width=\textwidth]{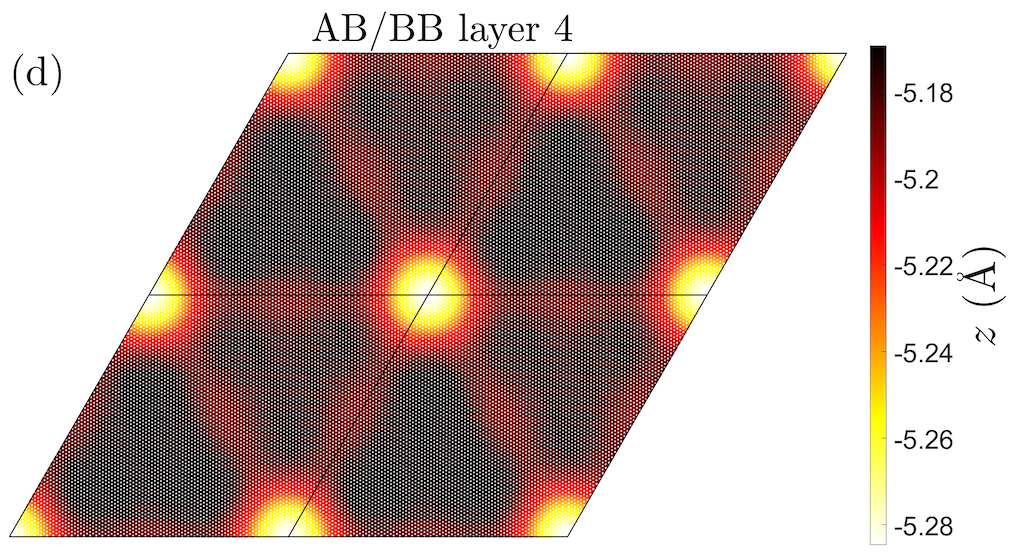}
    \end{subfigure}
    \begin{subfigure}[b]{0.47\textwidth}   
        \centering 
        \includegraphics[width=\textwidth]{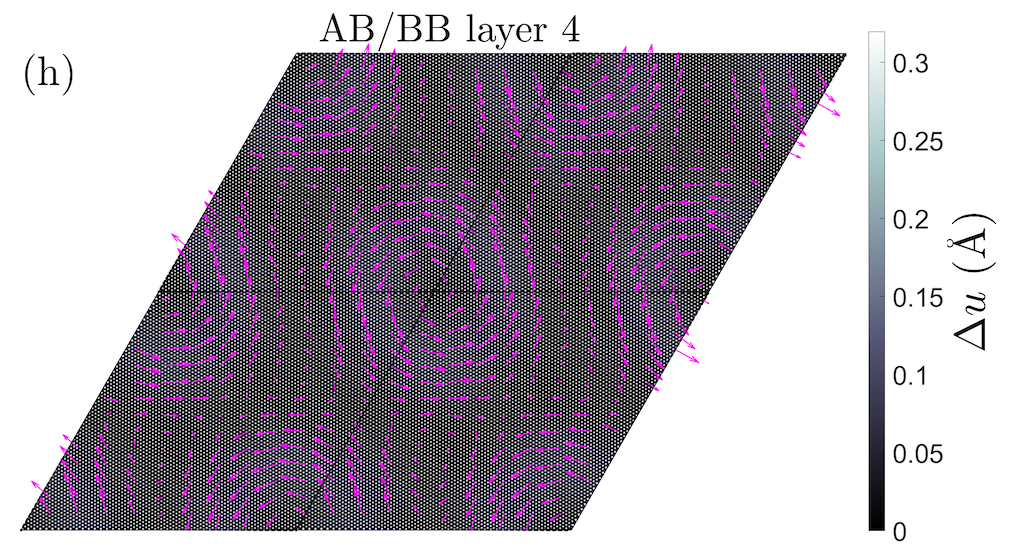}
    \end{subfigure}
    \caption{Out-of-plane and in-plane relaxations of tAB/AA tDBLG for a twist angle of $\theta=0.73\degree$. [(a)-(d)] Out-of-plane displacements for layers 1 to 4, respectively;  [(e)-(h)] In-plane displacements for layers 1 to 4, respectively.}
    \label{Z-surface_AB/AA}
\end{figure*}

In Figure~\ref{Z-surface_AB/AA} we display the relaxations of the AB/AA system. The inner layers of this system relax in a similar way to tBLG, just as we showed for AB/AB and AA/AA in the main text. The outer layer of the AB bilayer follows the relaxations if the inner layer, in an attempt to retain the low energy AB stacking throughout the moir\'e unit cell. While the outer layer of the AA bilayer relaxes in a more complicated way, similarly to AA/AA described in the main text.

\section{Additional results: band structures at other twist angles}

\begin{figure*}[ht]
\begin{subfigure}{0.36212658\textwidth}
  \centering
  \includegraphics[width=1\linewidth]{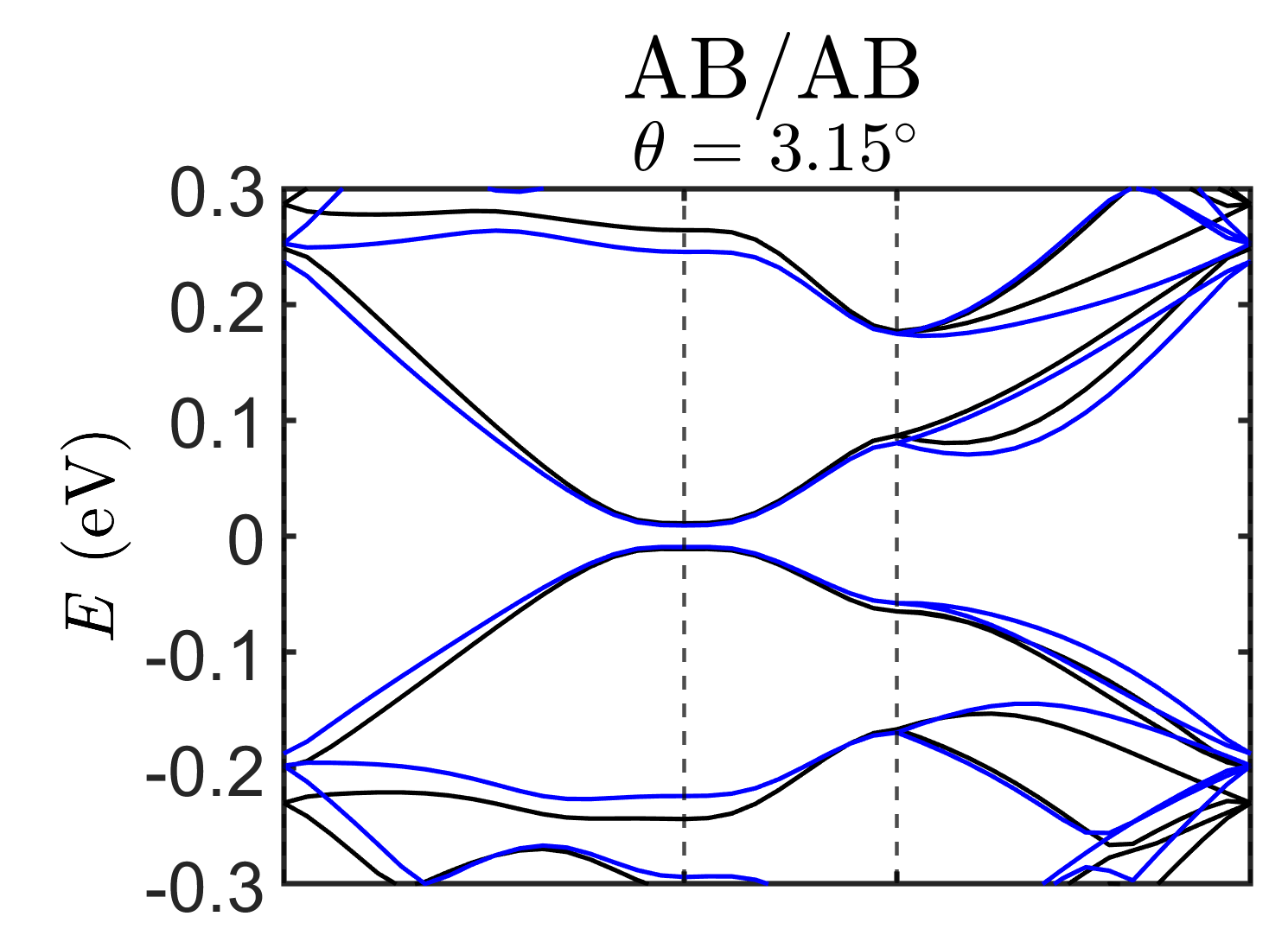}
\end{subfigure}
\begin{subfigure}{0.2989367\textwidth}
  \centering
  \includegraphics[width=1\linewidth]{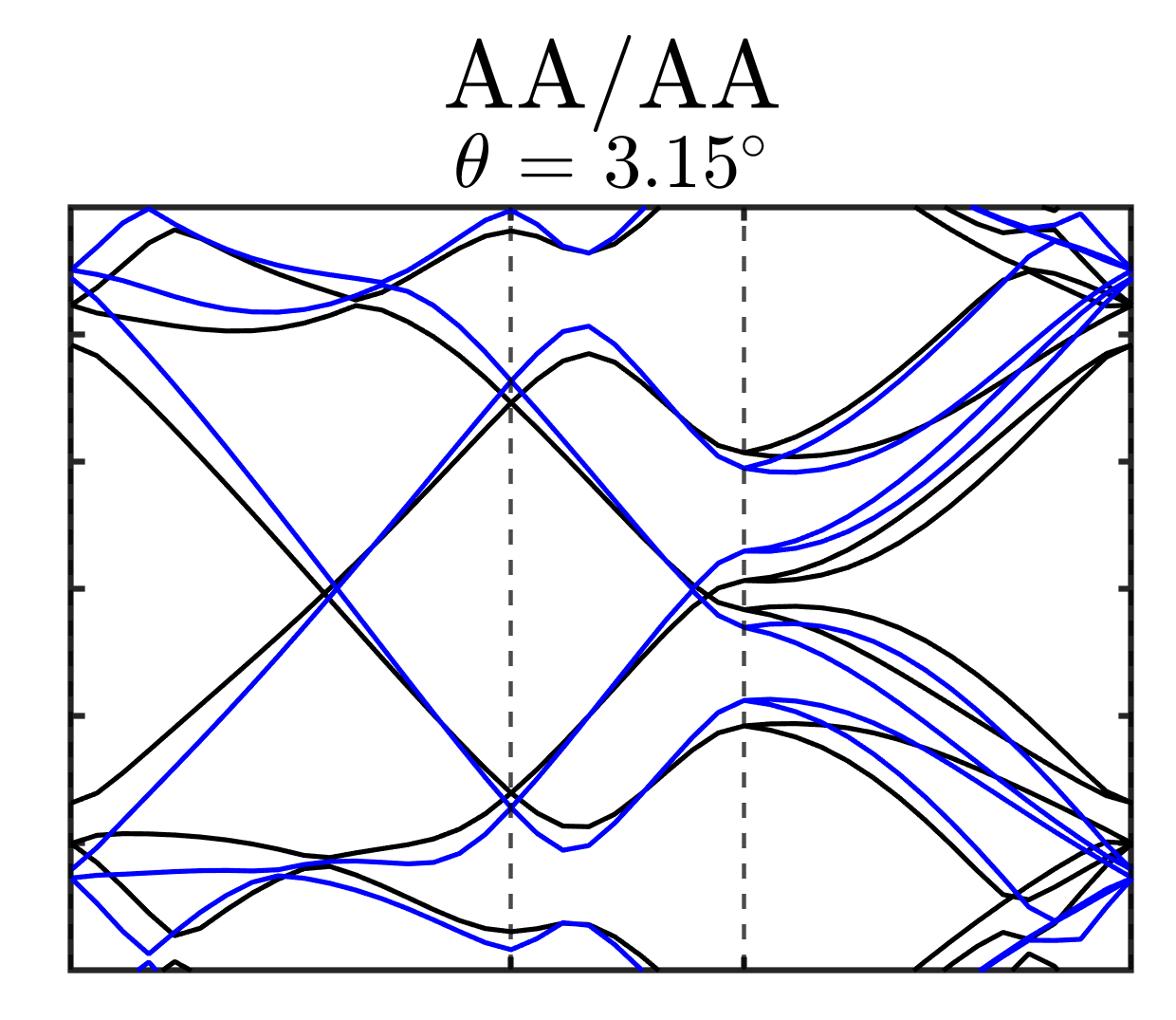}
\end{subfigure}
\begin{subfigure}{0.2989367\textwidth}
  \centering
  \includegraphics[width=1\linewidth]{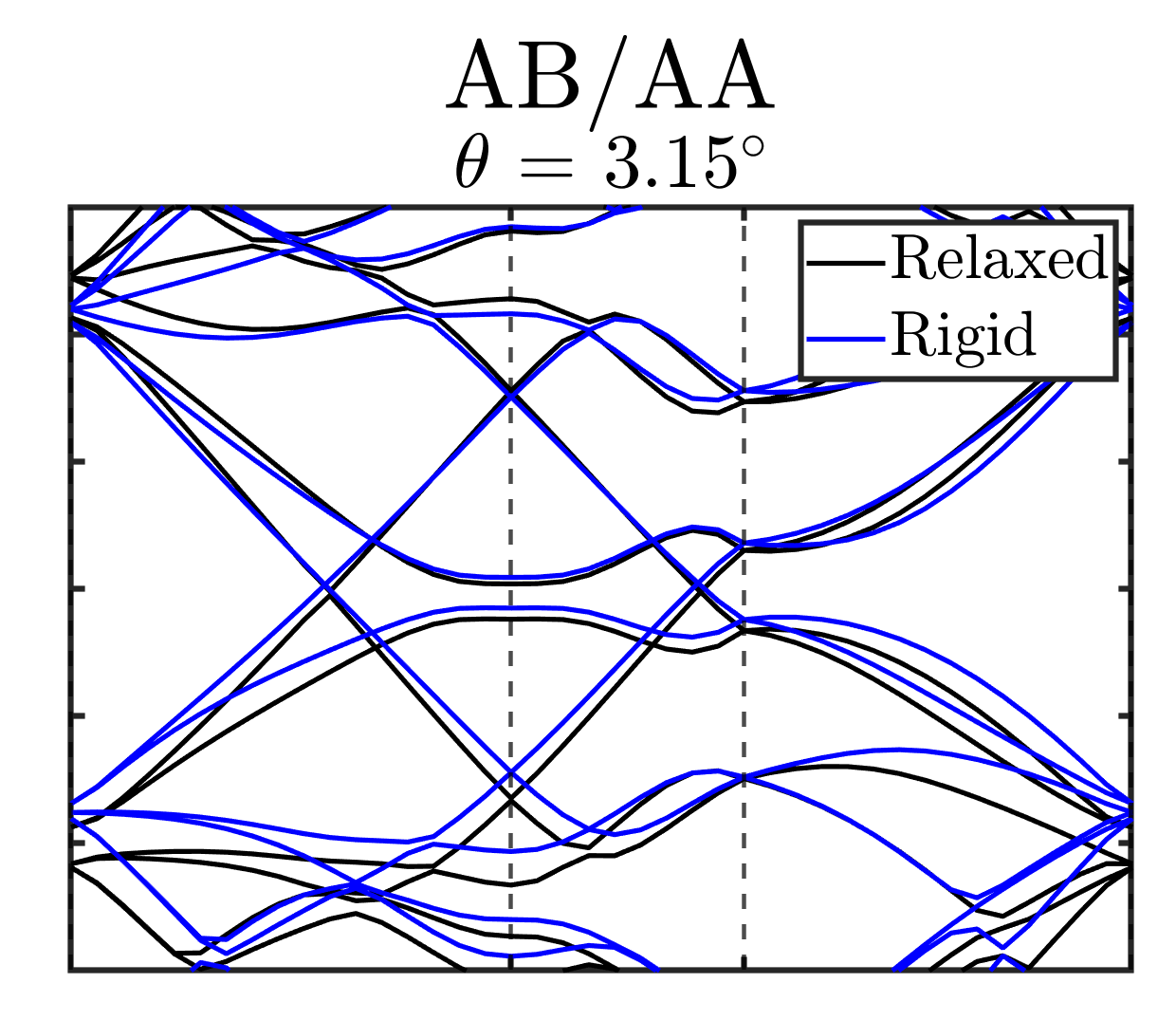}
\end{subfigure}
\begin{subfigure}{0.36212658\textwidth} 
  \centering
  \includegraphics[width=1\linewidth]{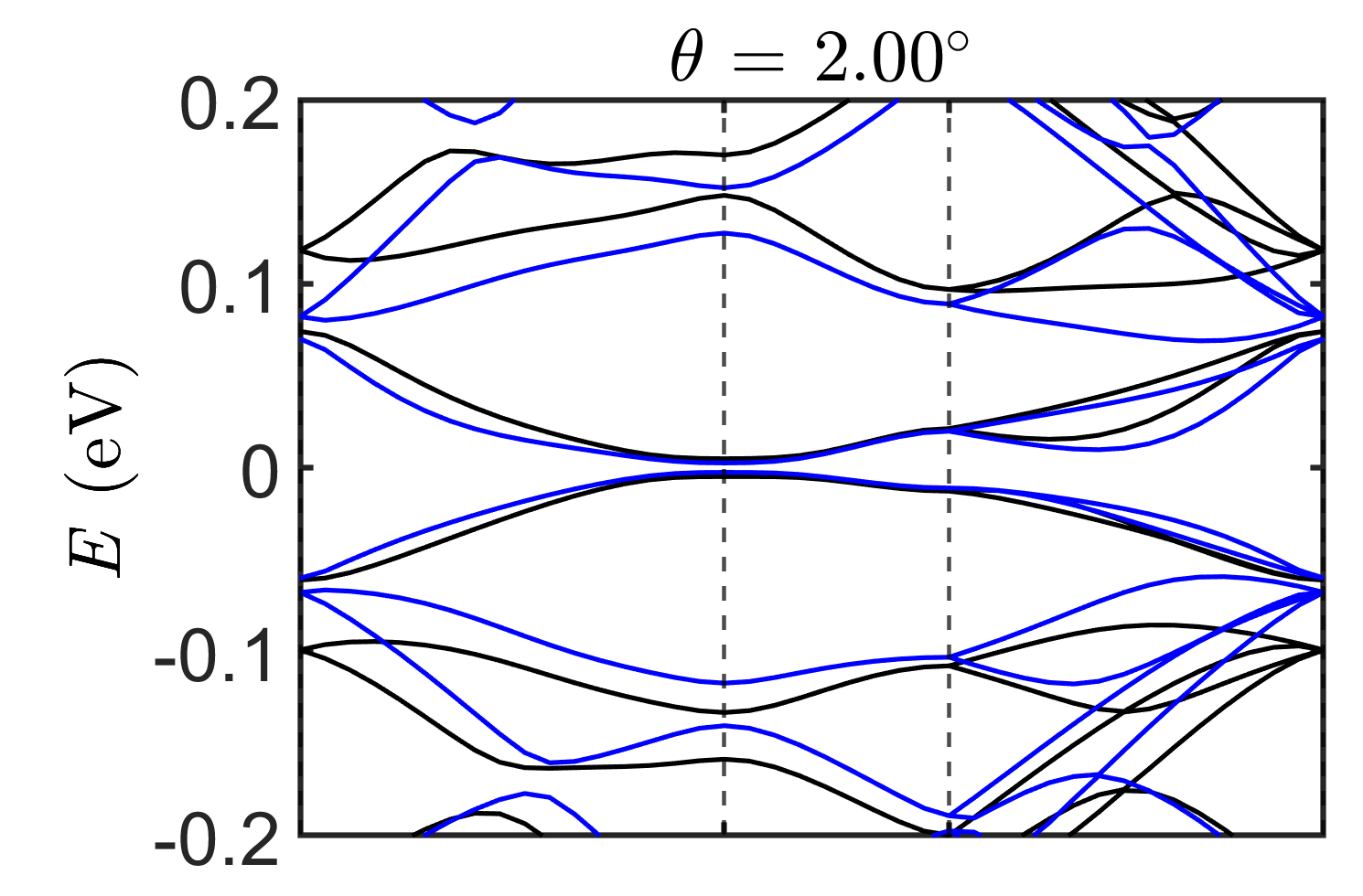}
\end{subfigure}
\begin{subfigure}{0.2989367\textwidth}
  \centering
  \includegraphics[width=1\linewidth]{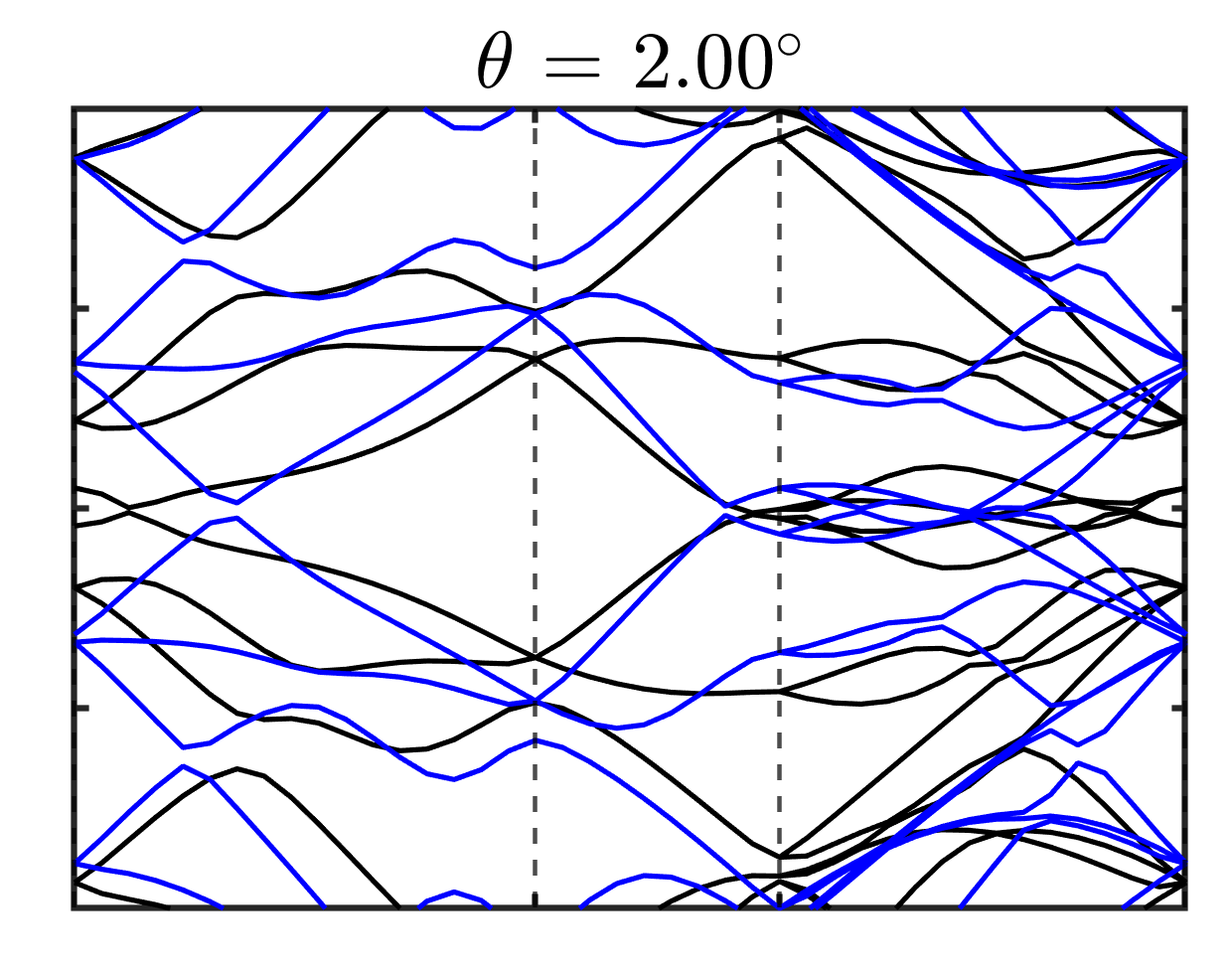}
\end{subfigure}
\begin{subfigure}{0.2989367\textwidth}
  \centering
  \includegraphics[width=1\linewidth]{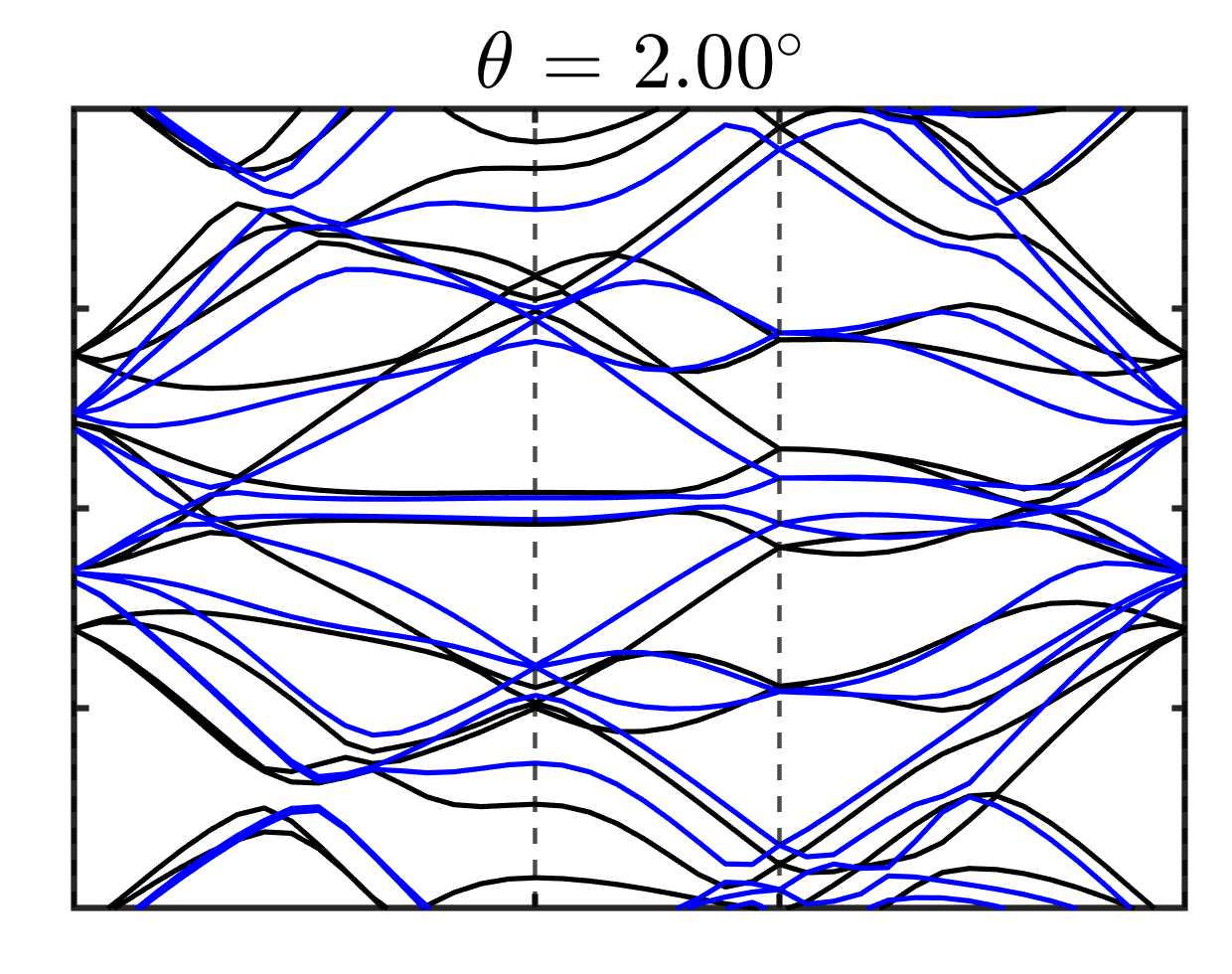}
\end{subfigure}
\begin{subfigure}{0.36212658\textwidth}
  \centering
  \includegraphics[width=1\linewidth]{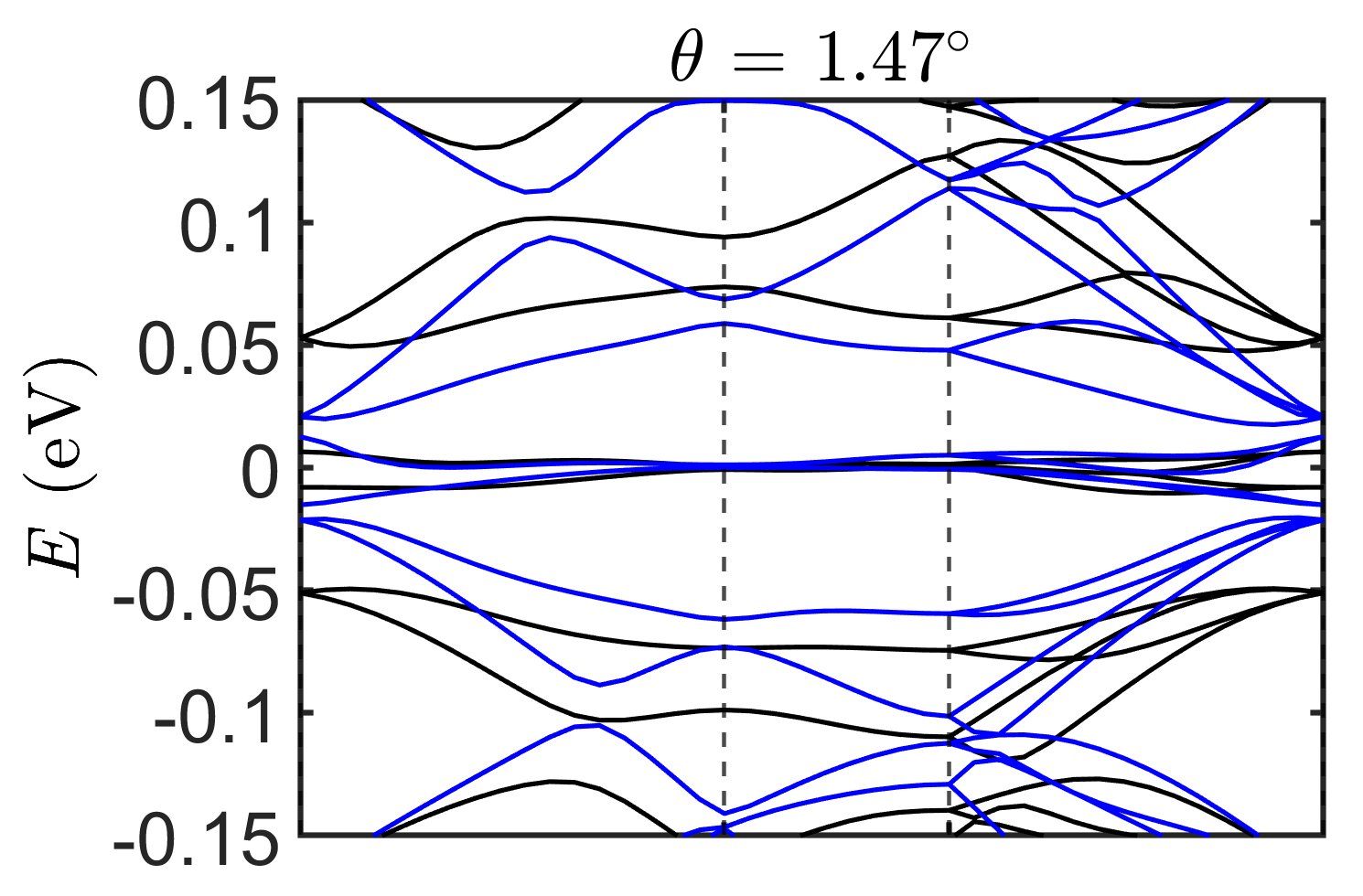}
\end{subfigure}
\begin{subfigure}{0.2989367\textwidth}
  \centering
  \includegraphics[width=1\linewidth]{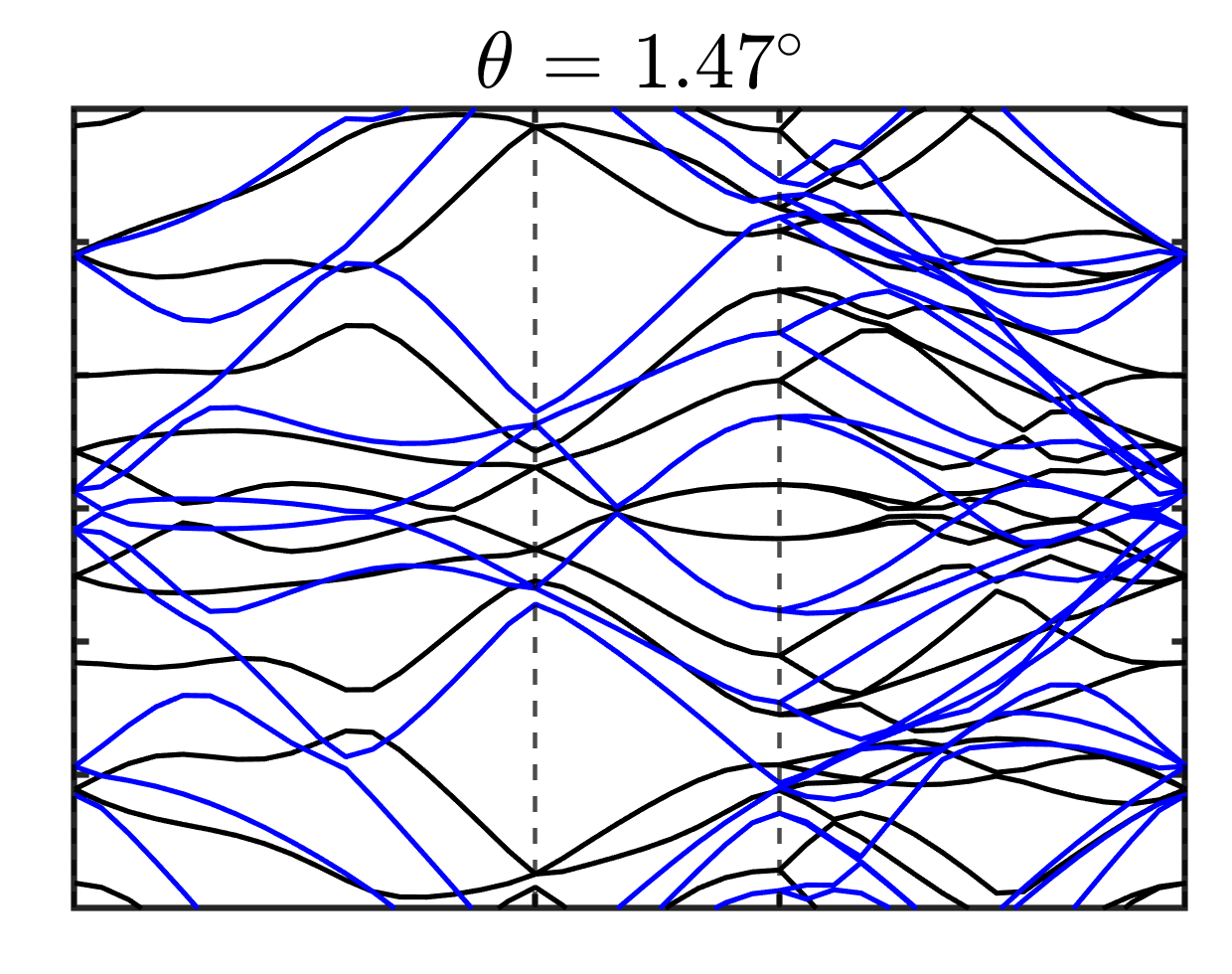}
\end{subfigure}
\begin{subfigure}{0.2989367\textwidth}
  \centering
  \includegraphics[width=1\linewidth]{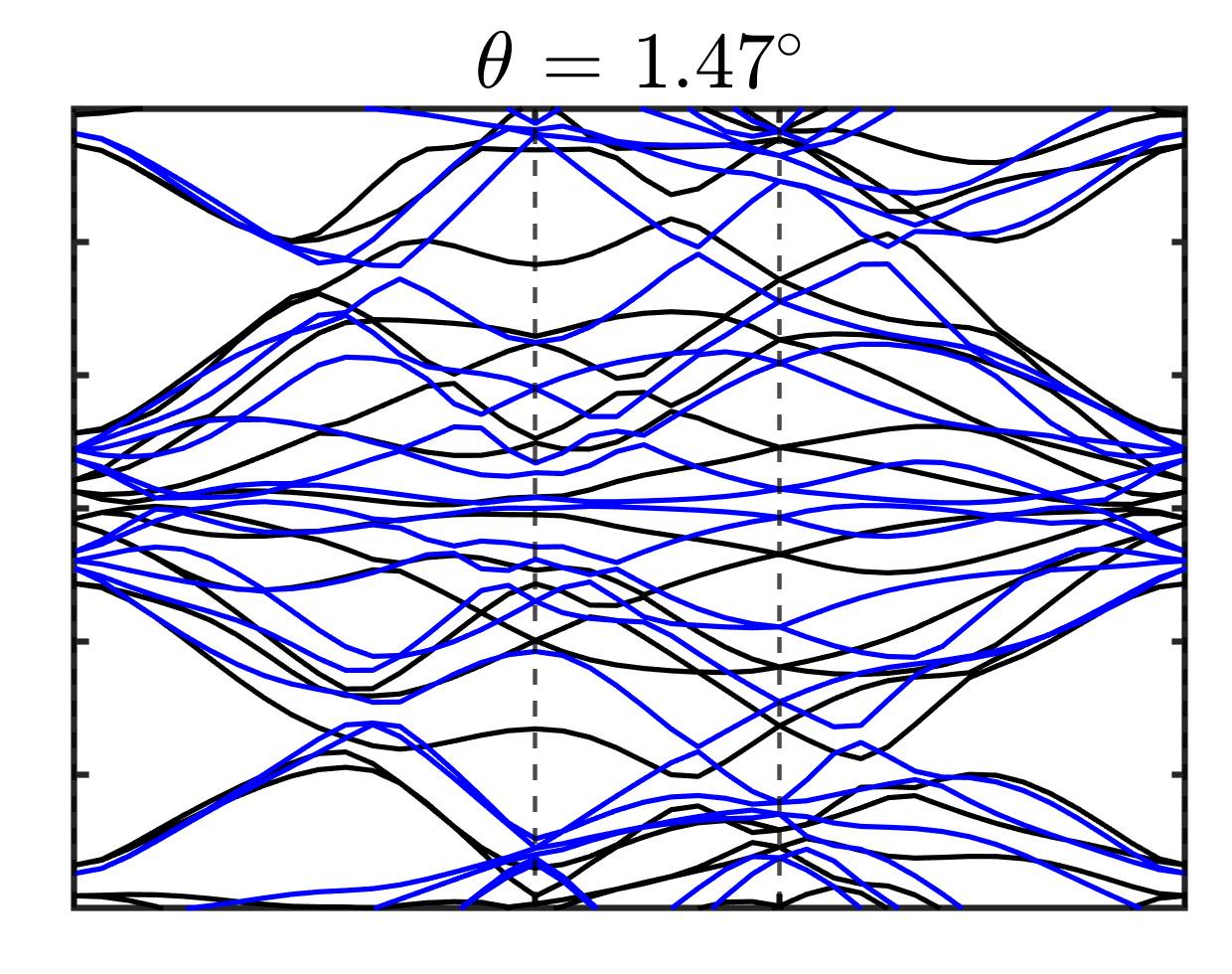}
\end{subfigure}
\begin{subfigure}{0.36212658\textwidth}
  \centering
  \includegraphics[width=1\linewidth]{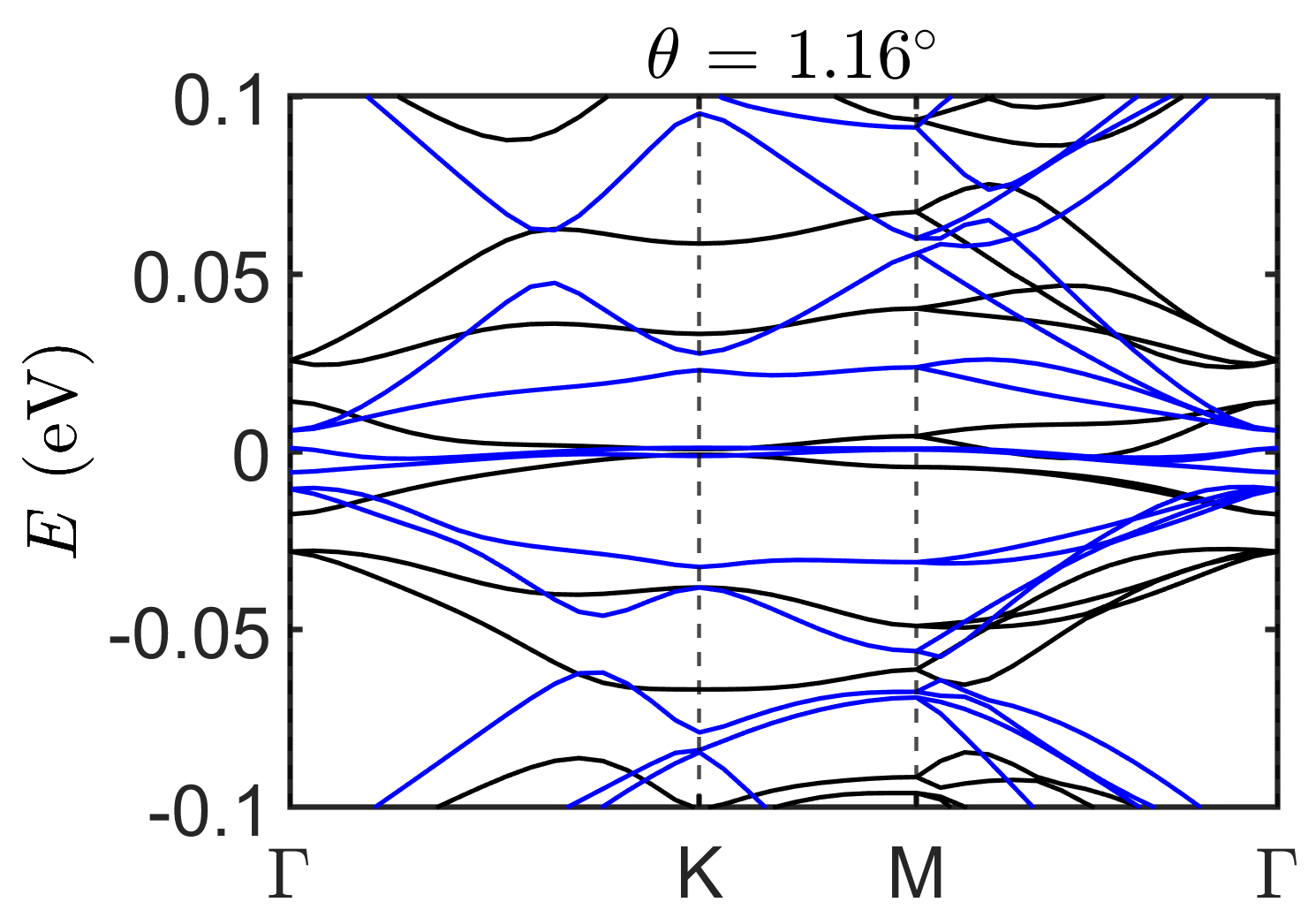}
\end{subfigure}
\begin{subfigure}{0.2989367\textwidth}
  \centering
  \includegraphics[width=1\linewidth]{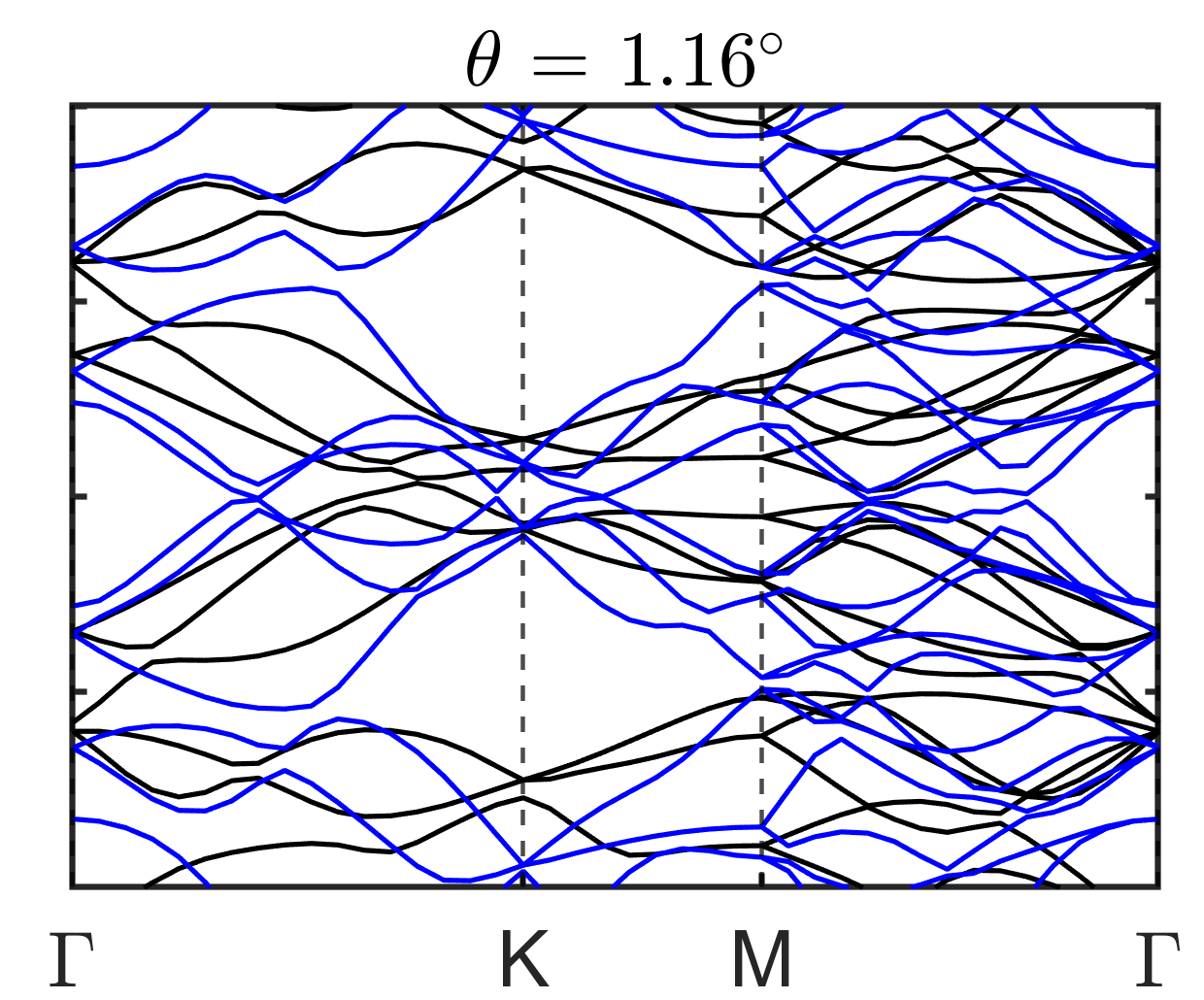}
\end{subfigure}
\begin{subfigure}{0.2989367\textwidth}
  \centering
  \includegraphics[width=1\linewidth]{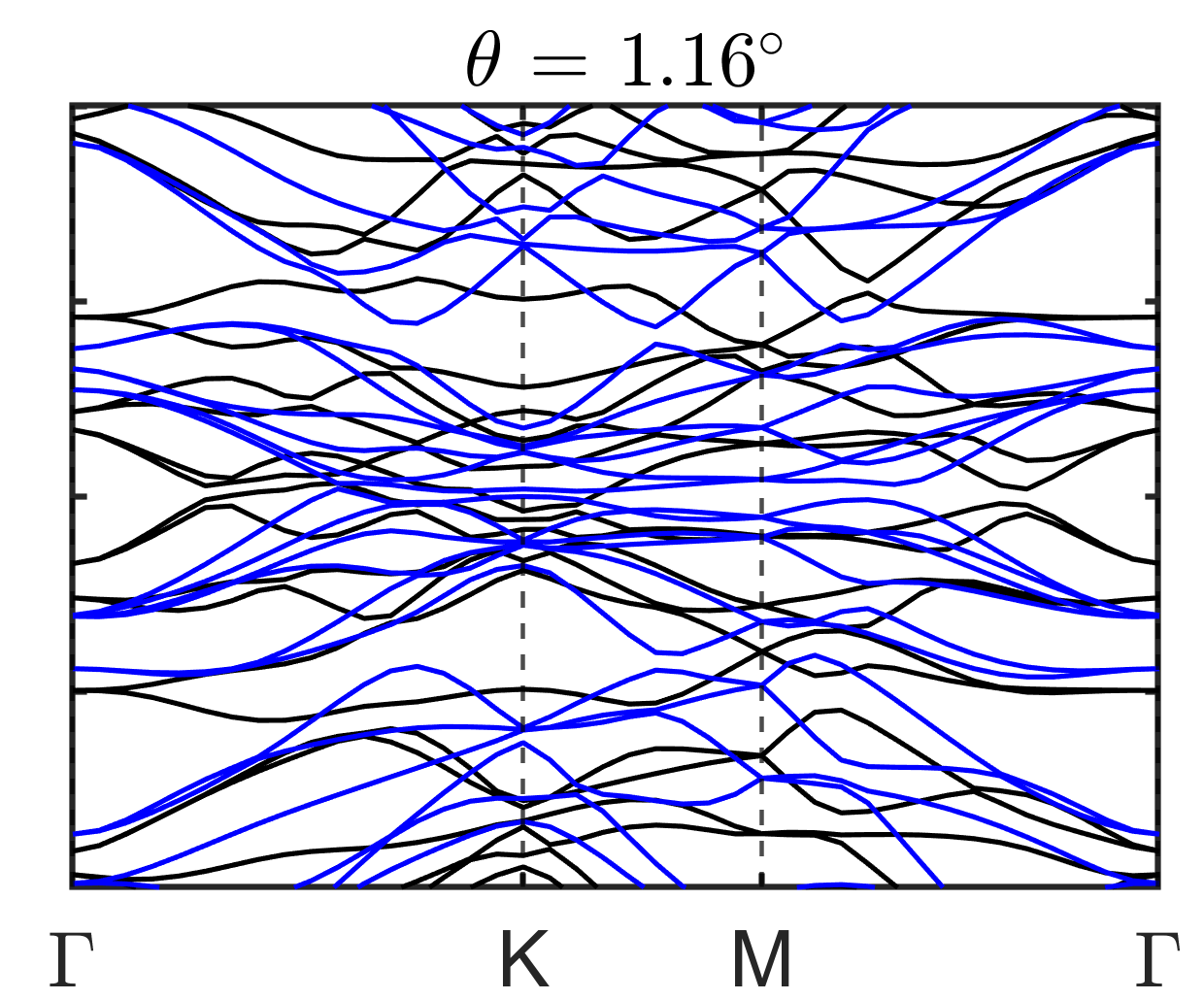}
\end{subfigure}
\caption{Band structures of relaxed (black) and unrelaxed (blue) AB/AB, AA/AA and AB/AA tDBLG. The Fermi energy of the undoped system is set to zero. Onsite potentials for the inner layer atoms were included in all calculations. For the unrelaxed structures, the following interlayer separations were used: for the AB/AB system all interlayer separations were set to 3.35$~\text{\AA}$; for the AA/AA system all interlayer separations were set to 3.6$~\text{\AA}$; for the AB/AA system all but the AB interlayer separation (which was set to 3.35$~\text{\AA}$) were set to 3.6$~\text{\AA}$. For the AB/AB system we find that relaxations increase the energy gaps between the flat bands and the remote bands -- similarly as in twisted bilayer graphene. The effects of relaxations are more complicated for the AA/AA and AB/AA systems. For all systems, the effect of relaxations becomes more important as the twist angle is reduced.}
\label{BS_SM}
\end{figure*}

In Figure~\ref{BS_SM} we display the band structures for AB/AB, AA/AA and AB/AA at a number of twist angles not shown in the main text. In these plots, we show band structures with pristine atomic positions and relaxed atomic positions. Note, the onsite potential, outlined in the Methods, was used for both pristine and relaxed structures. We find the lattice reconstruction opens up gaps between the flat bands and the bands adjacent to these, as is also well known for tBLG.

\clearpage

\section{Additional results: effect of out-of-plane relaxations}

In Figure 11 we display band structure calculations for AB/AB, AA/AA and AB/AA with fully relaxed positions and where only the $z$ positions are relaxed (obtained from setting the fully relaxed $x-y$ positions back to their pristine $x-y$ coordinates, while retaining the relaxed $z$ coordinates). Note the onsite potential of -30 meV on the inner layers was included. This highlights the effect of out-of-plane relaxations. We find that the out-of-plane relaxations dominate at large angles, with the effect of in-plane relaxations becoming more important at small angles.

\begin{figure*}[ht]
\begin{subfigure}{0.36212658\textwidth}
  \centering
  \includegraphics[width=1\linewidth]{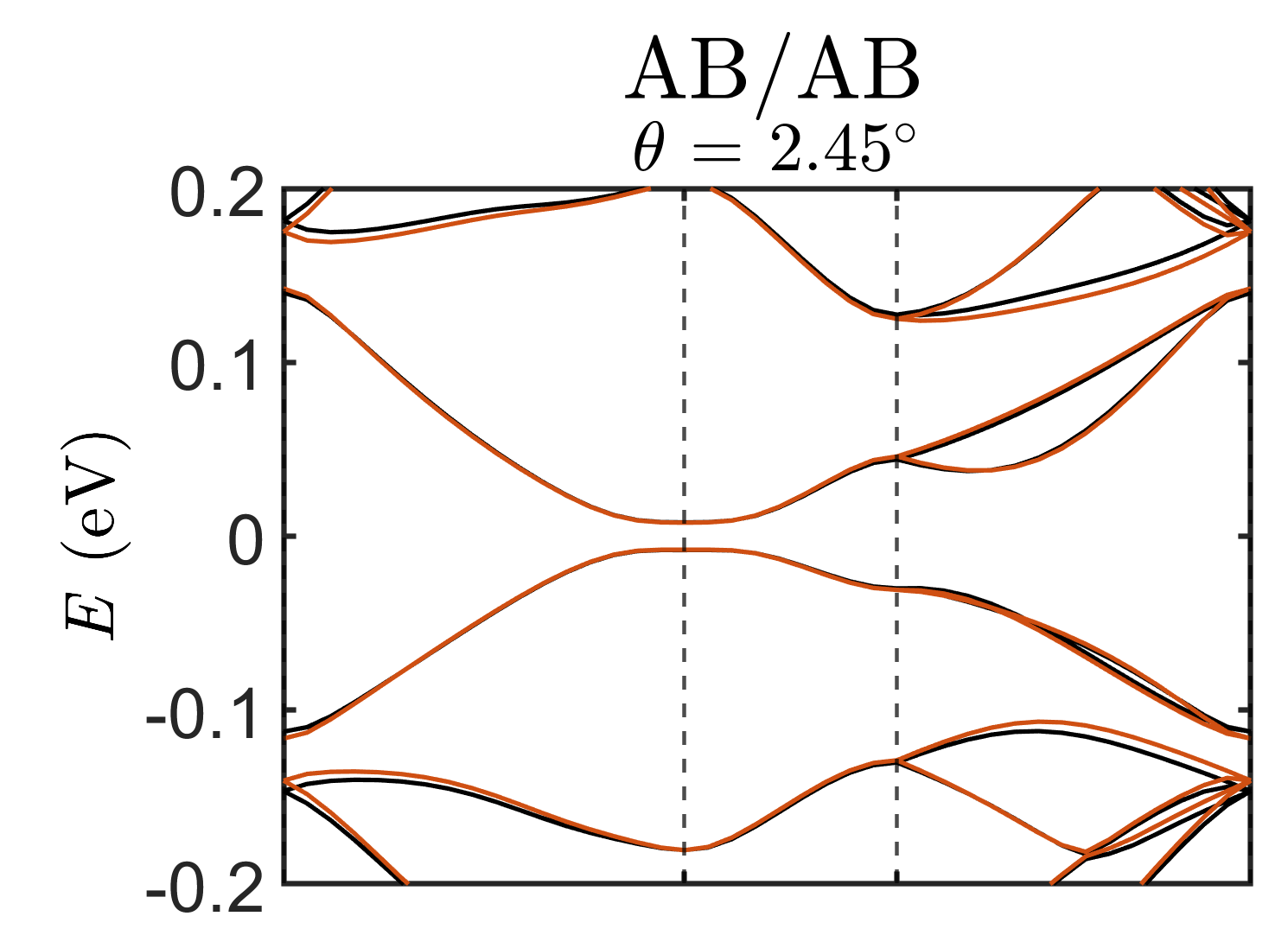}
\end{subfigure}
\begin{subfigure}{0.2989367\textwidth}
  \centering
  \includegraphics[width=1\linewidth]{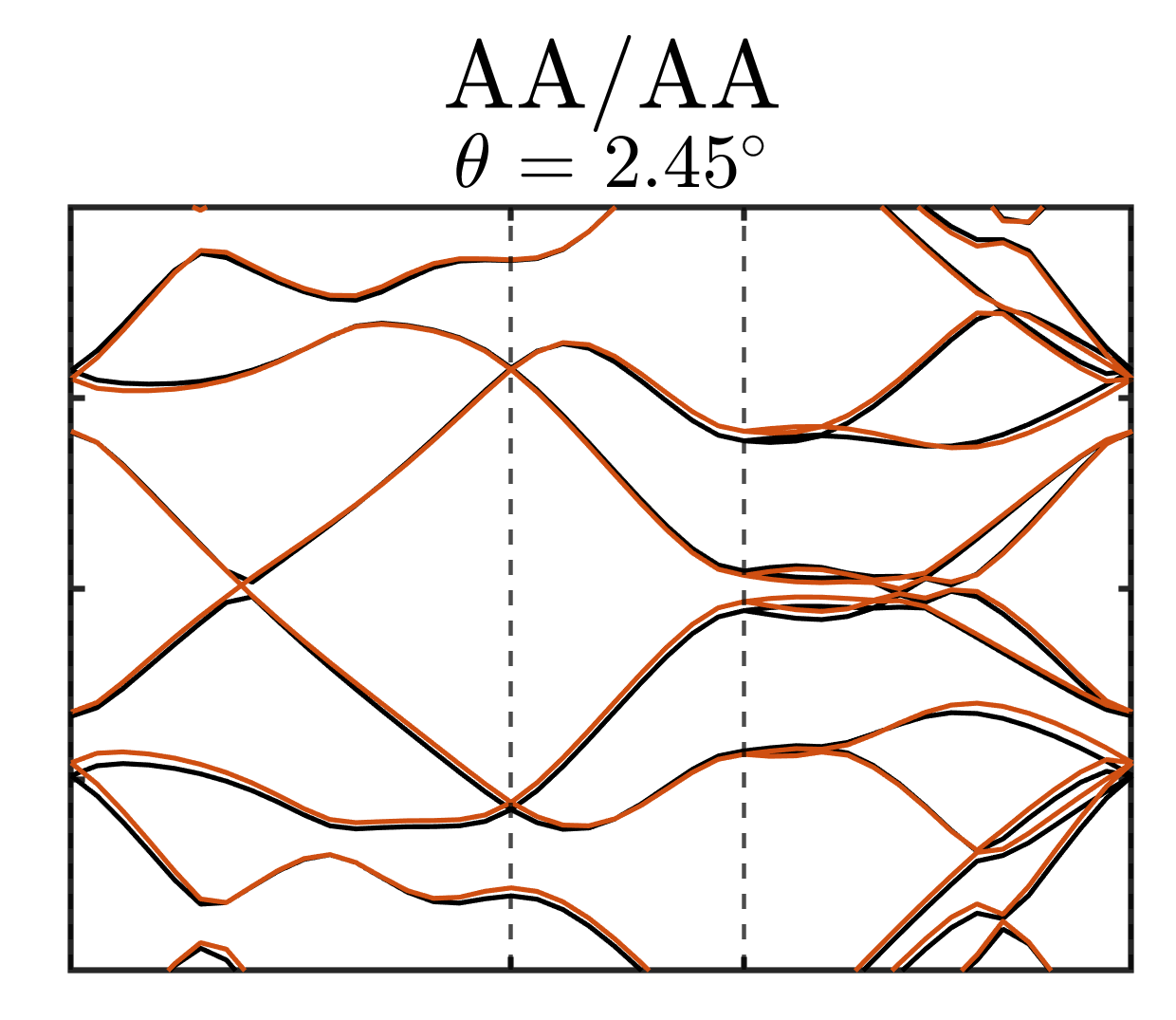}
\end{subfigure}
\begin{subfigure}{0.2989367\textwidth}
  \centering
  \includegraphics[width=1\linewidth]{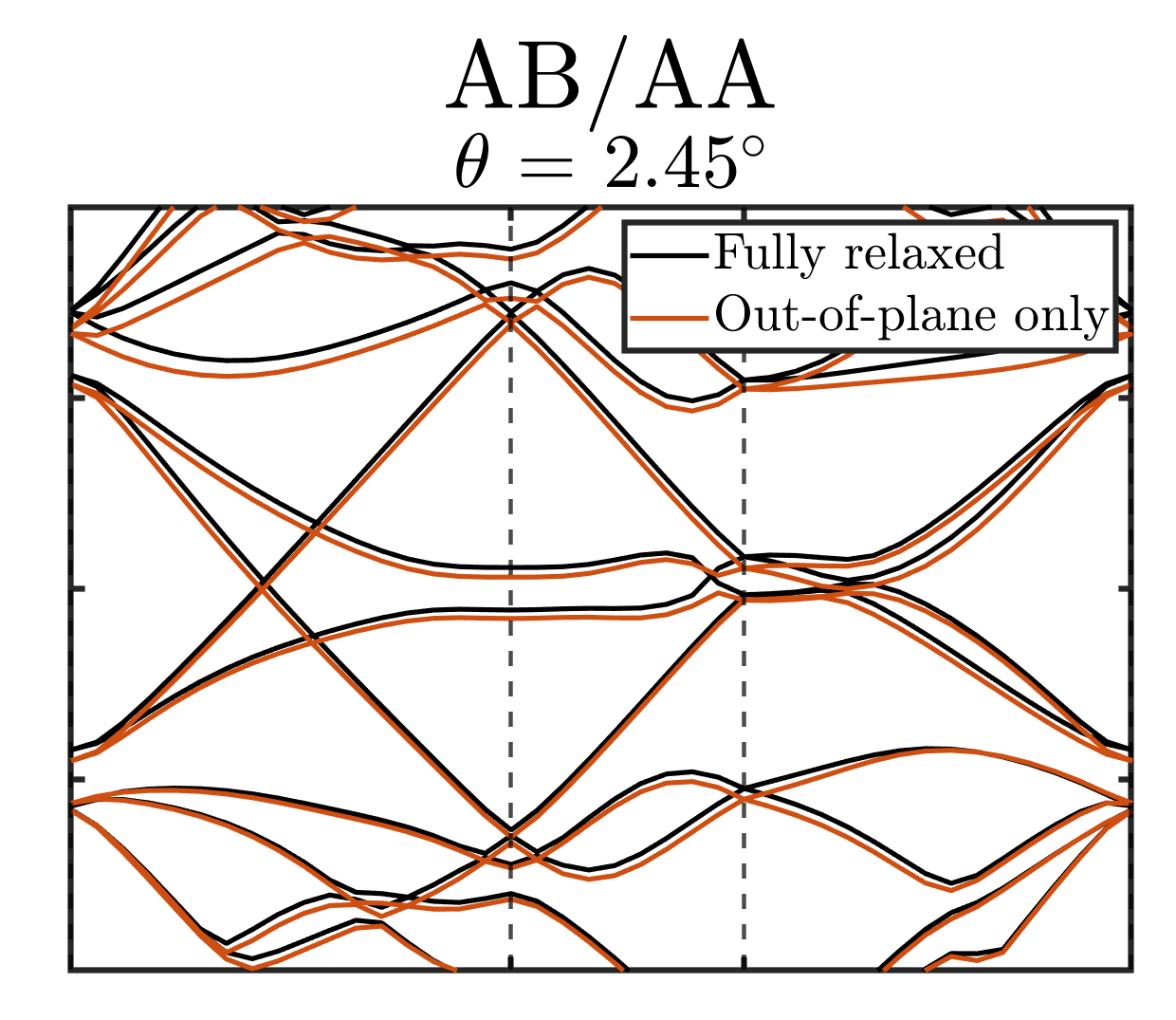}
\end{subfigure}
\begin{subfigure}{0.36212658\textwidth}
  \centering
  \includegraphics[width=1\linewidth]{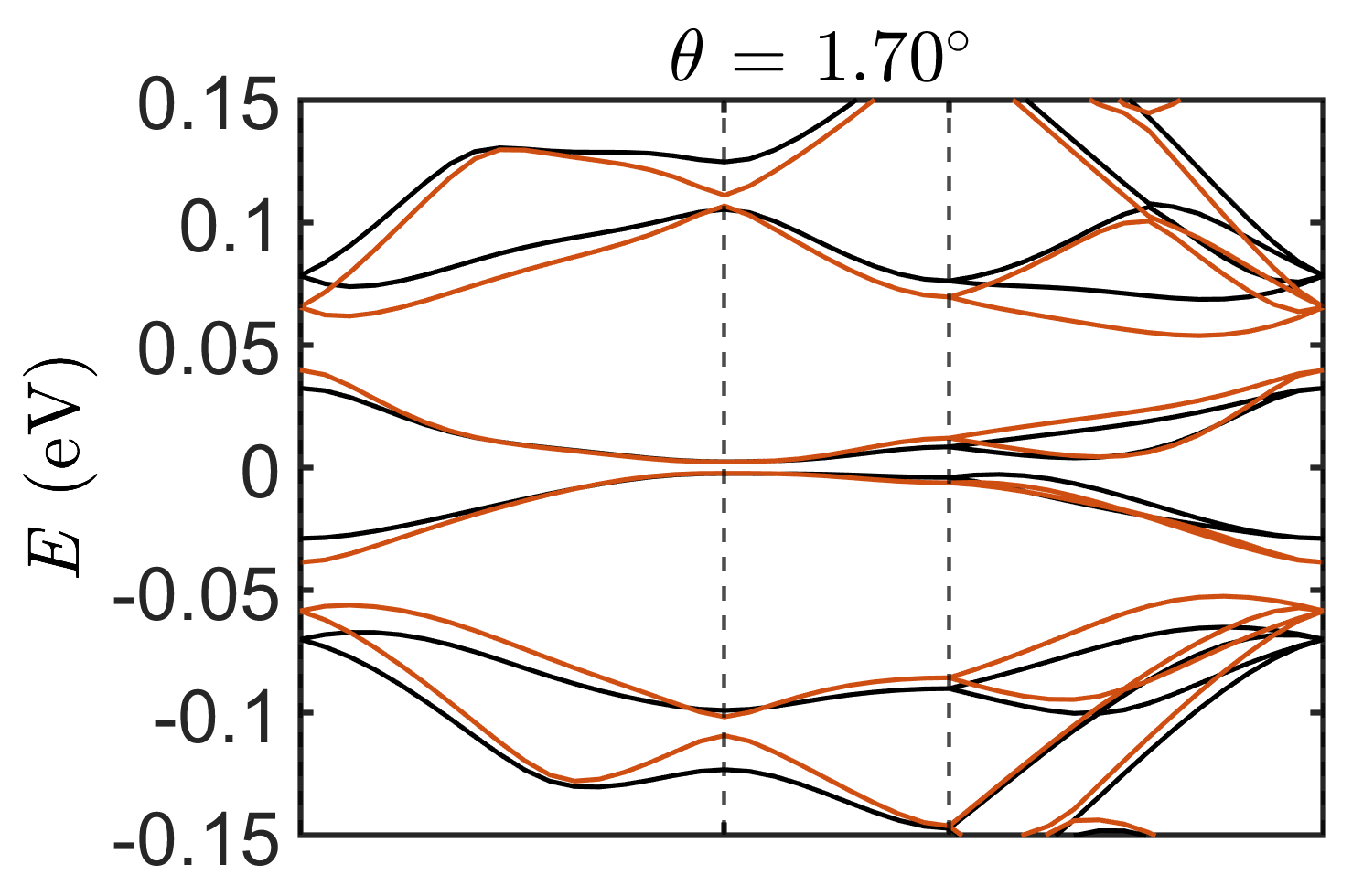}
\end{subfigure}
\begin{subfigure}{0.2989367\textwidth}
  \centering
  \includegraphics[width=1\linewidth]{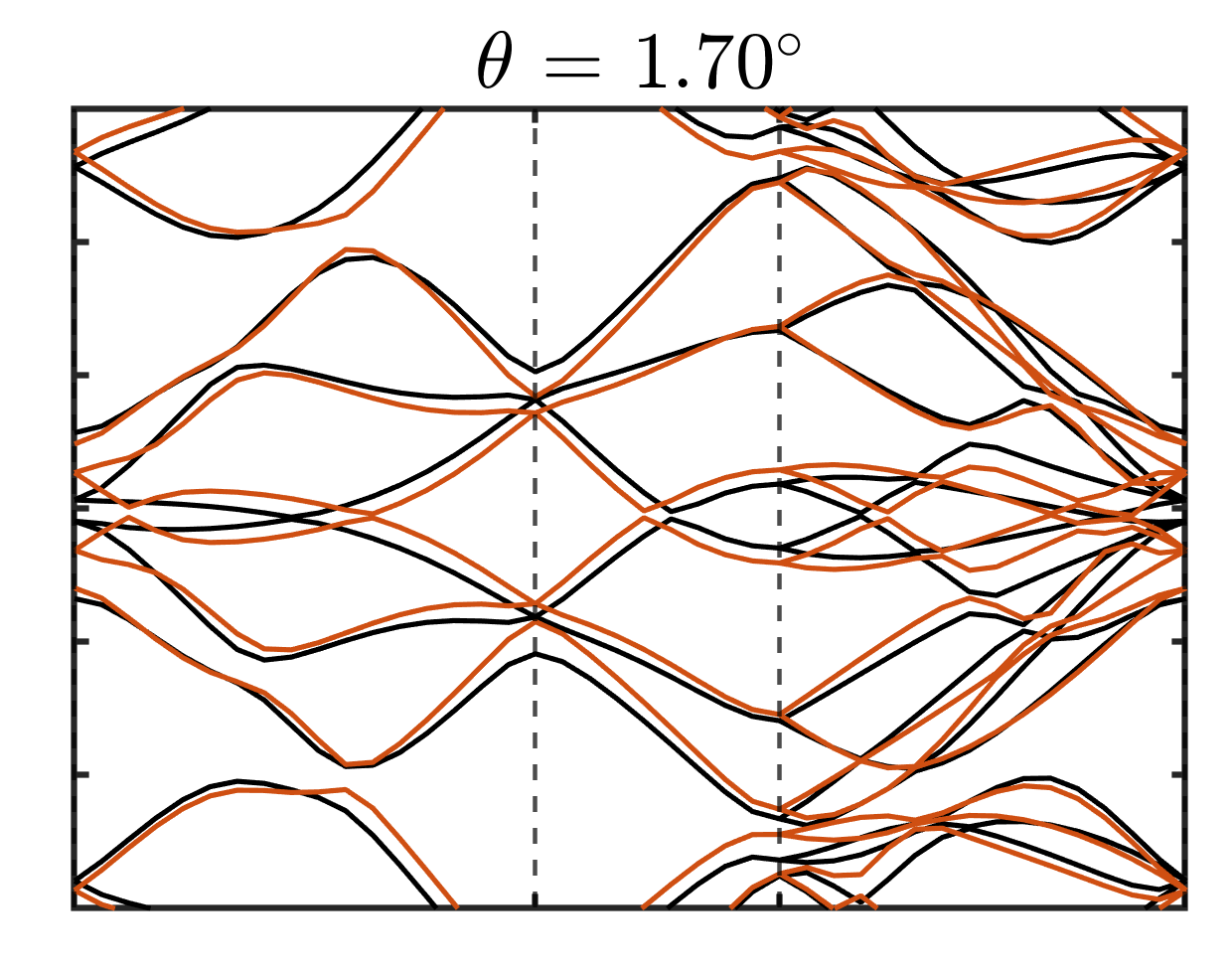}
\end{subfigure}
\begin{subfigure}{0.2989367\textwidth}
  \centering
  \includegraphics[width=1\linewidth]{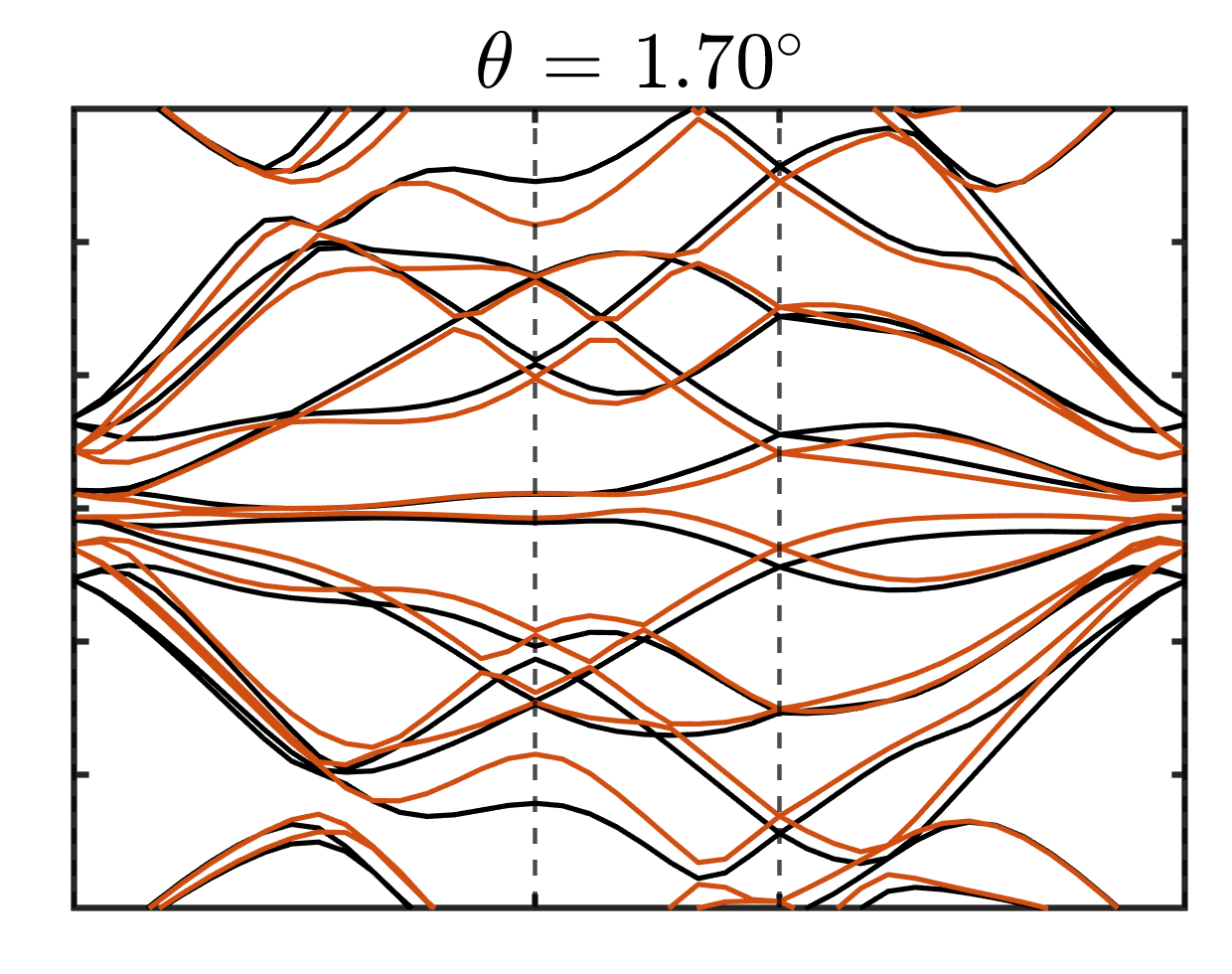}
\end{subfigure}
\begin{subfigure}{0.36212658\textwidth} 
  \centering
  \includegraphics[width=1\linewidth]{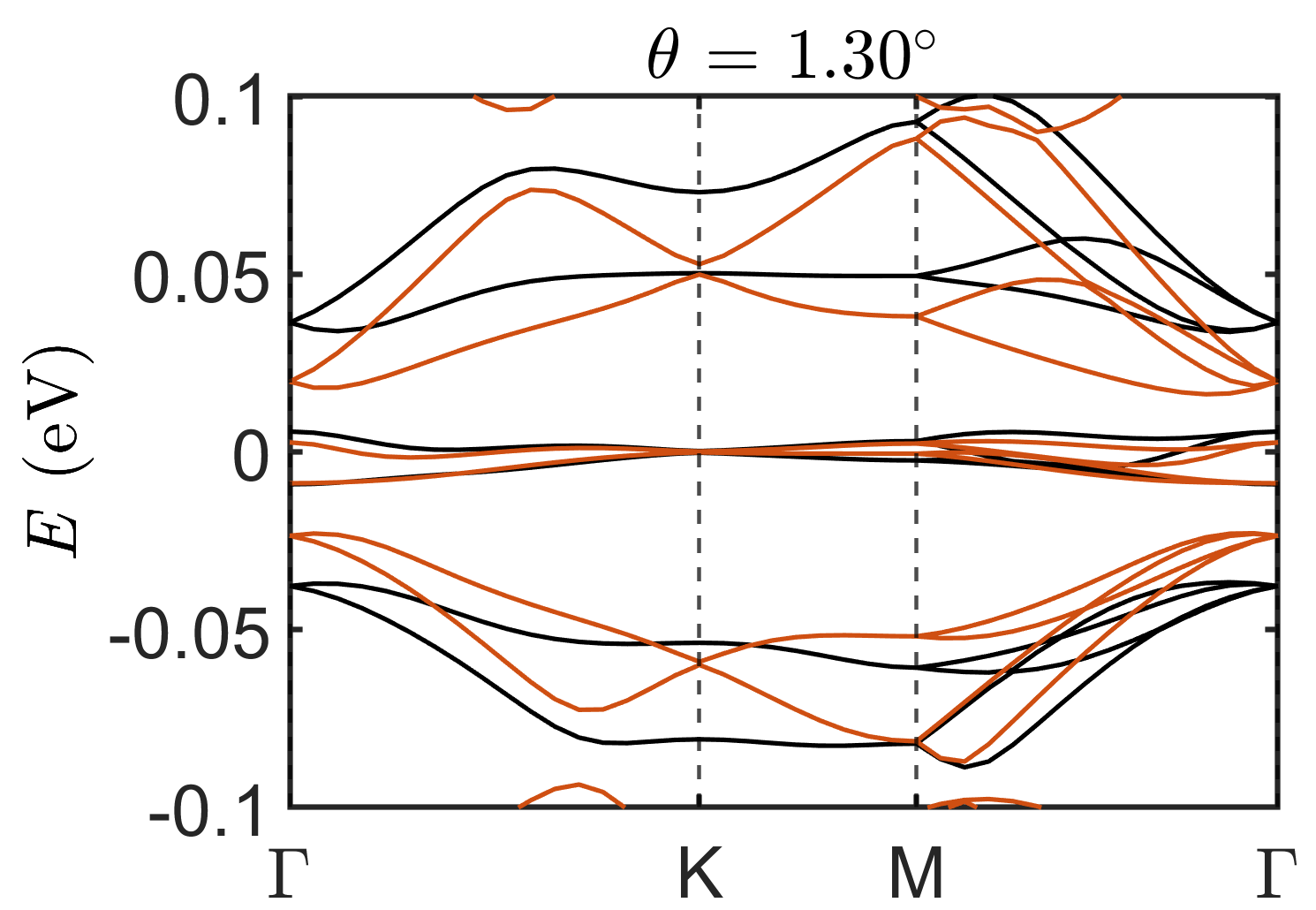}
\end{subfigure}
\begin{subfigure}{0.2989367\textwidth}
  \centering
  \includegraphics[width=1\linewidth]{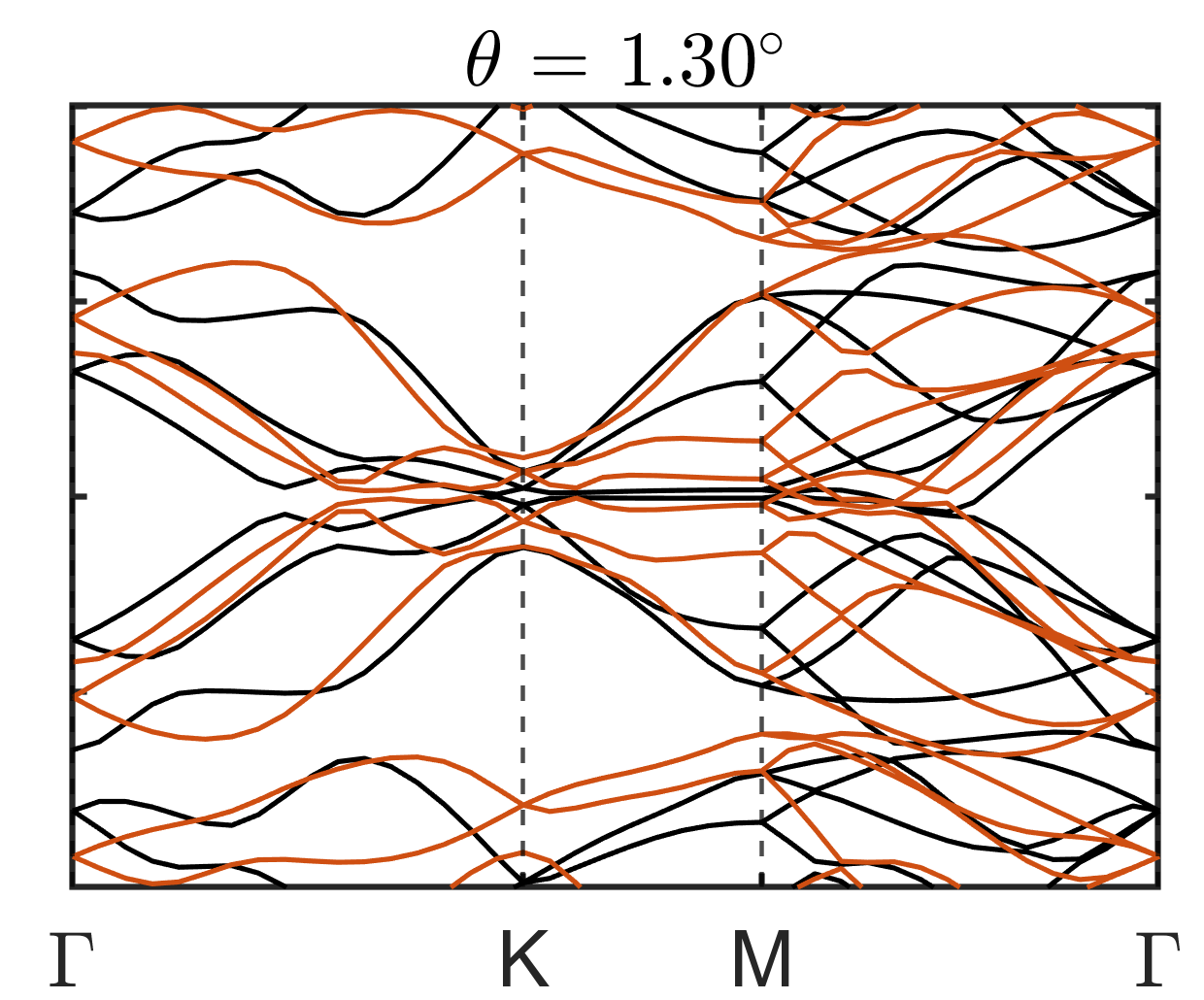}
\end{subfigure}
\begin{subfigure}{0.2989367\textwidth}
  \centering
  \includegraphics[width=1\linewidth]{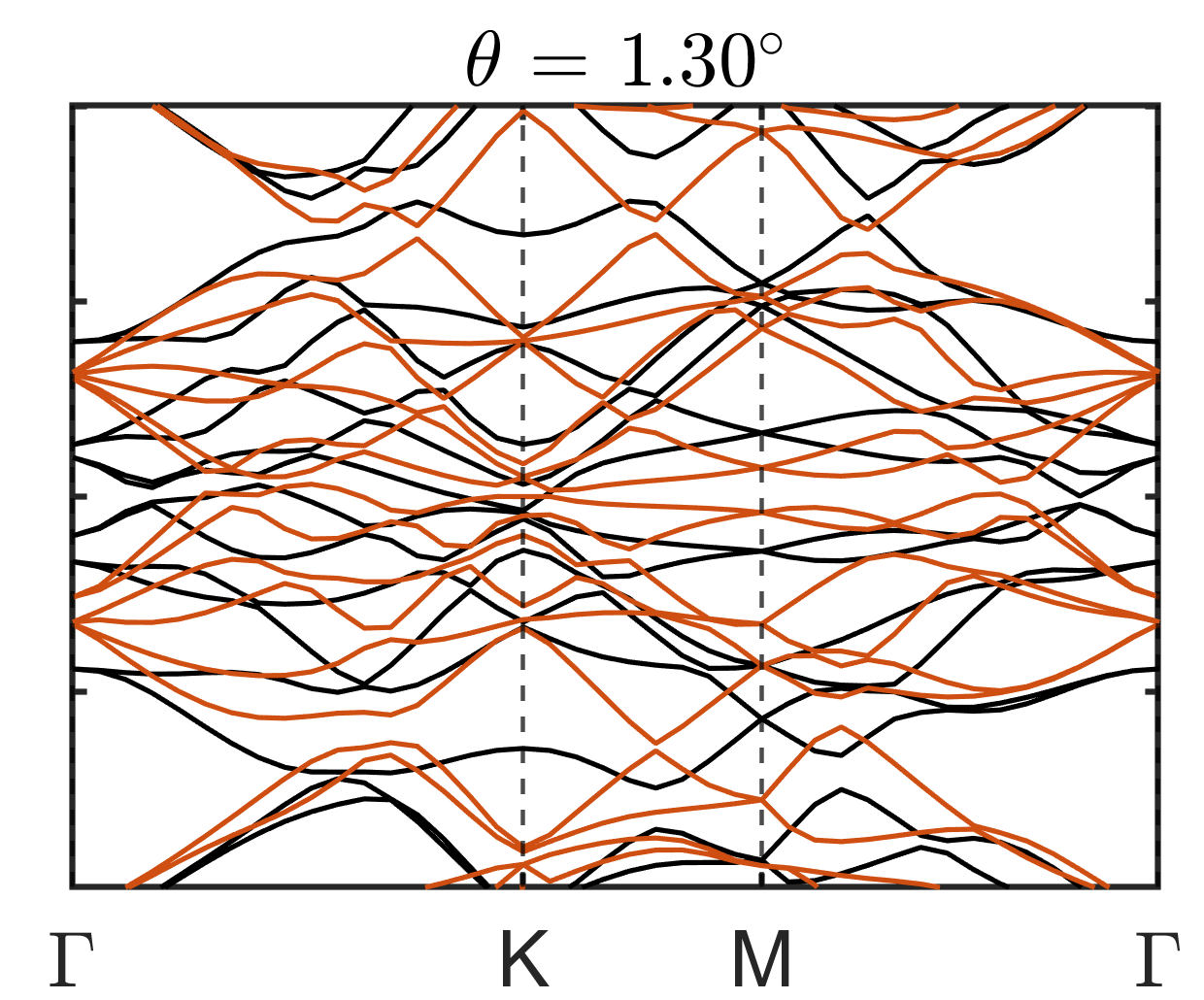}
\end{subfigure}
\caption{Comparison of band structures of fully relaxed (black) AB/AB, AA/AA and AB/AA tDBLG with systems that only include out-of-plane relaxations and no in-plane relaxations (dark orange). The Fermi energy of the undoped system is set to zero. Including only out-of-plane relaxations is sufficient to reproduce the band structures of the fully-relaxed systems at large twist angles. At small twist angles, however, significant differences are observed, demonstrating the importance of in-plane relaxations.}
\label{RelaxEffect}
\end{figure*}

\end{document}